\documentclass[letterpaper, 11pt]{article}

\usepackage{mathtools}
\usepackage{array}
\usepackage{enumitem}
\usepackage{amsmath,amsfonts,amssymb,bm}
\usepackage{titlesec}
\usepackage{pgfgantt,rotating}
\usepackage{float}
\usepackage{graphicx}
\usepackage{sidecap}
\usepackage{wrapfig}
\usepackage{bold-extra}
\usepackage{authblk}
\usepackage{appendix}
\usepackage{makecell}
\usepackage{placeins}
\usepackage{dirtytalk}
\usepackage{lscape}
\usepackage{booktabs}
\usepackage{cite}

\usepackage[margin=1in]{geometry}
\newcommand{\noin}{\noindent}
\newcommand{\beq}{\begin{equation}}
\newcommand{\eeq}{\end{equation}}
\newcommand{\bgqar}{\begin{eqnarray}}
\newcommand{\enqar}{\end{eqnarray}}
\newcommand{\bgqarn}{\begin{eqnarray*}}
\newcommand{\enqarn}{\end{eqnarray*}}
\newcommand{\bgary}{\begin{array}}
\newcommand{\enary}{\end{array}}
\newcommand{\df}{:=}
\newcommand{\bld}[1]{\mbox{\boldmath $#1$}}

\graphicspath{{figures/}}

\title{Time-varying Identification of Guided Wave Propagation under Varying Temperature via Non-Stationary Time Series Models}

\author{Shabbir Ahmed}
\author{Fotis Kopsaftopoulos\footnote{Corresponding author.}}

\affil{\small Intelligent Structural Systems Laboratory (ISSL) \\ Department of Mechanical, Aerospace and Nuclear Engineering \\ Rensselaer Polytechnic Institute, Troy, NY, USA \\ Email: \{ahmeds6,kopsaf\}@rpi.edu }

\date{\today}

\begin{document}

\maketitle


\begin{abstract}

Modern-day civil, mechanical, and aeronautical structures are transitioning towards a continuous, online, and automated maintenance paradigm in order to ensure increased safety and reliability. The field of structural health monitoring (SHM) is playing a key role in this respect and active sensing acousto-ultrasound guided-wave based SHM techniques have shown great promise due to their potential sensitivity to small changes in the structure. However, the methods' robustness and diagnosis capability become limited in the presence of environmental and operational variability such as changing temperature. In order to circumvent this difficulty, in this paper, a novel stochastic time series-based framework was adopted to model guided wave propagation under varying temperatures. Different stochastic time-varying time series models, such as Recursive Maximum Likelihood Time-varying Auto-Regressive (RML-TAR) and Recursive Maximum Likelihood Time-varying Auto-Regressive with Exogenous Excitation (RML-TARX) models are put forward to model and capture the underlying dynamics of guided wave propagation under varying temperatures. The steps and facets of the identification procedure are presented and clearly explained. Then the identified models are used to perform one-step-ahead prediction as well as ``simulation" of the guided wave signals. In order to gain insight from a physics perspective, high-fidelity finite element (FE) models were also established to model the effect of temperature variation on guided wave propagation. Finally, surrogate models are formulated through the use of stochastic time-dependent RML-TARX models and compared with the FE models under varying temperatures. It is shown that by using only a limited number of data sets and interpolating the model parameters, it is possible to ``simulate" guided wave signals for a wide range of temperatures. It is concluded that RML-TARX- based time series models can be used as a surrogate for FEM to simulate guided wave signals under varying temperatures, and therefore, saving a significant amount of computational time.

\end{abstract}


\clearpage

\section*{Important conventions and symbols}

\noin Definition is indicated by $\df$. Matrix transposition is indicated by the superscript $T$.

\noin Bold-face upper/lower case symbols designate matrix/column-vector quantities, respectively.
 
\noin A functional argument in parentheses designates function of a real variable; for instance $P(x)$ is a function of the real variable $x$.

\noin A functional argument in brackets designates function of an integer variable; for instance $x[t]$ is a function of normalized discrete time $(t=1,2,\ldots)$. The conversion from discrete normalized time to analog time is based on $(t-1)T_s$, with $T_s$ designating the sampling period.

\noin A hat designates estimator/estimate; for instance $\widehat{\bld{\theta}}$ is an estimator/estimate of $\bld{\theta}$.


\section*{Acronyms}

\noin\begin{tabular}{lcl} 
ACF  & : & Autocorrelation function  \\
AIC  & : & Akaike information criteria \\
AR   & : & Autoregressive  \\
ARMA & : & Autoregressive moving average \\
ARMAX  & : & Autoregressive moving average with exogenous excitation  \\

BIC  & : & Bayesian information criterion  \\
BSS  & : & Baseline signal stretch  \\
CCF  & : & Cross-correlation function  \\
EOC  & : &  Environmental and operational conditions  \\
ESS  & : & Error sum of squares  \\
FEM  & : & Finite element method \\
FRF  & : & Frequency response function  \\
iid  & : & identically independently distributed  \\
MA   & : & Moving average \\
OBS  & : &  Optimal baseline selection  \\
OLS  & : & Ordinary least squares  \\
PCA  & : & Principal component analysis  \\
PEM   & : & Prediction error method  \\
PSD    & : & Power spectral density \\
PZT  & : & Lead zirconate titanate \\
RML  & : & Recursive maximum likelihood  \\
RSS  & : & Residual sum of squares  \\
SEM  & : &  Spectral element method  \\
SHM  & : & Structural health monitoring  \\
SSS  & : & Signal sum of squares  \\
STFT  & : &  Short-time Fourier transform  \\
SVD  & : & Singular value decomposition  \\
TAR  & : &  Time-varying autoregressive  \\
TARMA  & : &  Time-varying autoregressive moving average  \\
TARX  & : &  Time-varying autoregressive with exogenous excitation  \\
TARMAX  & : &  Time-varying autoregressive moving average with exogenous excitation  \\
ToF  & : &  Time of flight  \\
WLS  & : & Weighted least squares  \\
X    & : & Exogenous

\end{tabular} 

\newpage\pagebreak 

\tableofcontents 


\section{Introduction} \label{sec:intro}

Ultrasonic guided waves (GW) are widely used in the field of structural health monitoring (SHM), which pertains to diagnosing damage within a structure with the help of permanently installed sensors/actuators in the structure \cite{Ihn-Chang04a,Ihn-Chang04b}. The choice of using GWs in SHM is motivated by the fact that they are easy to generate and detect within thin structures, hollow tubes, etc. with the help of piezo-electric transducers, can travel long distance without negligible attenuation, sensitive to minor changes in structural features and material properties, can access easily remote and hard-to-reach areas. In addition, GW-based SHM offers the possibility of monitoring service-induced damage initiation and subsequent damage progression in near real-time. This has been made possible by the advancement of required hardware development and the data interpretation and modeling techniques for the health management of a structure. This ability to diagnose damage in near real-time is causing a paradigm shift in the current structural maintenance practices from rigid scheduled-based maintenance towards online, automated condition-based maintenance, with a significant saving in both the operating and maintenance cost. 

However, GWs are susceptible to changes in environmental and operating conditions (EOC) such as a change in temperatures, loading and boundary effects, humidity, variation in material properties, etc. \cite{roy2015load,roy2014novel,ahmed2019uncertainty,Janapati-etal16}. The presence of EOC affects the guided wave propagation, providing a false indication of the presence of damage. Among the EOCs, the effect of temperature on guided waves is well investigated and it changes both the amplitude and phase of the propagating signal \cite{liu2015robust,mujica2020considering,cross2012features,el2018sparse,qiu2019enhanced,fendzi2016data,ren2019gaussian}. This has been recognized as a major research challenge and the SHM community is addressing the problem by proposing different temperature compensation techniques based on experimentally recorded sensor data. However practical implementation of the data-driven temperature compensation strategies may suffer from the requirements of collecting a lot of sensor data from baseline conditions of a structure. Additionally, in order to apply the concept of digital twin in the context of SHM, where the physical system would be controlled and monitored by a computational model, it is thus necessary to develop approaches so that the knowledge of underlying physical aspects of a system can be coupled with appropriate system identification models. 

When building a mathematical model of a physical system (or response signal), one may follow the path of analytical approach (physics-based), where the laws of physics are applied to describe the dynamic behavior of the system or may stick to the ``instrumental" models (data-based), where the form of the model is to some extent arbitrary and parameters are determined from experimental data using statistical procedures \cite{akaike1977entropy,akaike1978newer}. The coefficients of a ``physical" model usually have a clear physical meaning as they are functions of system parameters such as mass, density, resistance, or environmental parameters such as temperature, coefficient of friction, etc. On the other hand, instrumental models are not phenomenologically justified and therefore, their coefficients have no physical significance \cite{niedzwiecki2000identification}. However, due to their relative simplicity, they allow one to formulate and solve problems in a mathematically tractable way.    

Guided waves, also known as Lamb waves, are elastic waves and propagate within thin structures. The first analytical solution of guided wave propagation was proposed by Horace Lamb, where he assumed the geometry of wave propagation as a thin infinite plate whose surfaces are free of stress \cite{lamb1917waves}. He mentioned that for this kind of wave to propagate in solid media, the particle motion takes place in two dimensions and the waves are guided by the two stress-free surfaces. In the limiting case, when the wavelength becomes very short (at high frequencies), the solution converges to that of the Rayleigh waves or the surface waves. On the other hand, when the wavelength is large compared to the thickness of the plate, the solution converges to that of ordinary compressional waves. The first experimental proof on the existence of guided waves was provided by Worlton \cite{worlton1961experimental}. There are two types of propagating guided wave modes within a thin plate: one is called the symmetric mode and the another is called the anti-symmetric mode \cite{graff2012wave,kundu2019mechanics,giurgiutiu2007structural}. Again, the symmetric and anti-symmetric modes can manifest themselves in an infinite number of ways. However, for the purpose of damage diagnosis, only the first symmetric and anti-symmetric modes are used referred to as the $S_0$ and $A_0$ mode, respectively.  

Apart from thin plates, guided waves can also exist in hollow cylinders and pipes and an analytical solution was provided by \cite{barshinger2004guided,silk1979propagation}. 
For complex geometries, the analytical solutions cease to exist and numerical modeling of guided wave propagation is performed by finite element methods (FEM)  \cite{kumar2006finite} or spectral element methods (SEM) \cite{lonkar2014modeling,roy2014novel}. However, only the first two modes, the $S_0$ and $A_0$ modes have been properly modeled with these methods \cite{ostachowicz2010wave,hayashi2000sequence}. The rest of the signal is difficult to model and does not properly match with the experimental signal as there are reflections from the boundary. In order to understand the effect of temperature on guided wave propagation, a number of studies were performed showing both experimental and numerical results \cite{raghavan2008effects,lanza2008temperature,clarke2009evaluation}. However, a typical guided wave response signal (piezo-sensor response) is affected not only by the changes in the mechanical properties of the base substrate but also by the variation in material properties of the PZT transducer and interfacial adhesive layer. Ha et al. \cite{ha2010adhesive} carried out numerical simulation studies based on SEM to understand the effect of the adhesive layer on the piezo-sensor response at elevated temperatures. The paper concluded that adhesive type and thickness along with piezoelectric and base substrate material play critical roles in determining the sensor response at elevated temperatures.

Upon gaining insight into the nature of temperature variation on guided wave propagation, both physics-based and data-based temperature compensation techniques were proposed. Roy et al. \cite{roy2014novel} proposed physics-based temperature compensation techniques using analytical as well as numerical techniques. A linear system identification model was used to relate the changes in the signal projection coefficients to the changes in the material properties affected by temperature variation. Andrews et al. \cite{andrews2008lamb} have proposed an analytical approach based on the decomposition of the frequency component of the signal to reconstruct the first fundamental symmetric mode ($S_0$ mode). Among the data-based temperature compensation techniques, baseline modification methods such as optimal baseline selection (OBS)\cite{croxford2007strategies,konstantinidis2007investigation}, baseline signal stretch (BSS)\cite{clarke2009evaluation}  and their variants \cite{putkis2013continuous,herdovics2019compensation,mariani2020compensation}, where the baseline signal is selected, modified or stretched in such a way so that the effect of temperature change is minimized in the scatter signal are prominent. Apart from baseline modification methods, projection-based methods such as singular value decomposition (SVD)\cite{liu2015robust}, where the effect of damage and temperature are separated into different singular vectors; principal component analysis (PCA) \cite{mujica2020considering,cross2012features}, where the signal matrix is projected onto a lower-dimensional space and the effect of temperature and damage is separated by orthogonal distance (OD); sparse estimation-based method\cite{el2018sparse}, where the current signal is estimated based on a few reference signal with similar temperature (EOC), and the estimation error is used as an indication of the presence of damage are also noteworthy for the compensation of temperature. Harley et al.\cite{harley2012scale} used advanced signal processing techniques to compensate for the effect of temperature. Other data-based temperature compensation techniques can be found in references \cite{qiu2019enhanced,fendzi2016data,ren2019gaussian}.

However, these methods achieve partial compensation of temperature, and only the $S_0$ or $A_0$ mode of guided wave signal can be compensated. Compensation of temperature for guided wave signal beyond $A_0$ mode is an extremely challenging task using the physics-based model as guided waves are dispersive in nature, that is, their frequency content change with time. From a data-based perspective, these are non-stationary signals which are difficult to model. Over a short time interval, most of the processes or systems can be satisfactorily approximated by linear dynamic time-invariant models, but over longer time intervals, they reveal non-stationary features or characteristics. Hence, they call for models with time-varying coefficients.  

In order to tackle these challenges, system identification techniques employing stochastic time-varying (non-stationary) time series models appear as a promising approach to study the stochastic and non-stationary nature of guided wave propagation. In the context of the vibration-based damage diagnosis process, principles of system identification have been widely used to detect structural damage via both stationary and non-stationary time series models \cite{spiridonakos2013fs,kopsaftopoulos2013functional,fassois2007time,kopsaftopoulos2010vibration,spiridonakos2014non}. These methods are data-based rather than physics-based and fundamentally of the inverse type. Being statistical in nature, they may offer the advantage of assimilating inherent uncertainty and obviate the need for formulating detailed physics-based F/SEM models. Unlike physics-based methods, they are capable of capturing different uncertainties without having recourse to subjectively assuming the uncertainty in structural parameters. The available methods of signal analysis may be broadly classified as non-parametric or parametric.  Non-parametric model structures are characterized by the property that the resulting models are curves or functions that do not explicitly employ a finite-dimensional parameter vector. Non-parametric methods include time-domain models such as the auto-correlation (ACF) and cross-correlation (CCF) \cite{kay1993fundamentals,box2011time} functions as well as frequency-domain models, such as the power spectral density (PSD) and the frequency response function (FRF) \cite{kay1988modern}. The above-mentioned non-parametric methods provide information either on the time or frequency domain. They do not provide any information on how time and frequency changes simultaneously. Time-frequency analysis provides a suitable means to analyze non-stationary signals in the time and frequency domain simultaneously. Such methods include the widely used spectrogram based on the short-time Fourier transform (STFT) \cite{hammond1996analysis} and its ramifications, distributions such as the Wigner-Ville and Choi-Williams that are unified under the Cohen class of distributions \cite{cohen1995time}.  On the other hand, parametric model structures are parameterized in terms of a parameter vector to be estimated from the available signals. Parametric model structure can be time-invariant as well as time-varying. Time-varying parametric methods are based upon parameterized representations of the time-dependent auto-regressive moving average (TARMA) or related types and their extensions. These representations differ from their conventional, stationary, counterparts in that their parameters are time-dependent \cite{poulimenos2006parametric,spiridonakos2014non,petsounis2000non,owen2001application,grenier1989parametric}. The methods based upon them are known to offer a number of potential advantages \cite{mrad1998polynomial,conforto1999spectral,owen2001application} such as representation parsimony, as models may be potentially specified by a limited number of parameters, improved accuracy, resolution, and tracking of the time-varying dynamics, flexibility in analysis, as parametric methods are capable of directly capturing the underlying structural dynamics responsible for the non-stationary behavior, and flexibility in synthesis (simulation) and prediction. Parametric time-varying methods may be further classified according to the type of structure imposed upon the evolution of the time-varying model parameters, namely: (i) the class of unstructured parameter evolution methods\cite{sotiriou2016adaptive,draper1998applied}, which impose no particular structure on the evolution of time-varying model parameters; (ii) the class of stochastic parameter evolution methods\cite{kitagawa1985smoothness,gersch1985time}, which impose \say{stochastic} structure on the evolution of the time-varying parameters; (iii) the class of deterministic parameter evolution methods\cite{fouskitakis2001estimation,spiridonakos2014non,poulimenos2009output}, which impose deterministic structure upon the evolution of time-varying parameters. From the above discussion on the literature of temperature compensation of guided waves for damage detection and a brief overview of time-varying models, it is clear that those models have not been used in the context of guided wave-based modeling and SHM, although they seem to offer a number of important characteristics. 

Therefore, in this work, in order to address the effect of temperature on guided wave propagation and to devise an effective and robust temperature compensation framework for the complete guided wave signal, i.e. the $S_0$, $A_0$, and the reflection part of the signal, a novel stochastic modeling and analysis framework is proposed for the effective representation of the non-stationary guided wave propagation signals via the use of time-varying autoregressive (TAR) and time-varying autoregressive with exogenous excitation (TARX) models. A series of laboratory experiments on an aluminum plate inside an environmental chamber outfitted with a network of piezoelectric sensors was performed under varying temperatures so that the effects of temperature could be reflected in the recorded signals and modeled properly. In addition, high fidelity FEM simulations were performed to investigate the effect of temperature variation on guided wave propagation and as a means to verify the proposed surrogate modeling approach. In the context of non-stationary signals, the model's predictive capability and simulation performance under varying temperatures using a single model structure have been presented and discussed. Model-based time-dependent frequency response function (FRF) and the effect of temperature variation on time-varying modal parameters have also been presented. The main contributions of the paper are:   



%
\begin{itemize}
\item[(a)] Investigate the stochastic nature of guided wave propagation in an aluminum plate inside an environmental chamber for a range of different temperatures by means of experimental investigation. Finite element modeling of guided wave propagation under varying temperatures was also performed to gain insight from the physics perspective; 

\item[(b)] Introduction of a novel data-based stochastic time-varying identification framework for tracking the dynamics of non-stationary guided wave signals under varying temperatures;  
\item[(c)] Perform mathematical modeling for one-step-ahead prediction of guided wave propagation signals under varying temperatures using a single RML-TAR model;

\item[(d)] Perform mathematical modeling for the simulation of guided wave propagation signals under varying temperatures using RML-TARX model;

\item[(e)] Using only a limited number of data sets at some specific temperatures, show that guided waves can be simulated at temperatures where data is not available;

\item[(f)] To show that RML-TARX based model can be used as a surrogate for the FEM model for the simulation of guided wave propagation under varying temperatures.

\end{itemize}

The remainder of the paper is organized as follows: non-stationary signal representations with the help of RML-TAR and RML-TARX are presented in Section 2. The theory and the steps of the identification procedure such as model parameter estimation and model structure selection are discussed in Section 3. The process of data generation with the help of experimental setup and finite element model is discussed in Section 4. Model-based analysis of guided wave propagation under varying temperatures, mainly in terms of the time-dependent frequency response function and modal parameters are discussed in Section 5. Finally, concluding remarks are summarized in Section 6.

\section{Stochastic Identification of Guided Wave Propagation} \label{sec:TARX-models}

The proper parametric representation of time-varying wave propagation requires the use of appropriate non-stationary stochastic model structures.  For a systematic review of both non-parametric and parametric non-stationary models, the reader is referred to \cite{poulimenos2006parametric}. Typical models are of the TARMA, i.e. time-varying autoregressive moving average type, or proper extensions (for instance TARMAX representations - that is TARMA representations with exogenous excitations, which additionally account for measurable excitation \cite{poulimenos2006parametric}). In this study, the parametric time-varying autoregressive (TAR) model structure has been used to represent guided wave propagation signals under varying temperatures due to its simple but effective form and parameter estimation method. However, for the ``simulation" of guided wave signals under varying temperatures and to formulate a surrogate models for FEM, the time-varying autoregressive with exogenous excitation (TARX) model has been used. TARX models are parameterized representations of their conventional, stationary ARX counterparts with the significant difference being that they allow their parameters to depend upon time. A recursive maximum likelihood time-varying auto-regressive with exogenous excitation model ( RML-TARX$(na,nb)$), with $na$ and $nb$ designating its autoregressive (AR) and exogenous excitation (X) order, respectively, is thus of the form:
\begin{equation}
    y[t] + \sum_{i=1}^{na} a_i[t] \cdot y[t-i] = \sum_{i=0}^{nb} b_i[t] \cdot x[t-i] + e[t] 
    \qquad e[t] \sim \, \mbox{iid} \, \mathcal{N} \bigl( 0,\sigma^2_e[t] \bigr) \label{eq:ar-model} 
\end{equation}
with $t$ designating its discrete time, $y[t]$ the time-varying signal to be modeled, $x[t]$ the actuation signal, $e[t]$ an (unobservable) uncorrelated (white) innovations sequence with zero mean and time dependent variance $\sigma^2_e[t]$ that generates $y[t]$ and $a_i[t]$, the model's time dependent $AR$ parameters. $\mathcal{N} (\cdot , \cdot)$ stands for normal distribution with the indicated mean and variance \cite{poulimenos2006parametric,sotiriou2016adaptive,Ljung99}.

It can be shown that the minimum mean square error (MMSE) one-step-ahead prediction $\widehat{y}[t/t-1]$ of the signal value $y[t]$ made at time $t-1$ (that is for given values of the signal up to time t-1) is \footnote{note that the hat more generally designates estimate/estimator of the indicated quantity.}:
\begin{equation}
     \widehat{y}[t/t-1] =- \sum_{i=1}^{na} a_i[t] \cdot y[t-i] + \sum_{i=0}^{nb} b_i[t] \cdot x[t-i] \label{eq:prediction} 
\end{equation}
Comparing this with RML-TARX model of Equation (\ref{eq:ar-model}), it is evident that the one-step ahead prediction error is equal to e[t], that is:
\begin{equation}
     \widehat{e}[t/t-1] \triangleq y[t] - \widehat{y}[t/t-1] = e[t]  \label{eq:prediction error} 
\end{equation}
This is an important observation, as it indicates that, just as in the stationary case, the model's one-step-ahead prediction error (also referred to as the ``residual") coincides with the (uncorrelated) innovations ``generating" the signal. This is valid as long as the true model parameters of Equation (\ref{eq:ar-model}) are used in the predictor equation of Equation (\ref{eq:prediction}).

Using the backshift operator $\mathcal{B} (\mathcal{B}^i \cdot y[t]\triangleq y[t-i], \mathcal{B}^i \cdot x[t]\triangleq x[t-i])$, the TARX representation of Equation (\ref{eq:ar-model}) is compactly re-written as: 
\begin{equation}
    y[t] + \sum_{i=1}^{na} a_i[t] \cdot \mathcal{B}^i \cdot y[t] = \sum_{i=0}^{nb} b_i[t] \cdot \mathcal{B}^i \cdot x[t] + e[t] 
     \label{eq:ar-model compact} 
\end{equation}
%
\begin{equation}
       A[\mathcal{B},t] \cdot y[t] = B[\mathcal{B},t] \cdot x[t] + e[t] \label{eq:ar-model compact 2}  
\end{equation}
 with
 \begin{equation*}
       A[\mathcal{B},t] = 1 + \sum_{i=1}^{na} a_i[t] \cdot \mathcal{B}^i \qquad  B[\mathcal{B},t] = \sum_{i=0}^{nb} b_i[t] \cdot \mathcal{B}^i \label{eq:ar-model compact 3}  
\end{equation*}

The polynomial $A[\mathcal{B},t]$, $B[\mathcal{B},t]$ is referred to as the $AR$ and $X$ time dependent polynomial operator, respectively. Also note that one may also define $a_0[t]=1$

\section{Non-stationary TARX model identification}

Given a single, $N$- sample long, non-stationary signal record (realization) $y^N \triangleq \{y[1], y[2], \cdots y[N]\}$ and the actuation signal $x^N \triangleq \{x[1], x[2], \cdots x[N]\}$, the TARX identification problem may be stated as the problem of selecting the corresponding model ``structure", the model AR parameters $a_i[t]$, the model X parameters $b_i[t]$, and the innovations variance $\sigma^2_e[t]$ that ``best" fit the available measurements. Model ``fitness" may be understood in various ways, a common approach being in terms of predictive ability. This implies that the ``best" model is the one characterized by minimal one-step-ahead prediction error. The methods based upon this principle minimize a function of the prediction error sequence (typically the residual sum of squares (RSS)) and are referred to as the prediction error methods (PEM).

In the case of TARX model, the model structure selection consists of determining the AR order $na$ and X order $nb$. More formally, the identification problem may be defined as the selection of the best fitting model from the set $\mathcal{G}$ of TARX model corresponding to a particular class

\begin{equation}
     \mathcal{G} \triangleq \{ \mathcal{M}(\bld{\theta}^N,(\sigma^2_e)^N):  y[t] + \sum_{i=1}^{na} a_i[t,\bld{\theta}[t]] \cdot y[t-i] = \sum_{i=0}^{nb} b_i[t,\bld{\theta}[t]] \cdot x[t-i] + e[t,\bld{\theta}^t]  \label{eq:model}  
\end{equation}

\begin{equation*}
    \sigma^2_e[t,\bld{\theta}^t] = E\{e^2[t,\bld{\theta}^t]\} ;  t=1,2,...N\} 
\end{equation*}

In this expression, $E\{\cdot\}$ designates statistical expectation and $\bld{\theta}[t]$ designates the instantaneous AR and X parameter vector at $t$ time instant \footnote{Bold-face upper/lower case symbols designate matrix/column-vector quantities, respectively; matrix transposition is designated by the superscript $T$}:

\begin{equation}
    \bld{\theta}[t] = [a_1[t]  \; a_2[t]  \; \ldots \;  a_{na}[t] \; \vdots \; b_0[t]  \; b_1[t]  \; \ldots \;  b_{nb}[t] ]_{(na+nb) \times 1}^T
\end{equation}
While $\bld{\theta}^t$ stands for all AR and X parameters up to time $t$, that is,

\begin{equation}
    \bld{\theta}^t \triangleq [\bld{\theta}^T[1]  \;\bld{\theta}^T[2] \; \bld{\theta}^T[3] \; \ldots \bld{\theta}^T[t] ]_{(na+nb)t \times 1}^T
\end{equation}
Moreover, $\bld{\theta}^N$ designates all AR and X parameters at all time instants, and, similarly, $(\sigma^2_e)^N$ the residual variance at all time instants:

\begin{equation}
    (\sigma^2_e)^N \triangleq \{ \sigma^2_e[1], \; \sigma^2_e[2], \;  \sigma^2_e[3], \;\ldots \sigma^2_e[N]\}
\end{equation}

The TARX model to be estimated is explicitly parameterized in terms of the specific parameters to be estimated. Thus, both the ARX parameters and the one-step-ahead prediction error (residual) signal are designated as functions of these parameters. For the residual signal, in particular, this signifies the fact that it is obtained based upon the current model parameters, the available guided wave signal $y^N$, and the actuation signal $x^N$ (using the TARX model expression in Equation (\ref{eq:model})). Thus, the ``best" model may be selected/estimated as the model with parameters that minimizes the prediction error. With the absence of the actuation signal $x^N$, the model becomes simplified, and the TAR model is used instead of the TARX model.

The model identification problem is usually distinguished into two subproblems: (i) the \textit{parameter estimation} subproblem, and (ii) the \textit{ model structure selection} subproblem. These are separately treated in section \ref{sec: model par est} and \ref{sec: model struc sel}, respectively.

\subsection{Model parameter estimation \label{sec: model par est}}

Model parameter estimation refers to the determination, for a given model structure, of the AR and X parameter vector $\bld{\theta}[t]$ and the residual variance $\sigma^2_e[t]$. For a review of the main estimation methods for non-stationary stochastic time series models please see \cite{poulimenos2006parametric}. In the case of TARX models, recursive or adaptive estimation methods provide an accurate and robust parameter estimate. TARX/TARMAX models estimated recursively are also referred to as RARX/RARMAX models. 

Although there are several variations of recursive methods, in the present work we employ a method based on the following exponentially weighted prediction error criterion \cite{poulimenos2006parametric,Ljung99}:
\begin{equation}
     \widehat{\bld{\theta}}[t] = \arg \min_{\bld{\theta}[t]} \sum_{\tau=1}^t \lambda^{t-\tau} \cdot e^2[\tau,\bld{\theta}^{\tau-1}],  \label{eq:WLS} 
\end{equation}
with 
\begin{equation}
      e[t,\bld{\theta}^{t-1}] \triangleq y[t] - \sum_{i=1}^{na} a_i[t-1] \cdot y[t-i] + \sum_{i=0}^{nb} b_i[t-1] \cdot x[t-i] \approx  e[t,\bld{\theta}^{t}]\label{eq:RAR-resid}  
\end{equation}

In these expressions arg min designates minimizing argument, and $e[t,\bld{\theta}^{t-1}]$ the model's one-step-ahead prediction error made at time $t-1$ without knowing the model parameter values at time $t$ as it would be normally necessary. Of course, as indicated in the expression above, $e[t,\bld{\theta}^{t-1}] \approx e[t,\bld{\theta}^{t}]$ for slow parameter evolution. The term $\lambda^{t-\tau}$ is a window or weighting function that assigns more weight to more recent errors where $\lambda \in (0,1)$ \cite[pp. 378-379]{Ljung99}. The quantity $\lambda$ is referred to as the ``forgetting factor". The smaller the value of $\lambda$, the faster older values of the error (and thus of the signal) are forgotten, thus increasing the estimator's adaptability, i.e. its ability to track the evolution of the dynamics.

The recursive estimation of $\bld{\theta}[t]$ based upon the above criterion is accomplished via the recursive maximum likelihood (least squares) method \cite{poulimenos2006parametric,Ljung99}, thus the models are designated as RML-TAR$(na)_{\lambda}$ for the TAR case, and RML-TARX$(na,nb)_{\lambda}$ for the TARX case. The estimator update at time $t$ based on the previous time instant $t-1$ is given as:

\begin{equation}
    \widehat{\bld{\theta}}[t] = \widehat{\bld{\theta}}[t-1] + \bld{k}[t] \cdot \widehat{e}[t|t-1]
\end{equation}
with $\widehat{e}[t|t-1]$ designating the prediction error:

\begin{equation}
    \widehat{e}[t|t-1] = y[t]- \widehat{y}[t|t-1] = y[t] - \bld{\phi}^T[t] \cdot \widehat{\bld{\theta}}[t-1]
\end{equation}
and \bld{k}[t] the adaptation gain:

\begin{equation}
    \bld{k}[t] = \frac{\bld{P}[t-1] \cdot \bld{\phi}[t] }{\lambda + \bld{\phi}^T[t] \cdot \bld{P}[t-1] \cdot \bld{\phi}[t]}
    \label{eq: gain}
\end{equation}
In the Equation (\ref{eq: gain}), $\bld{P}[t]$ designates the parameter covariance matrix at time $t$ that is updated based on the following recursive form:

\begin{equation}
    \bld{P}[t] =\frac{1}{\lambda}\left[ \bld{P}[t-1] - \frac{\bld{P}[t-1] \cdot \bld{\phi}[t]  \cdot \bld{\phi}^T[t] \cdot \bld{P}[t-1] }{\lambda + \bld{\phi}^T[t] \cdot \bld{P}[t-1] \cdot \bld{\phi}[t]}\right]
\end{equation}
with $\bld{\phi}[t]$ corresponding to the regression matrix of the RML-TARX model, defined as
\begin{equation*}
    \bld{\phi}[t] = \left[ -y[t-1] -y[t-2] \ldots -y[t-na] \; \vdots \; x[t] \; x[t-1] \ldots x[t-(nb+1)] \right]^T.
\end{equation*}
The method may be initiated with a zero parameter vector $\widehat{\bld{\theta}}[0] = \bld{0}$ and unity covariance matrix $ \bld{P}[0] = \alpha\bld{I}$, with $\alpha$ designating a ``large" positive number. The innovations, i.e. one-step-ahead prediction error, time-varying variance $\sigma^2_e[t]$ may be estimated via a window of length $2M+1$ centered at each time instant $t$ as follows:

\begin{equation}
\widehat{\sigma}_e^2[t] =\frac{1}{2M+1}\sum_{\tau=t-M}^{t+M}\widehat{e}^2[\tau/\tau-1]
\label{eq: variance}
\end{equation}   

\subsection{Model structure selection \label{sec: model struc sel}}

Model structure selection refers to the selection of the TARX model order $na$, $nb$, and the forgetting factor $\lambda$. It is generally based on trial-and-error, or successive fitting schemes\cite{poulimenos2003estimation}, according to which models corresponding to various candidate structures are estimated, and the one providing the best fitness to the non-stationary signal is selected. The fitness function may be the Gaussian log-likelihood function of each candidate model. The particular model that maximizes it is the most likely to be the actual underlying model responsible for the generation of the measured signal, in the sense that it maximizes the probability of having provided the measured signal values, and is thus selected. A problem with this approach is that the log–likelihood may be monotonically increasing with increasing model orders, and as a result, the over fitting of the measured signal occurs. For this reason, criteria such as the AIC (Akaike information criterion\cite{akaike1977entropy}) or the BIC (Bayesian information criterion\cite{schwarz1978estimating}) are generally used and can be represented as follows:

\begin{equation}
    \text{AIC} = -2 \cdot \ln{\mathcal{L}(\mathcal{M}(\bld{\theta}^N,(\sigma^2_e)^N)|x^N, y^N)} + 2\cdot d
\end{equation}
\begin{equation}
\text{BIC} = -\ln{\mathcal{L}(\mathcal{M}(\bld{\theta}^N,(\sigma^2_e)^N)|x^N, y^N)} +\frac{\ln{N}}{2}\cdot d
\end{equation}
with $\mathcal{L}$ designating the model likelihood, $N$ the number of signal samples, and $d$ the number of independently estimated model parameters. As it may be observed, both criteria consist of a superposition of the negative log-likelihood function and a term that penalizes the model order or structural complexity and thus discourages model over fitting. Accordingly, the model that minimizes the AIC or the BIC is selected.
The Gaussian log-likelihood function of the model structure,  $\mathcal{M}(\bld{\theta}^N,(\sigma^2_e)^N)|x^N, y^N)$,  given the signal sample $y^N$ and actuation $x^N$ is given by:

\begin{equation}
    \ln{\mathcal{L}(\mathcal{M}(\bld{\theta}^N,(\sigma^2_e)^N)|x^N,y^N)} = \ln{f(\mathcal{M}(x^N,y^N|\bld{\theta}^N,(\sigma^2_e)^N))}
\end{equation}
$$= \ln{\prod_{t=1}^{N} f(e[t]|\bld{\theta}^N,(\sigma^2_e)^N))}$$

$$= \sum_{t=1}^{N} \ln{\left((2\pi\sigma^2_e[t])^{-\frac{1}{2}}\cdot \exp\{-(\frac{e^2[t]}{2\sigma^2_e[t]})\}\right)}$$

\begin{equation}
     \ln{\mathcal{L}(\mathcal{M}(\bld{\theta}^N,(\sigma^2_e)^N)|x^N,y^N)} = -\frac{N}{2}\cdot\ln{2\pi} - \frac{1}{2}\sum_{t=1}^{N} \left( \ln{\sigma^2_e[t]}+ \frac{e^2[t]}{\sigma^2_e[t]}\right)
\end{equation}
As a result, the BIC equation can be written as:

\begin{equation}
\text{BIC} = -\frac{N}{2}\cdot\ln{2\pi} - \frac{1}{2}\sum_{t=1}^{N} \left( \ln{\sigma^2_e[t]}+ \frac{e^2[t]}{\sigma^2_e[t]}\right)  +\frac{\ln{N}}{2}\cdot d
\end{equation}
The innovations (one-step-ahead prediction error) variance $\sigma^2_e[t]$ may be estimated via Equation (\ref{eq: variance}).

The ratio of the residual sum square versus the signal sum square (RSS/SSS) may also be used as another fitness criteria for the selection of the best model. For determining the best time-varying model, in addition to searching for the best candidate model structure, it is also necessary to search for a suitable ``forgetting factor" for which the minimum RSS/SSS occurs. Again, model order resulting from the least RSS/SSS may provide spurious natural frequencies and damping ratios. As a result, the frozen-time power spectral density for the non-stationary signal may provide additional insight in the selection of the best model structure.

\subsection{Modal characteristics \label{sec:modals}}

The notion of power spectral density (PSD) for stationary case has no direct counterpart for the non-stationary case. The frozen-time PSD of a system may be obtained by utilizing a sequence of ``frozen" stationary systems (corresponding to a specific time instant) for representing the non-stationary system \cite{grenier1989parametric,poulimenos2006parametric}. By analogy to the stationary case, the power spectral density for each time instant may be expressed based on the estimated TARX model as:

\begin{equation}
    S_F(\omega,t) = \left|\frac{\sum_{i=0}^{nb} b_i[t]\cdot e^{-j\omega T_s i}}{1+\sum_{i=1}^{na} a_i[t]\cdot e^{-j\omega T_s i}} \right|^2\cdot \sigma^2_e[t]
\end{equation}
Note that this would be the power spectral density of the response signal if the system were ``frozen" at the time instant $t$. As such, the information conveyed is very useful, because it represents the characteristics that the system would have if it became stationary (``frozen") with a specific configuration (corresponding to the considered time instant $t$).
The normalized TARX model's ``frozen" frequency response function can be given by the following equation: 
\begin{equation}
    H_F(e^{j\omega T_s},t) = \frac{\sum_{i=0}^{nb} b_i[t]\cdot e^{-j\omega T_s i}}{1+\sum_{i=1}^{na} a_i[t]\cdot e^{-j\omega T_s i}} \cdot \sigma_e[t]
\end{equation}
And the corresponding ``frozen" modes with natural frequencies and damping ratios may be given by the following equations\cite{jury1964theory}:

\begin{equation}
    \omega_{ni}[t] = \frac{\left|\ln{\lambda_i[t]}\right|}{T_s} \quad \text{(rad/time)}   \qquad  \zeta_i [t] = -\cos{ \bigl(\arg(\ln{\lambda_i[t]})\bigr)}
\end{equation}
with $\lambda_i$ designating the i-th discrete time frozen pole.

\subsection{Simulation with RML-TARX model}

The system on which guided waves propagate can be thought of as having a time-varying transfer function $G[\mathcal{B},t]$, and the system's response signal $y[t]$ can be represented by the following equation:
\begin{equation}
    y[t] = G[\mathcal{B},t] \cdot x[t] +e[t]
\end{equation}

Once the system's transfer function is properly identified, it is possible to ``simulate" the response signal $y[t]$. RML-TARX model can be used to estimate the system transfer function and to ``simulate" the response guided wave signal. The following form of the equation of RML-TARX model can be used for ``simulation" of response guided wave signal $y[t]$.
\begin{equation}
    y[t] = \frac{B[\mathcal{B},t]}{A[\mathcal{B},t]} \cdot x[t] + \frac{1}{A[\mathcal{B},t]} \cdot e[t]
    \label{eq: simulation}
\end{equation}
where the transfer function $G[\mathcal{B},t]$ is approximated with $\frac{B[\mathcal{B},t]}{A[\mathcal{B},t]}$. Note that on both the left and right hand sides of Equation (\ref{eq: simulation}) contains the same time index $t$ as opposed to the one-step-ahead prediction Equation (\ref{eq:ar-model}).

\section{Data Generation} 

\subsection{Laboratory Experimental Set-up} 
\begin{figure}[t]
    \centering
    \includegraphics[width=0.5\columnwidth]{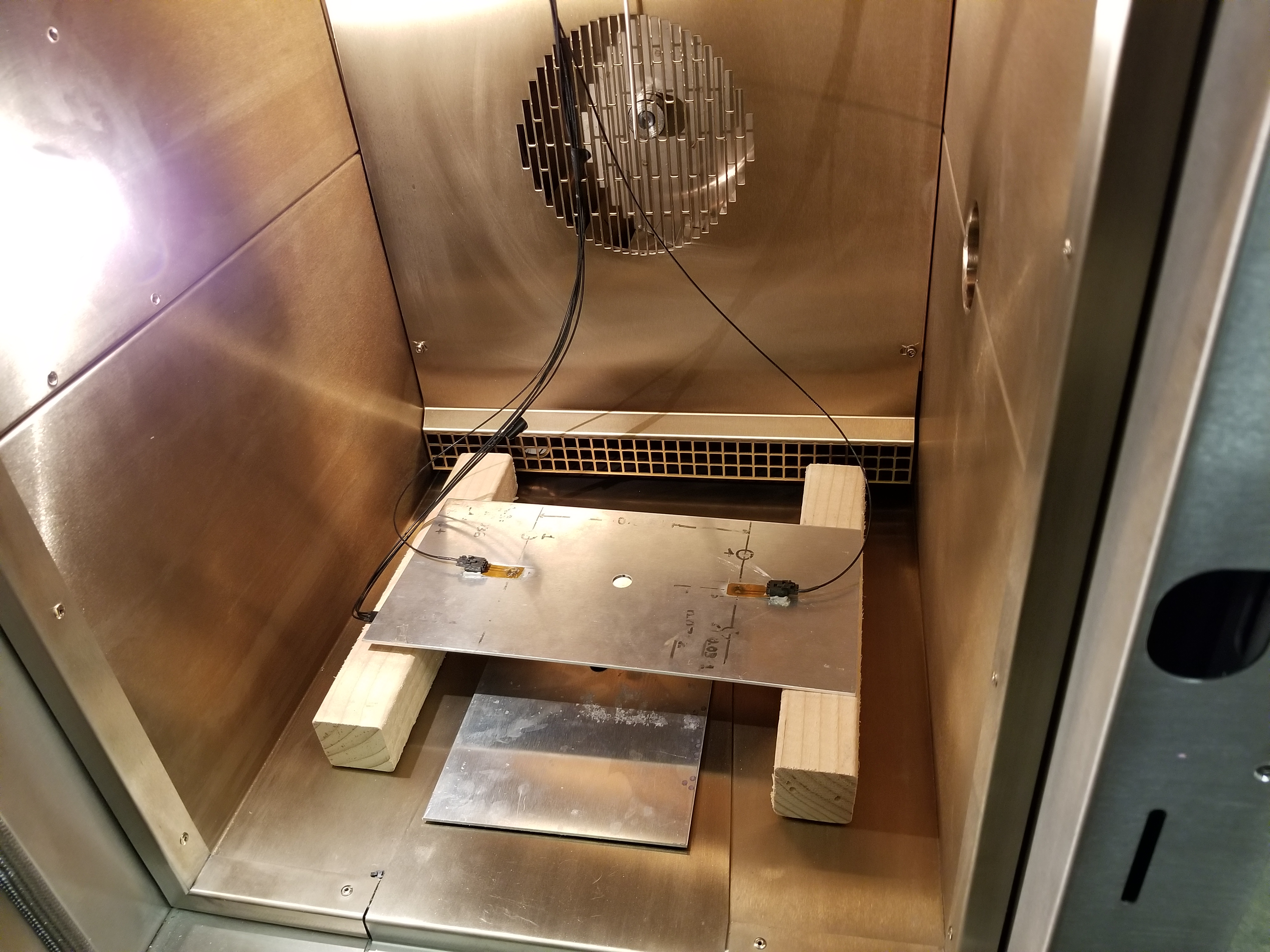}

    \caption{\label{experimental} An aluminum plate fitted with the PZT transducers in the pitch-catch mode inside an environmental chamber for maintaining a constant temperature.}
\end{figure}

The laboratory experimental set-up is shown in figure \ref{experimental}. It consists of an aluminum plate (Aluminum 6061) with a dimension of $304.8 \times 152.4 \times 2.286$ mm ($12 \times 6 \times 0.093$ inch)($L\times W\times H$) (McMaster Carr). The plate has a hole in the middle with a diameter of a $12$ mm ($0.5$ inch). Two single-PZT (Lead Zirconate Titanate) transducers were fitted to the aluminum coupon by using Hysol EA 9394 adhesive. The distance between the two PZTs is $152.4$ mm (6 inches). The PZTs are of type PZT-5A and acquired from Acellent Technologies, Inc. The PZT transducers are $0.2$ mm in thickness and $6.35$ mm ($1/4$ inch) in diameter.

\FloatBarrier

Actuation signals used in the PZT transducer were in the form of 5-peak tone bursts (5-cycle Hamming-filtered sine wave) signal having an amplitude of 90 V peak-to-peak. Various center frequencies of the 5-peak tone burst signal ranging from 200 to 700 kHz were generated in a pitch-catch configuration over each sensor consecutively. With a sampling frequency of 24 MHz, data were collected using a ScanGenie III data acquisition system (Acellent Technologies, Inc) for 11 different temperature points starting from $50^o$C to $100^o$C with an increment of $5^o$C (T = \{$50,55, \cdots ,100^o$C\}). In order to maintain the specified temperature, the aluminum plate was placed inside an environmental chamber. The dimensions of the aluminum plate can be found in Table \ref{tbl:dimension}. A thermocouple was mounted on the aluminum plate to record the temperature of each measured signal. Preliminary analysis was conducted, and a center frequency of 250 kHz was chosen for the complete analysis presented in this study based upon the best separation between the first two wave packets. In order to perform time series modeling, the actual data were down-sampled 12 times to have a sampling frequency of 2 MHz and a bandwidth of 1 MHz because there are no dynamics present beyond this frequency. The initial part of the signal (time of flight) which does not contain any dynamics have been discarded for the analysis. The number of data points was $N=$ 601 (after discarding the time of flight portion). All data sets were exported to MATLAB for analysis.\footnote{Matlab version R2019a}

\begin{table}[!t] 
\begin{minipage}{\columnwidth} 
\centering
\caption{Dimensions of aluminum plate, PZT sensors and adhesive.}
\label{tbl:dimension}
\begin{tabular}{lc} \hline \hline
Object   & Dimension \\ \hline
Thickness of aluminum plate & $2.286$ mm \\
Thickness of PZT sensor & $0.25$ mm \\
Diameter of PZT sensor & $6.35$ mm \\
Thickness of adhesive & $0.05$ mm \\
Length of aluminum plate & $304.8$ mm \\
Width of aluminum plate & $152.4$ mm \\ \hline \hline
\end{tabular} \end{minipage}
\end{table}

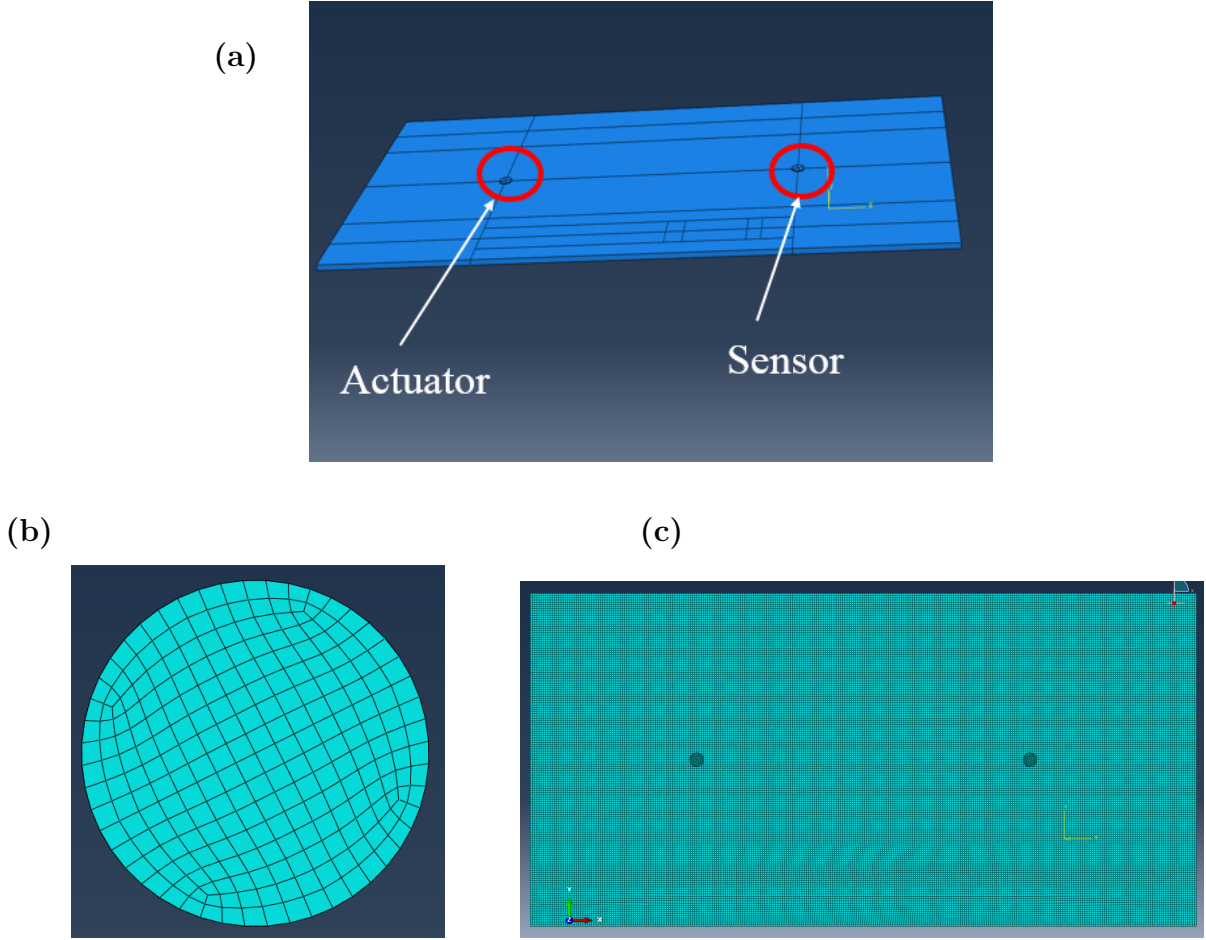
\begin{figure}[t!]
    \centering
    \begin{picture}(400,120)
    \put(70,-50){\includegraphics[width=0.55\columnwidth]{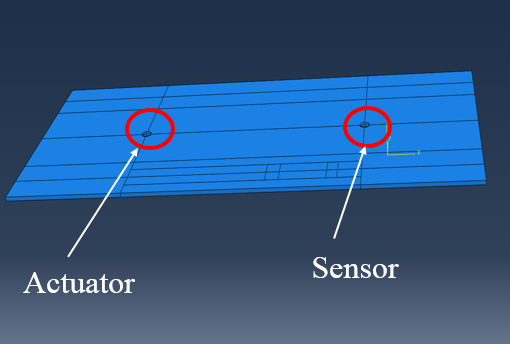}}
    \put(-20,-230){\includegraphics[width=0.3\columnwidth]{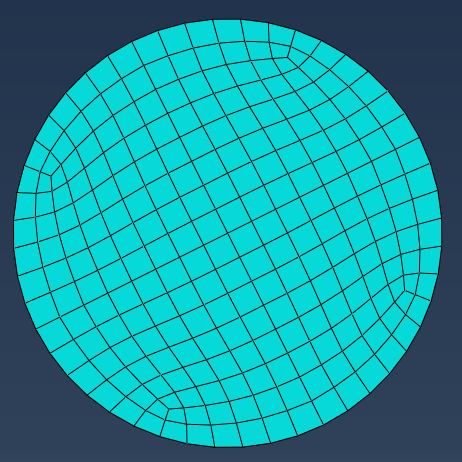}}
    \put(150,-230){\includegraphics[width=0.55\columnwidth]{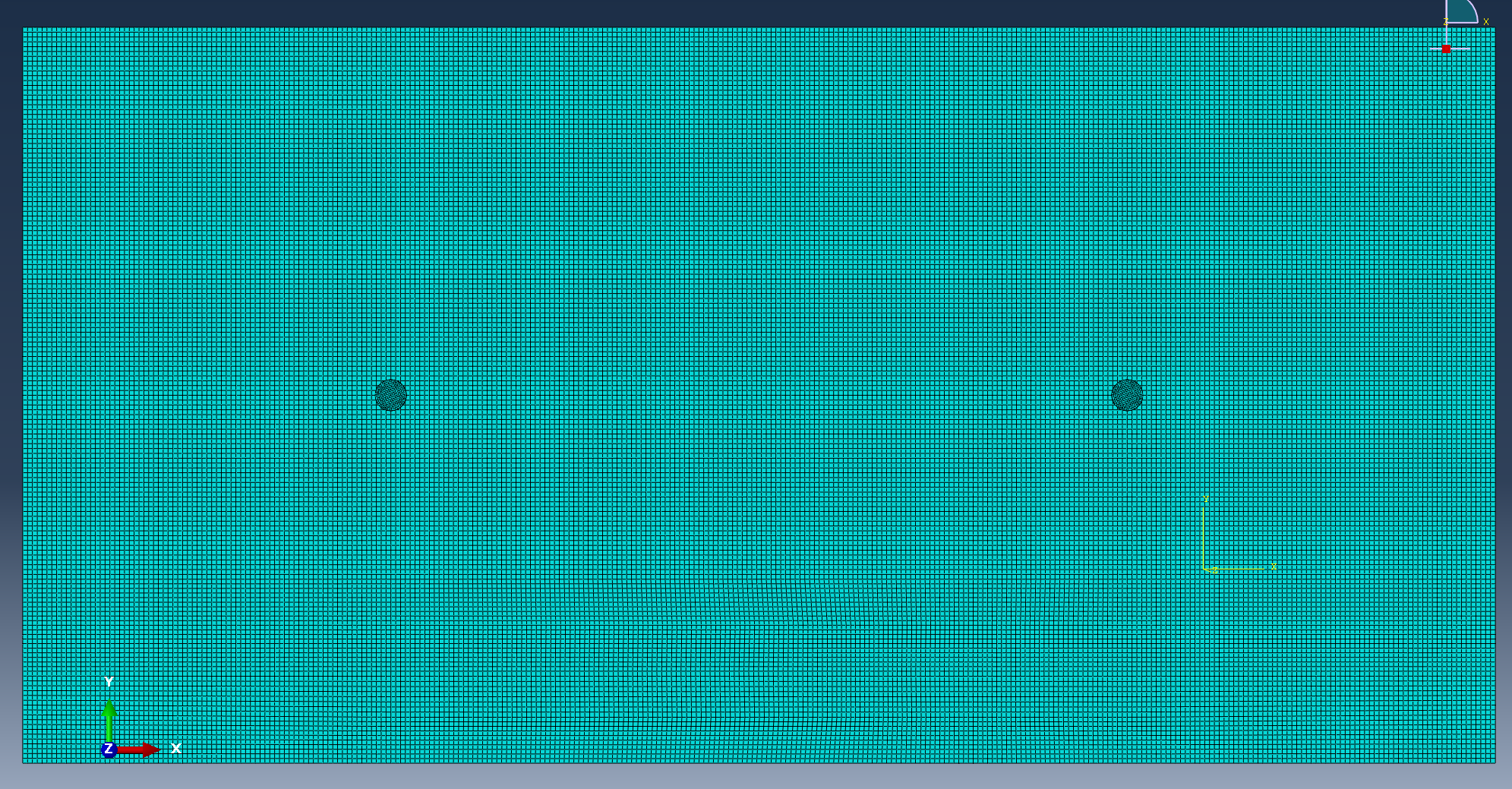} }
    \put(30,100){ \large \textbf{(a)}}
    \put(-45,-80){\large \textbf{(b)}}
    \put(195,-80){\large \textbf{(c)}}    
    \end{picture}
    \vspace{8cm}
    \caption{\label{fig:schematic simu} (a) FEM model of the aluminum plate with adhesive and PZT sensor; (b) Meshed view of the PZT sensor; (c) the meshed plate with adhesive and PZT sensor.} 
\end{figure} 
\subsection{Finite Element Model}

In order to simulate the effect of temperature on guided wave propagation, a high-fidelity finite element model was constructed by using commercial finite element software ABAQUS 2018. In the beginning, three separate parts were created namely: an aluminum plate, adhesive, and piezo-electric disk (PZT) following the dimension given in Table \ref{tbl:dimension}. Then material properties were defined for each part and assigned to the individual sections. The material properties of aluminum, adhesive and, PZT can be found in the appendix. Once the material properties were assigned to the specific sections, all separate parts are assembled together to form a unified part or model. Then, interactions between different parts were created by defining master and slave surfaces. The interaction between the PZT and adhesive was created by selecting the bottom surface of PZT as the master surface and the top surface of the adhesive as the slave surface. This must be done for each pair of adhesives and PZT disk. 


For the wave propagation simulations, a total time period of $0.0001$ s and a time step or increment of $1e-7$ s were used. For defining the actuation signal, a 5-peak-tone burst signal with a center frequency of 250 kHz was used. As boundary conditions, the two ends of the aluminum plate were kept fixed. Then an electric potential boundary condition was applied at the two end surfaces of the piezo-electric disk. An electric potential of zero volt and 100 volts was applied at the bottom and top surfaces of the piezoelectric actuator disk, respectively. This sets up an electric field between the two surfaces of the disk. For the sensor disk, zero voltage was applied only at the bottom surface and the top surface was kept free. 

\begin{figure}[t!]
    \centering
    \begin{picture}(400,160)
    \put(90,0){\includegraphics[width=0.52\columnwidth]{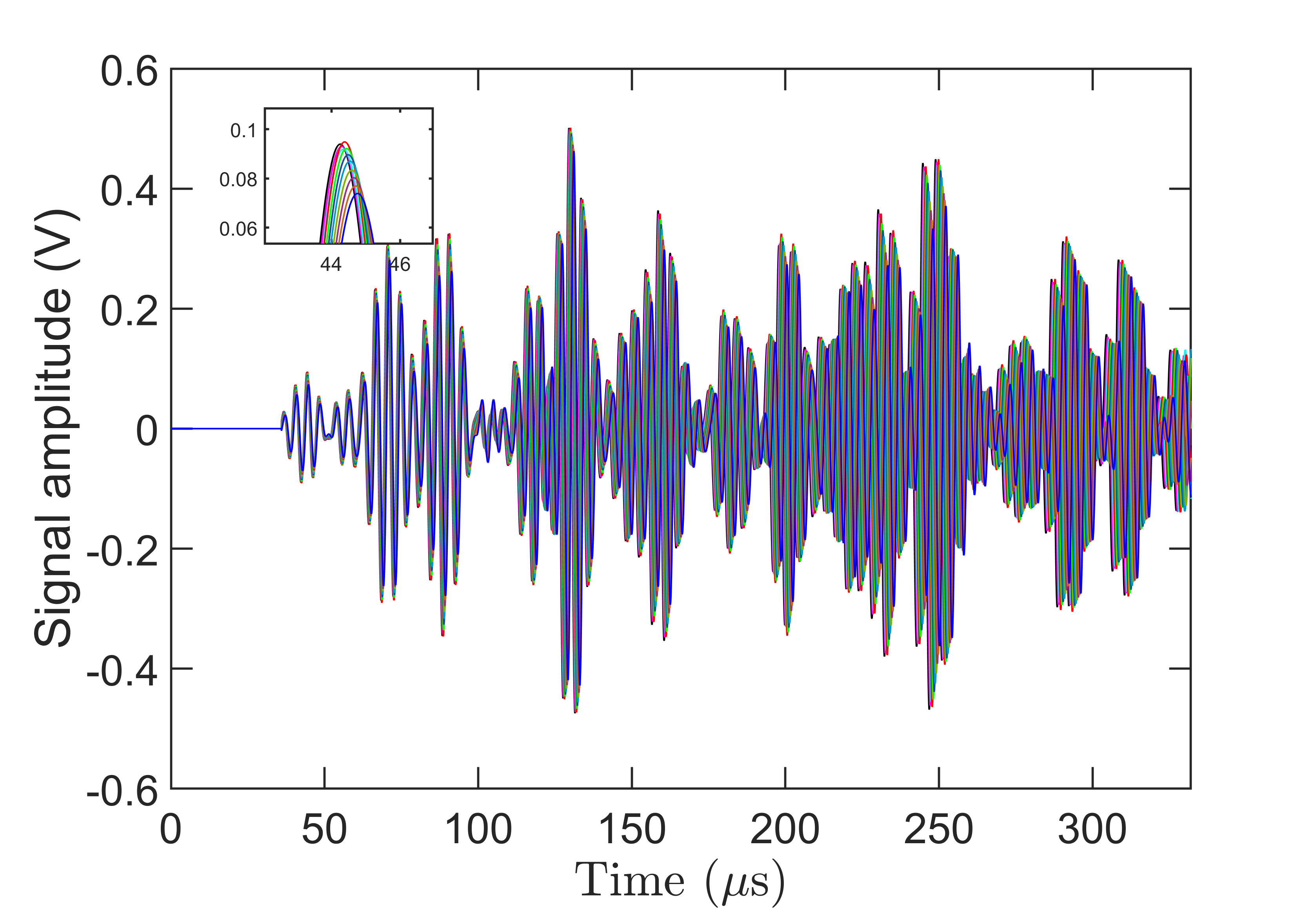}}
    \put(-30,-170){\includegraphics[width=0.52\columnwidth]{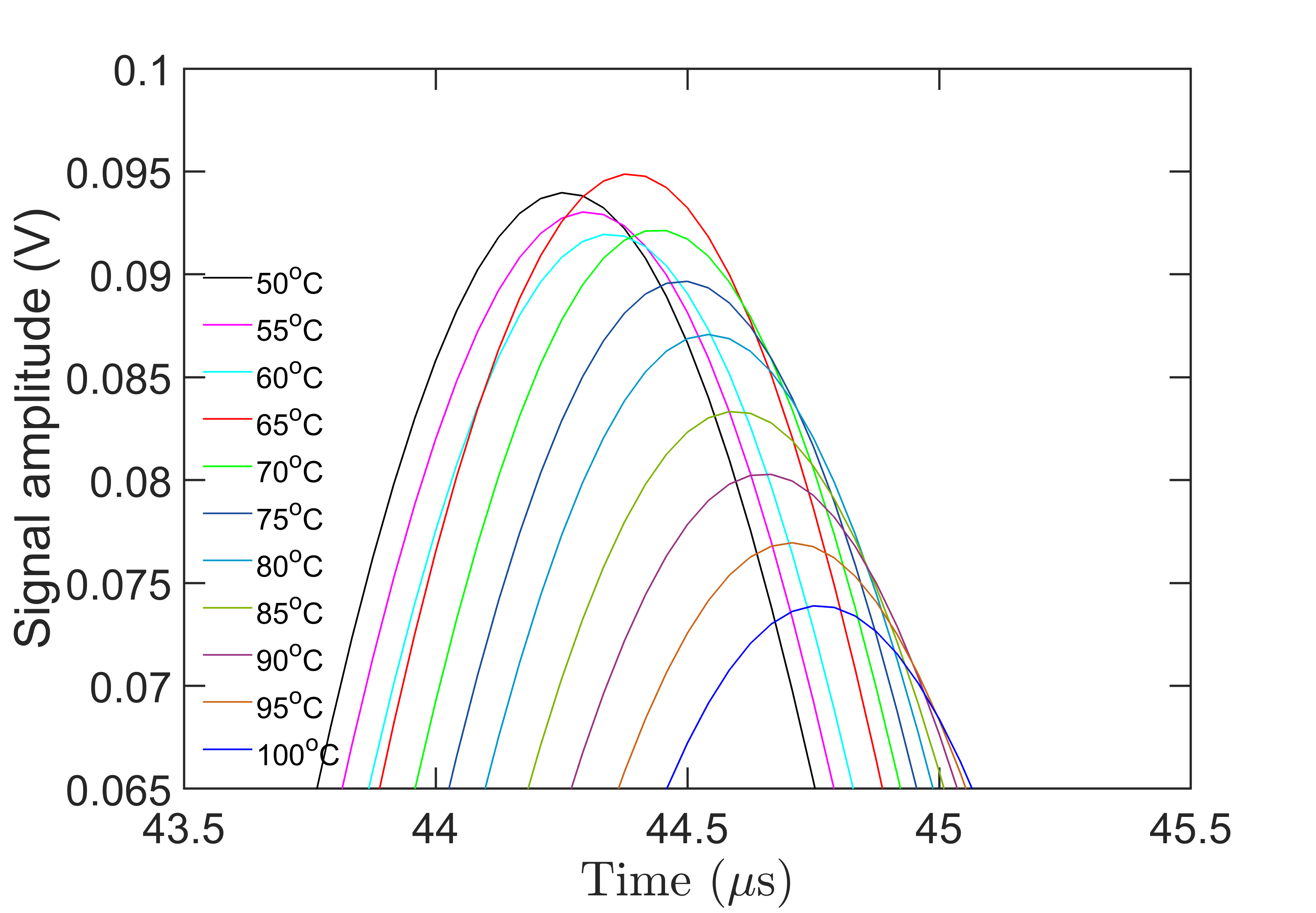}}
    \put(210,-170){\includegraphics[width=0.52\columnwidth]{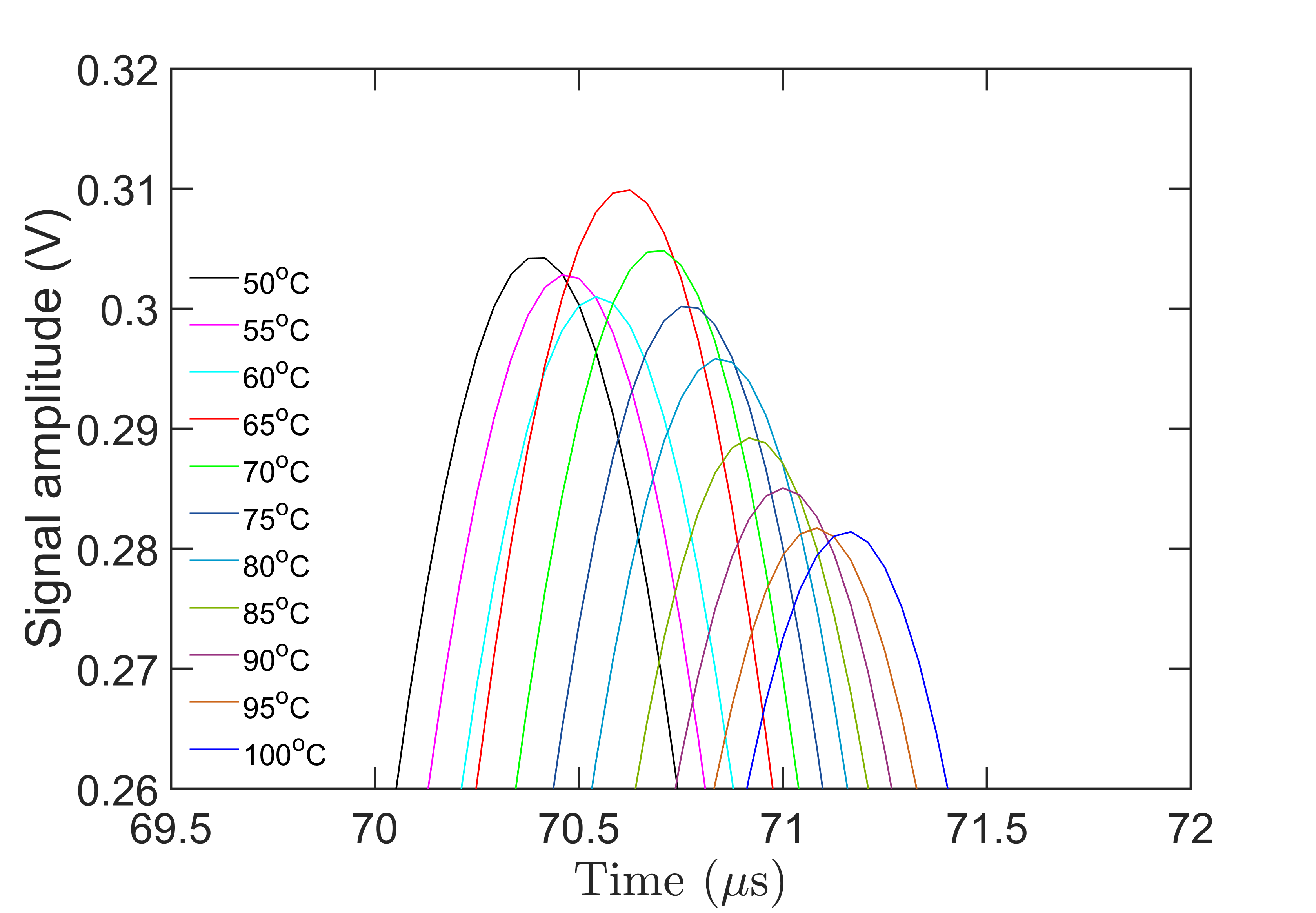}}
    \put(80,150){ \large \textbf{(a)}}
    \put(-35,-20){\large \textbf{(b)}}
    \put(200,-20){\large \textbf{(c)}}    
    \end{picture}
    \vspace{6cm}
    \caption{Experimental guided wave signal collected at different temperatures: (a) shows the mean signal collected at each temperature level for the entire time duration; zoomed-in view of the mean signal at different temperatures; (b) the highest peak of $S_0$ mode; (c) the highest peak of $A_0$ mode.} 
\label{fig: elevate} \vspace{-12pt}
\end{figure}
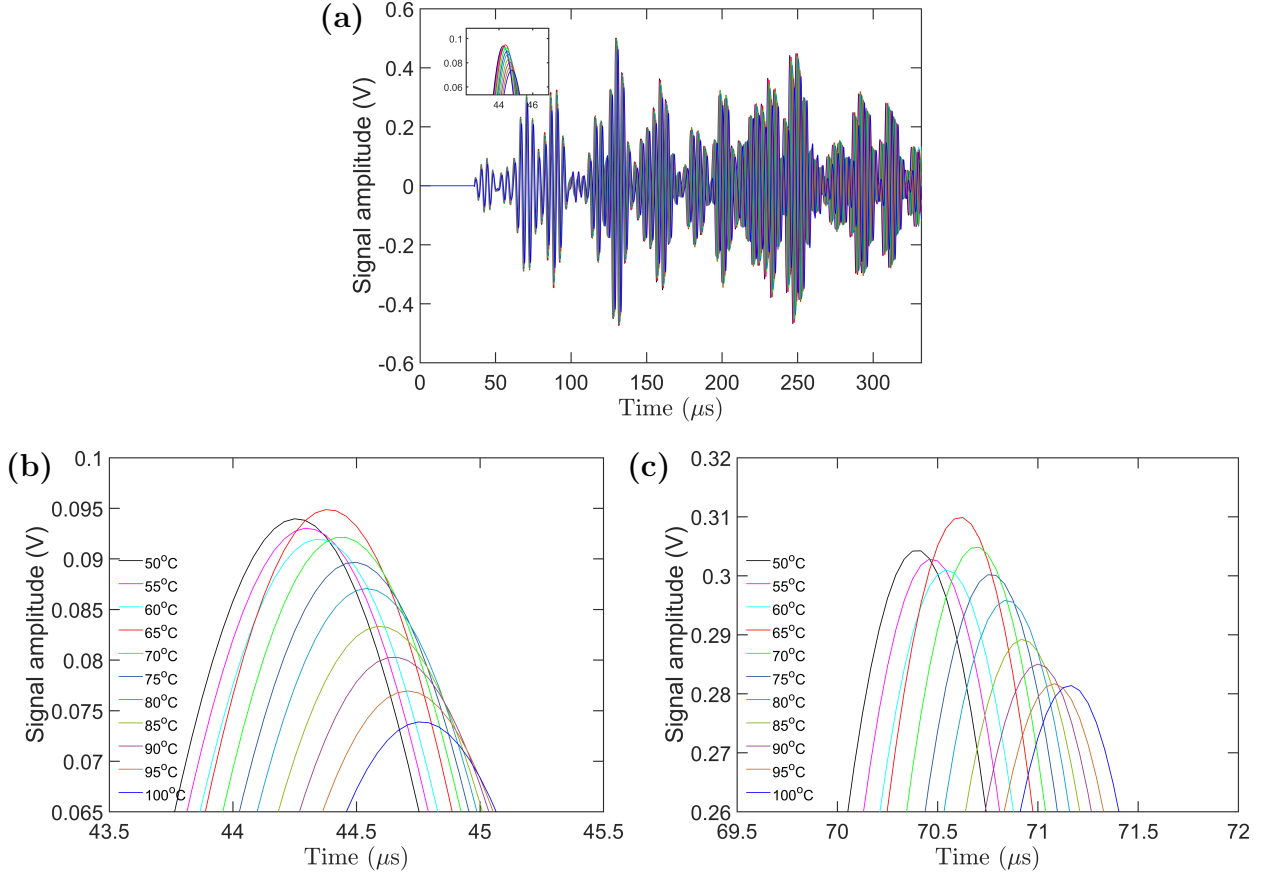 

The electric potential (EPOT) was selected as the response signal from the PZT disk. The output signal can be collected from any single node on the top surface of the PZT sensor. However, signals obtained from different nodes of the surface are different. Hence, it is necessary to obtain the average of all the signals from all the nodes on the top surface of the PZT sensor. This task of averaging can be performed by using equation constraints. Under an equation constraint, a single node is selected as a master node and all other nodes on the surface are constrained with respect to this master node. This forces the electric potential to be the same on all other nodes of the surface which is the same as of the master node. It was found that averaging the signals from all the nodes on the surface was equivalent to using the equation constraint where a master node was used to constrain all other nodes on the surface. The coefficient used for the master node and the other nodes were 1 and -1, respectively. The degrees of freedom used for PZT material was 9.

In order to facilitate meshing, partitioning was performed on all of the three parts, namely, aluminum plate, adhesive materials, and PZT disk. For all three parts, structured hexahedron mesh was used. In order to select the mesh size, a convergence study was performed. Based on this study, a global mesh size of 0.001 was chosen for the aluminum plate. The total number of elements in the plate was 93330. For the adhesive and PZT material, a global mesh size of 0.0004 was used. The total number of elements in the adhesive and PZT disk were 480. It was ensured that at least 20 elements exist per wavelength. The mesh size can be further reduced, however, doing so increases computational cost with no significant increase in accuracy.

Linear 3D stress elements (C3D8R) were selected for aluminum plate and adhesive materials. For the PZT disk, 8-node linear piezo-electric brick elements were used. With all these specifications, a single simulation takes about 6 hours. Figure  \ref{fig:schematic simu} shows the full FEM model of the aluminum plate as well as the mesh of the PZT disk and the plate.

\section{Results and Discussion} 

\subsection{Stochasticity in Guided Waves Under Varying Temperature}

\subsubsection{Experimental Guided Wave Signal Analysis}

In order to model guided wave propagation under varying temperatures, it is first necessary to understand how guided wave signals respond to temperature variation. In order to do so, experiments were performed inside an environmental chamber with an aluminum plate fitted with PZT actuators/sensors. Data were collected at 11 temperatures (T = \{$50,55, \cdots ,100^o$C\}). At each temperature, 20 signals were recorded. Figure \ref{fig: elevate} shows how guided wave signals change at those 11 temperatures. Figure \ref{fig: elevate}(a) shows the mean signals collected at each temperature. Until 33$\mu$s, the amplitude of the signal is zero which is the time of flight of the guided wave signal. The first wave packet extends from 33 $\mu$s to 50 $\mu$s and is referred to as the first symmetric mode ($S_0$ mode) of guided wave propagation. The second wave packet extends from 60 $\mu$s to 78 $\mu$s and is referred to as the first anti-symmetric mode ($A_0$ mode) of guided wave propagation. The rest of the signal is coming from the boundary reflections.

\begin{figure}[t!]
    \centering
    \begin{picture}(400,130)
    \put(-30,-50){ \includegraphics[width=0.5\columnwidth]{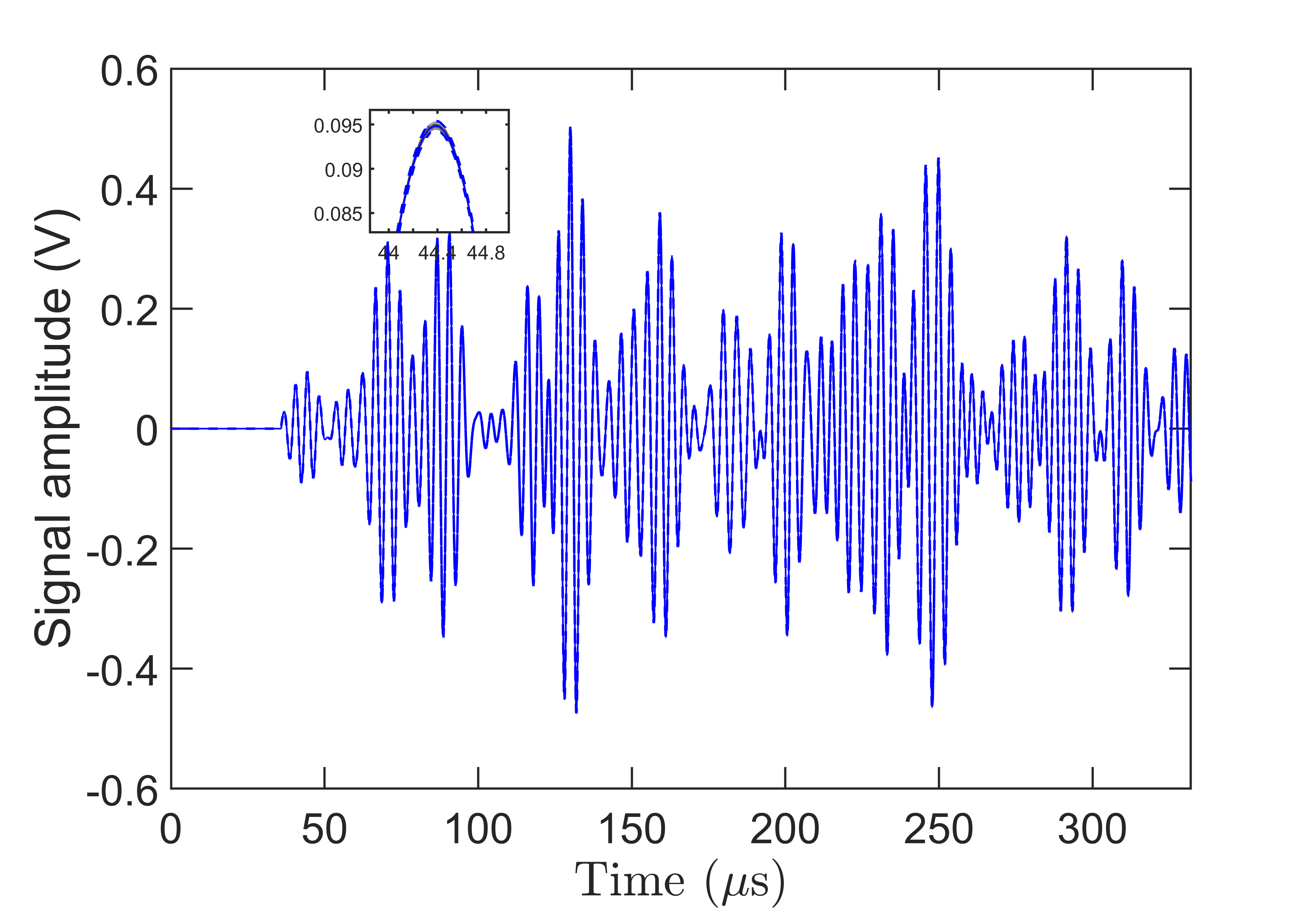}}
    \put(220,-50){\includegraphics[width=0.5\columnwidth]{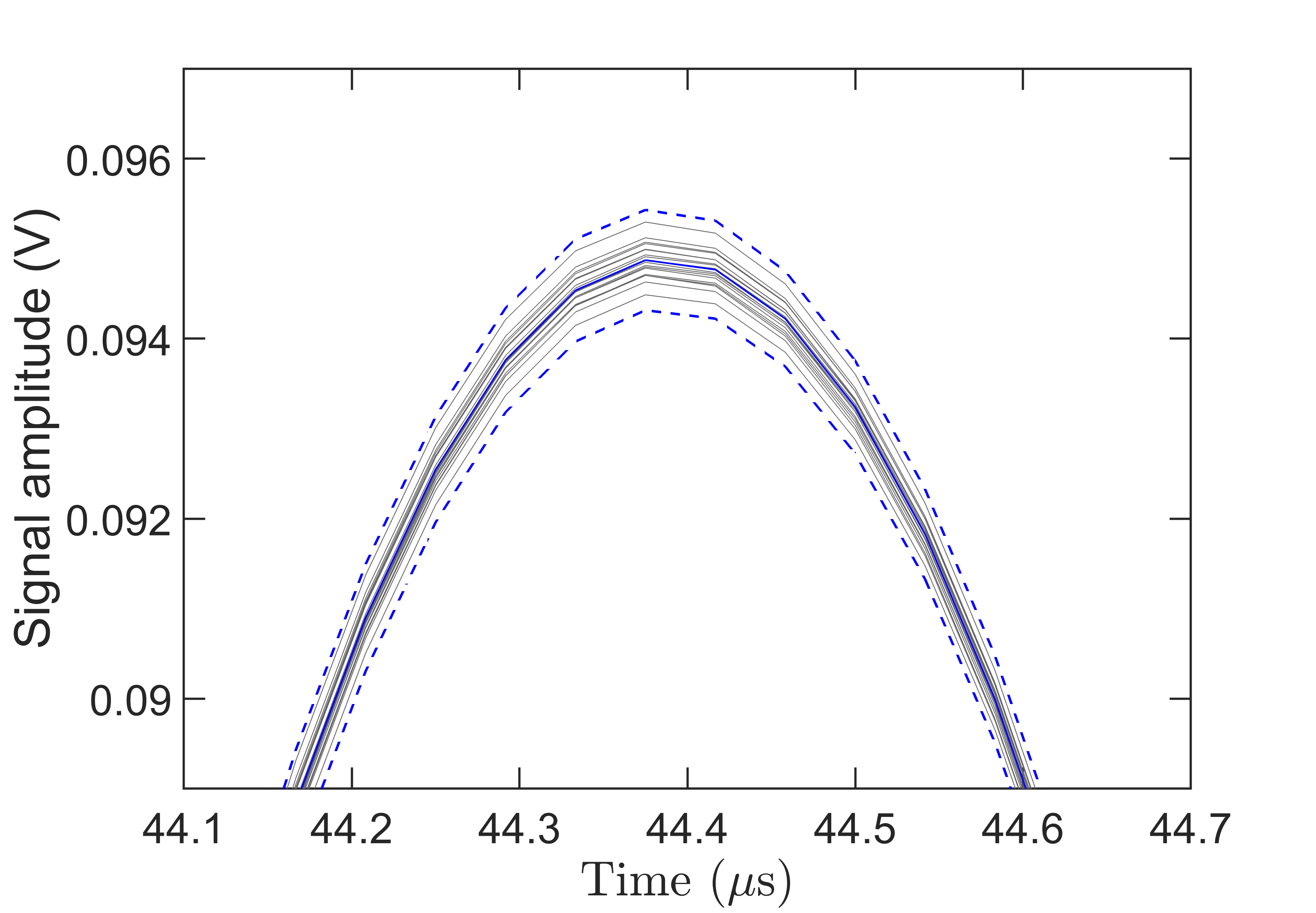}}
    
    \put(-35,100){\large \textbf{(a)}}
    \put(200,100){ \large \textbf{(b)}}
    \end{picture}
    \vspace{2cm}
    
    \caption{Variation of guided wave signals at a representative temperature: (a) shows the mean of the 20 experimental signals collected at $65^o$C and their $\pm$3 standard deviations; (b) zoomed-in view of the highest peak of the $S_0$ mode.} 
\label{fig: Stochastic} 
\end{figure} 

Figure \ref{fig: elevate}(b) shows a zoomed-in view of the highest peak of the $S_0$ mode at different temperatures. It can be observed that with the increase in temperature from $50^o$C to $60^o$C, the amplitude of the signal decreases. At $65^o$C, the amplitude increases suddenly, and then monotonically decreases from $70^o$C to $100^o$C. However, in each case, the signal is shifted to the right. This is due to the fact that with the increase in temperature, the aluminum plate expands, and the guided wave takes an extra little time to reach the sensor. Figure \ref{fig: elevate}(c) shows the zoomed-in view of the amplitude of the highest peak of the $A_0$ mode at different temperatures. It can be observed that the trend of the amplitude at different temperatures for $A_0$ mode is similar to the $S_0$ mode. However, at $100^o$C, the amplitude remains about the same as that of $95^o$C. Again, the signal shifts to the right similar to the $S_0$ mode.

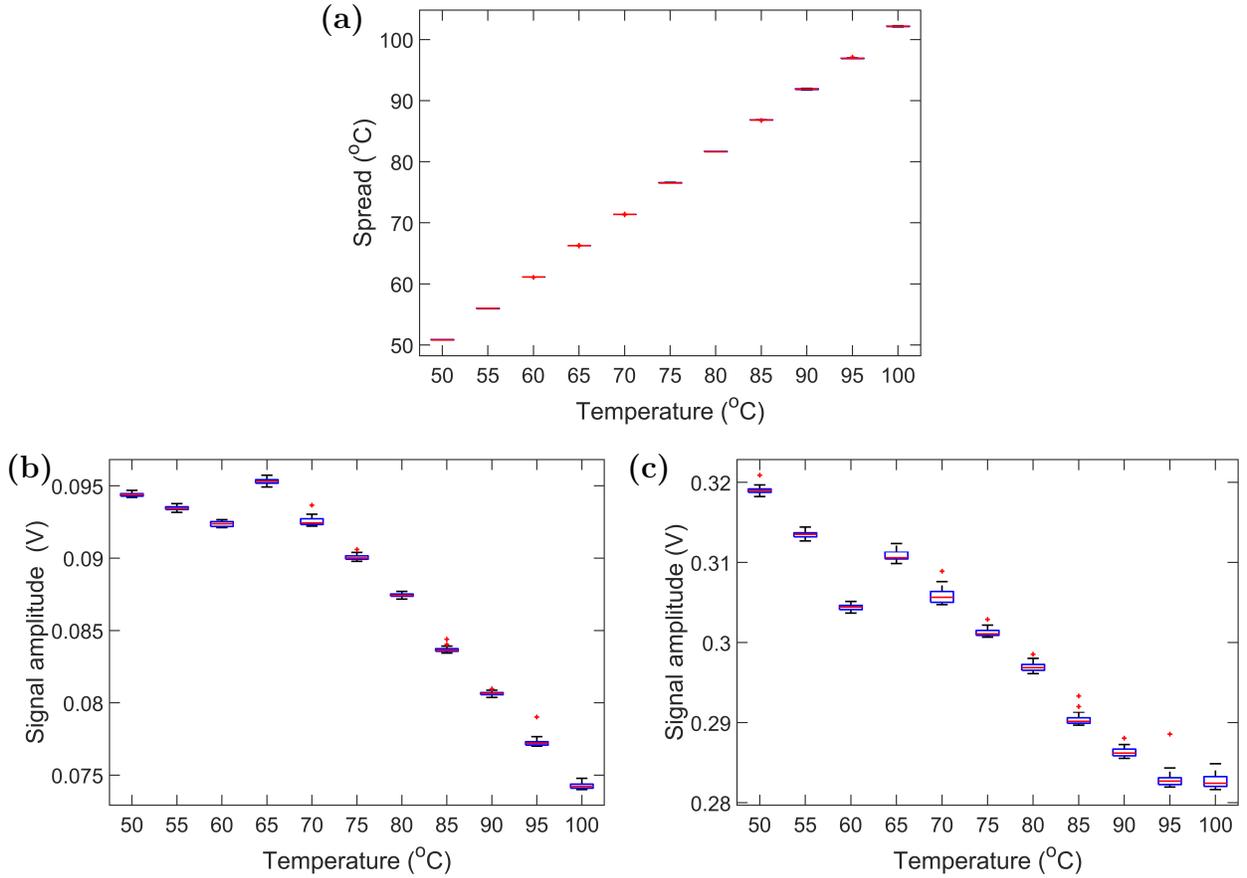
\begin{figure}[t!]
    \centering
    \begin{picture}(400,140)
    \put(90,0){\includegraphics[width=0.52\columnwidth]{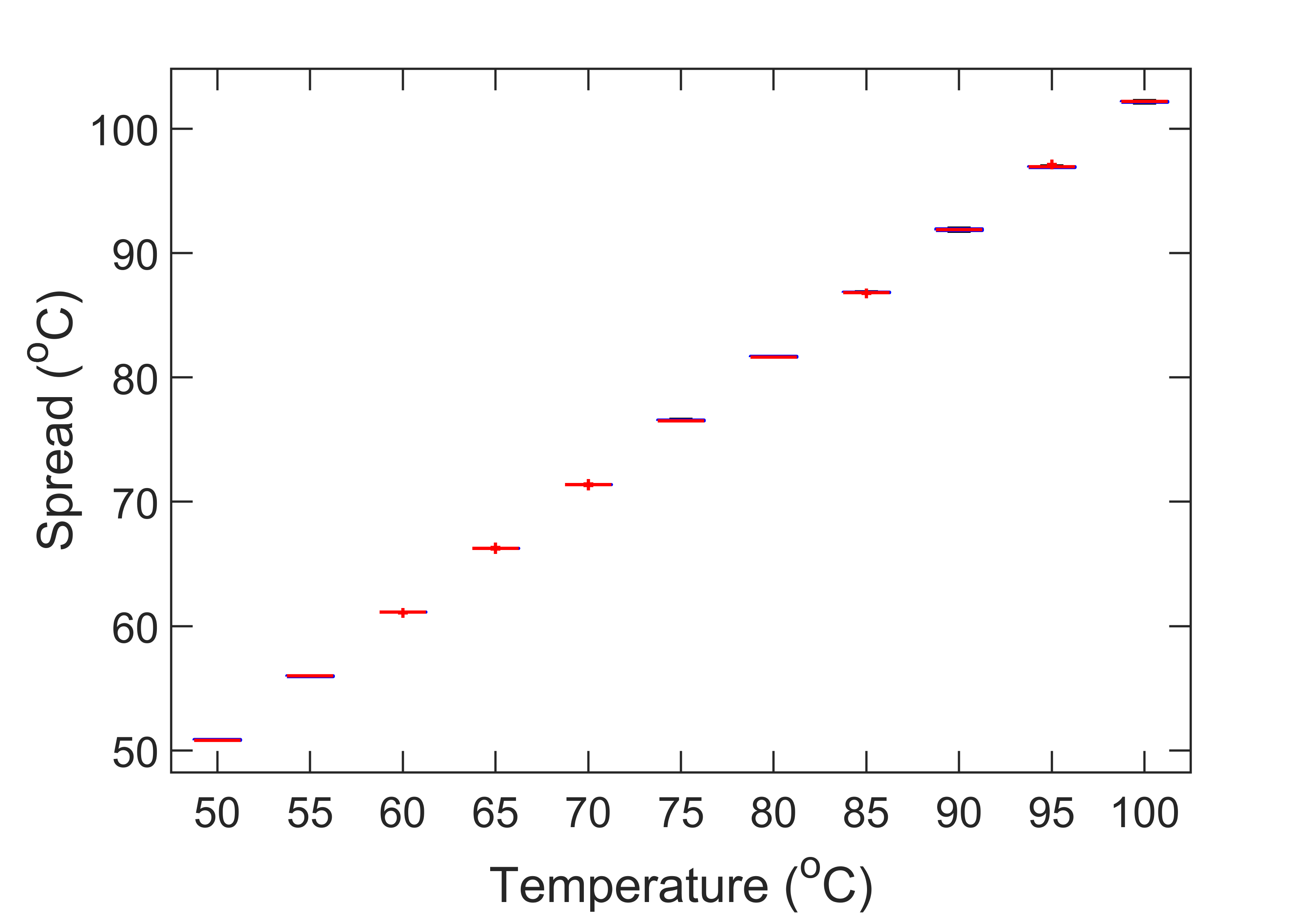}}
    \put(-30,-170){\includegraphics[width=0.52\columnwidth]{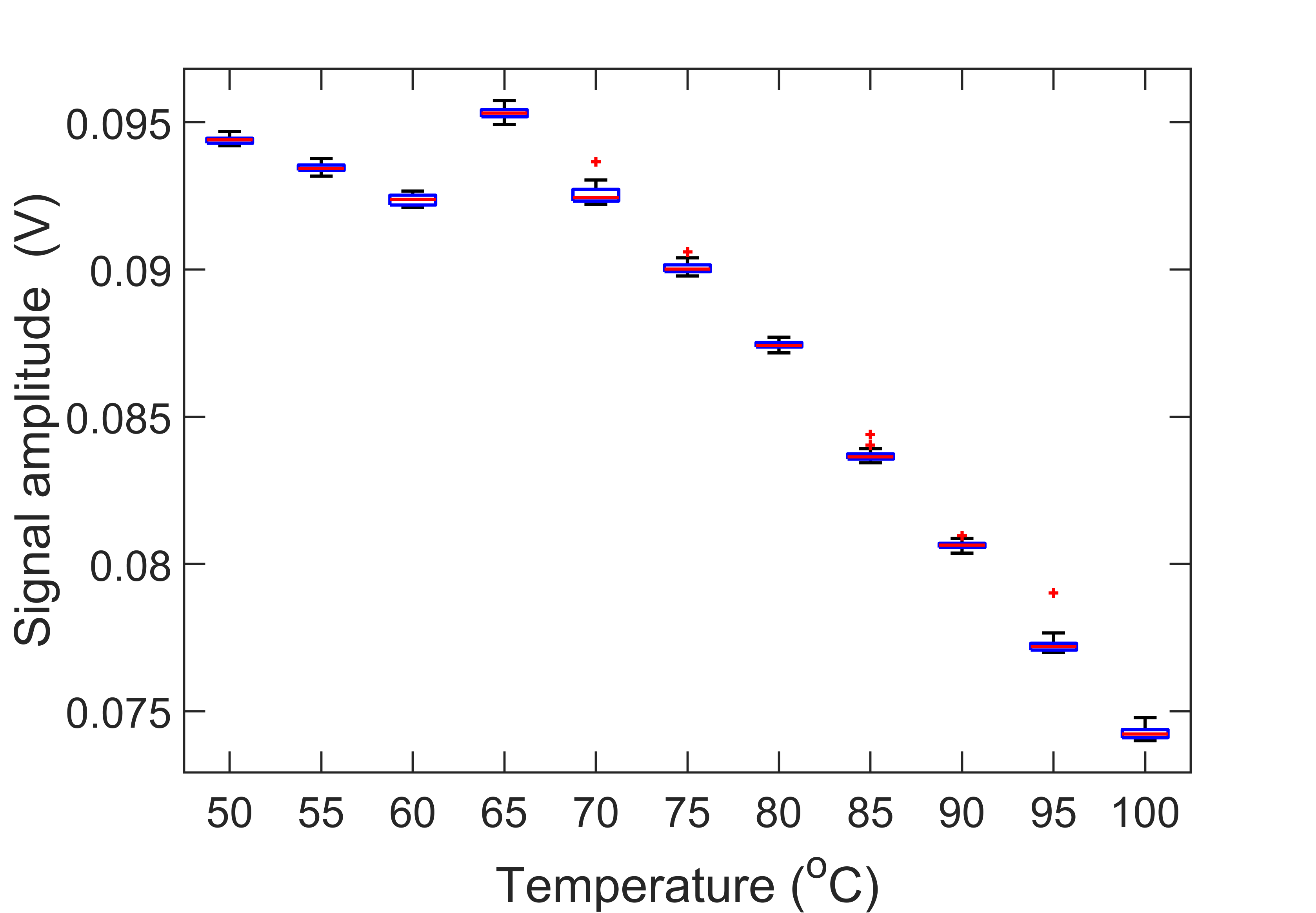}}
    \put(210,-170){\includegraphics[width=0.52\columnwidth]{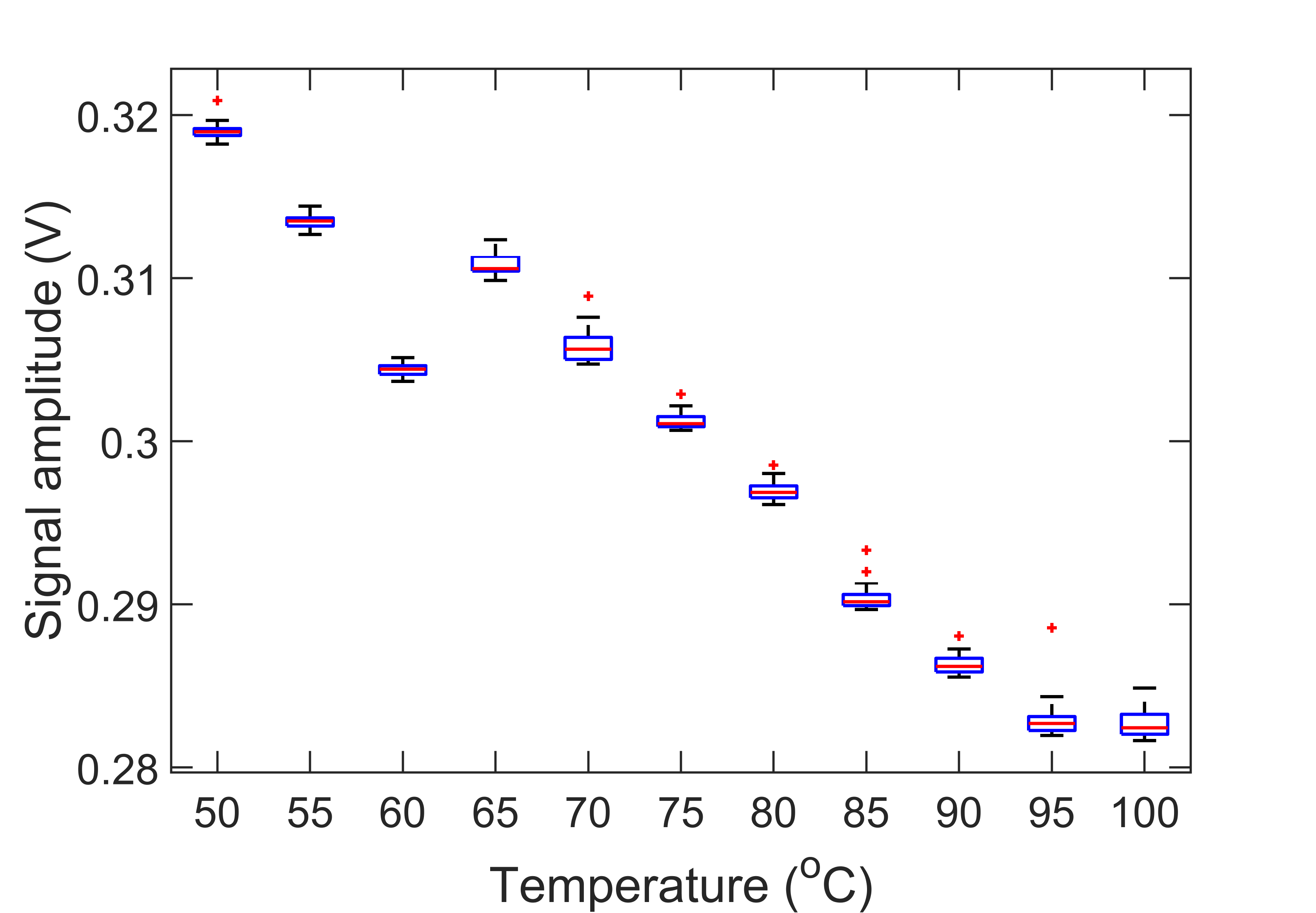}}
    \put(80,150){ \large \textbf{(a)}}
    \put(-35,-20){\large \textbf{(b)}}
    \put(200,-20){\large \textbf{(c)}}    
    \end{picture}
    \vspace{6cm}
    \caption{Temperature distribution at different temperature points and the corresponding variation in the signal amplitude: (a) box plot representation of the temperature distribution of all different temperatures within the chamber; box plot representation of the Hilbert transform of the highest peak of the (b) $S_0$ mode; (c) $A_0$ mode for all different temperatures.} 
\label{fig:  Exp boxplot} \vspace{-12pt}
\end{figure} 
%
Figure \ref{fig: Stochastic}(a) shows the plot of all the signals collected at a representative temperature (in this case $65^o$C) inside an environmental chamber along with their mean and standard deviations. Figure \ref{fig: Stochastic}(b) shows a zoomed in view of the third peak of the $S_0$ mode. The grey lines represent all the signals, the solid blue line represents the mean signal, and the dotted blue lines represent their $\pm3$ standard deviations. It can be observed that even though the experiment is performed inside an environmental chamber, there is variation between the signals. The amplitude of the third peak of the $S_0$ mode of the signals is seen to vary between 0.0943 V and 0.0954 V.

Figure \ref{fig:  Exp boxplot}(a) shows the box plot representation of the variation of temperature during the experiment. As for example, although the nominal temperature is kept constant at $50^o$C, there is a slight variation in temperature inside the environmental chamber which cannot be controlled. Figure \ref{fig:  Exp boxplot}(b) and (c) shows the corresponding variation of the amplitude of the signal at those specified temperatures. Figure \ref{fig:  Exp boxplot}(b) shows the distribution of the Hilbert transform of the signal amplitude of the third peak of the $S_0$ mode in the form of a box plot. The blue box represents the 25th and 75th percentile of the data or the interquartile range. The red line represents the mean of the data. The two whiskers at the end represent the range of the data. It can be observed that the amplitude distribution of the third peak of the $S_0$ mode decreases from $50^o$C to $60^o$C, then suddenly increases and then decreases monotonically until $100^o$C. Figure \ref{fig:  Exp boxplot}(c) shows the distribution of the Hilbert transform of the signal amplitude of the fifth peak of the $A_0$ mode in the form of a box plot. Similar trends can be observed as the $S_0$ mode, however, the range of the distribution of the signal amplitude is higher for $A_0$ mode than the $S_0$ mode. 

    

\subsubsection{Simulated Guided Wave Signal Analysis}

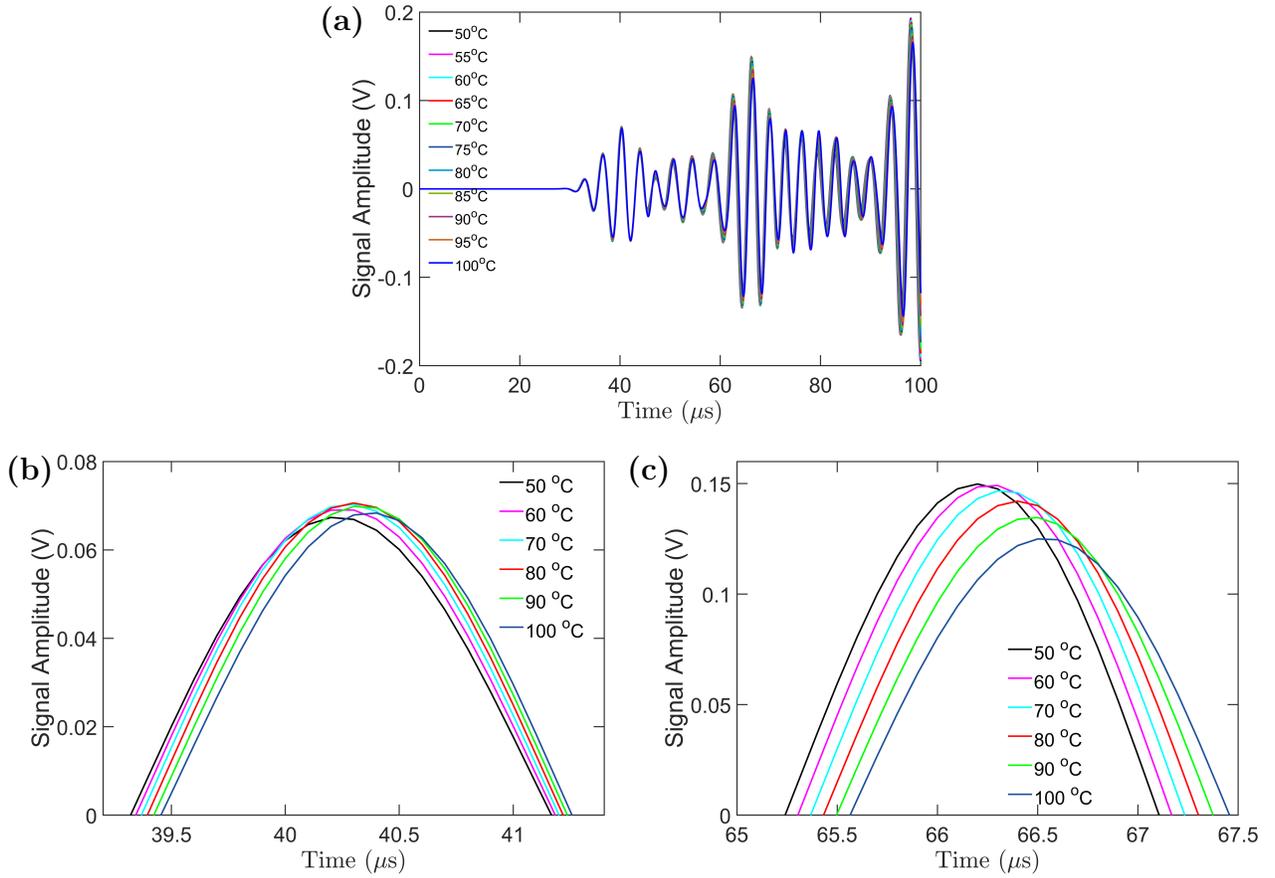
\begin{figure}[t!]
    \centering
    \begin{picture}(400,140)
    \put(90,0){\includegraphics[width=0.52\columnwidth]{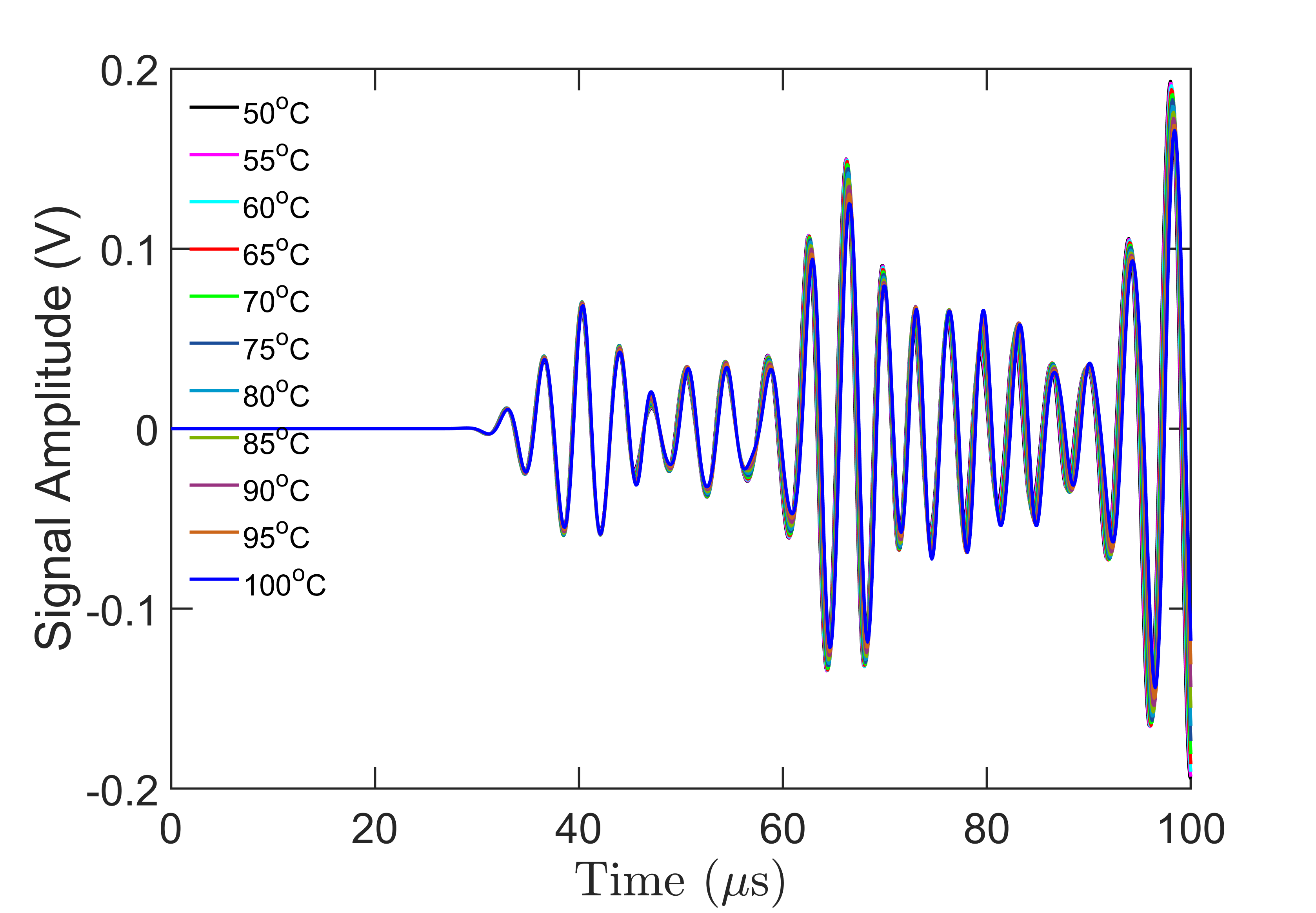}}
    \put(-30,-170){\includegraphics[width=0.52\columnwidth]{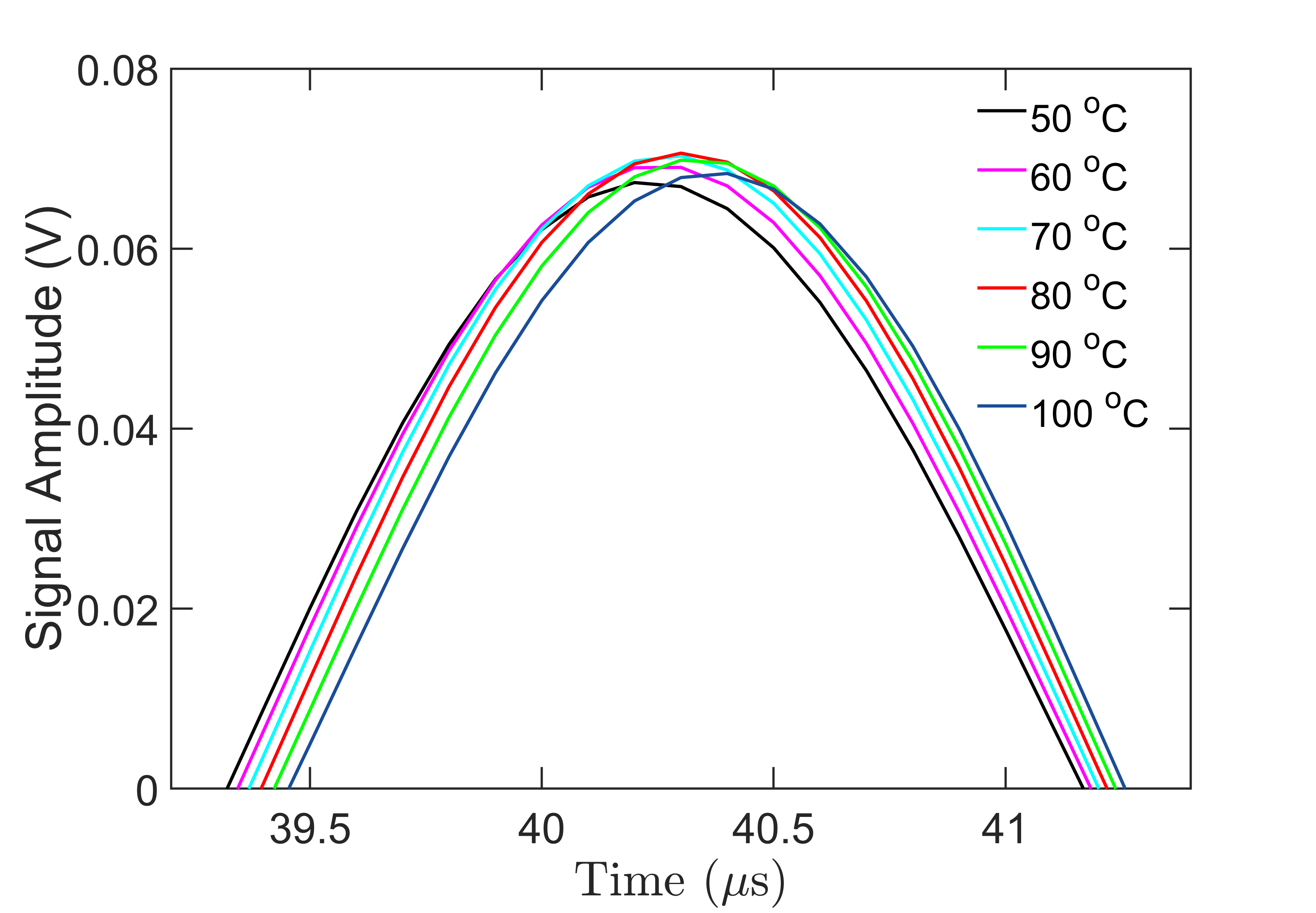}}
    \put(210,-170){\includegraphics[width=0.52\columnwidth]{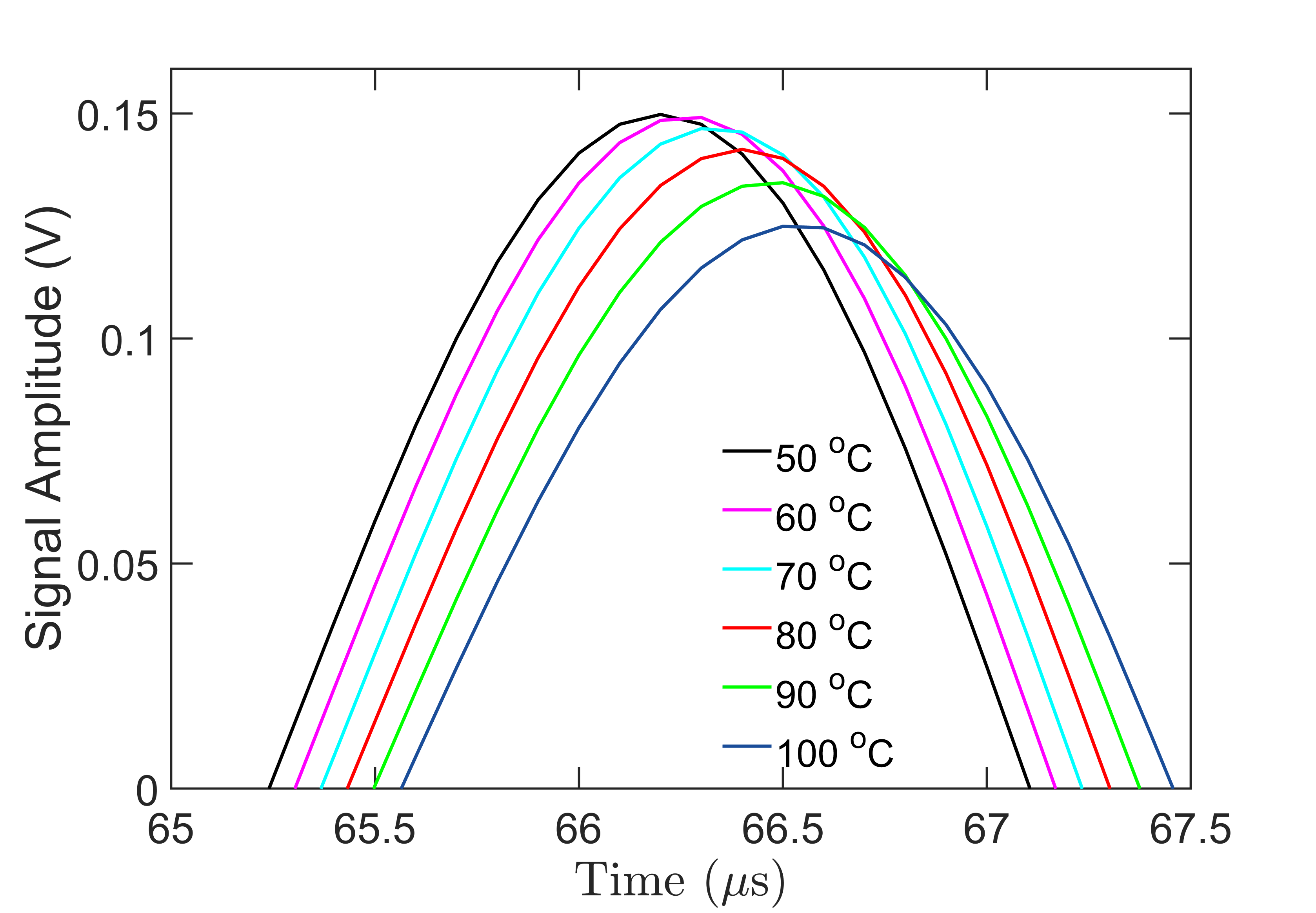}}
    \put(80,150){ \large \textbf{(a)}}
    \put(-35,-20){\large \textbf{(b)}}
    \put(200,-20){\large \textbf{(c)}}    
    \end{picture}
    \vspace{6cm}
    \caption{Simulated guided wave signal modeled at different temperatures: (a) shows the FEM signal collected at each temperature level for the entire time duration; (b) shows the zoomed-in view of the FEM signal at different temperatures for the highest peak of $S_0$ mode; (c) for the highest peak of $A_0$ mode.} 
\label{fig: FE signal} \vspace{-12pt}
\end{figure}

High fidelity finite element modeling was also performed to simulate guided wave propagation under temperature variation. A temperature range of $25^o$C to $100^o$C was considered with an increment of $5^o$C. As such 16 temperature points were considered. Figure \ref{fig: FE signal}(a) shows the plot of simulated guided wave signal for all 16 temperatures. The time of flight of the simulated guided wave signal is 31 $\mu$s which is very close to the experimental time of flight (33 $\mu$s). The part of the signal from 31 $\mu$s to 50 $\mu$s is considered as the $S_0$ mode and from 57 $\mu$s to 74 $\mu$s is considered as the $A_0$ mode. The rest of the signal is coming from the boundary reflections.

Figure \ref{fig: FE signal}(b) shows the zoomed-in view of the third peak of the $S_0$ mode for different temperatures. It can be observed that the amplitude of the signal increases from $50^o$C to $80^o$C and then decreases until $100^o$C. This is in contrast to the experimental case, where the amplitude decreases with increasing temperature. However, similar to the experimental case, the signals get shifted to the right with increasing temperature. Figure \ref{fig: FE signal}(c) shows the zoomed-in view of the highest peak of the $A_0$ mode. It can be observed that as the temperature increases from $50^o$C to $100^o$C, the amplitude of the guided wave signal monotonically decreases, and the signals get shifted to the right. This case aligns with the experimental observation.

\FloatBarrier

    
    

\subsection{Non-parametric Analysis}

Short-time-Fourier-transform (STFT), also referred to as the ``spectrogram", is a widely used method for analyzing non-stationary signals. It provides information on the non-stationary signal's time and frequency content at the same time. The idea is to break down the non-stationary signal into smaller chunks within which the signal may be considered as stationary. Then perform the Fourier transform of the signal's smaller chunks and sum them together to get the time and frequency information simultaneously. However, due to the ``uncertainty principle", an arbitrarily small time signal cannot be used because it gives a very bad frequency resolution. As a result, there is a compromise between the time and frequency resolution of the non-stationary signal \cite{cohen1995time}.

Keeping in mind the above-mentioned compromising facts, the non-parametric analysis of the non-stationary guided wave signals was performed by ``spectrogram.m" function of MATLAB \footnote{MATLAB 2019a}. The analysis is based on 601 sample long response signals with a sampling frequency of 2 MHz obtained from the piezoelectric sensor mounted on the aluminum plate. A 30 sample long Hamming data window (frequency resolution $\Delta f = 666.66$ Hz) with 98\% overlap has been used for spectrogram analysis. The number of FFT points used were 100 times the window size.

\begin{figure}[t!]
    \centering
    \begin{picture}(400,120)
    \put(-32,-30){\includegraphics[width=0.5\columnwidth]{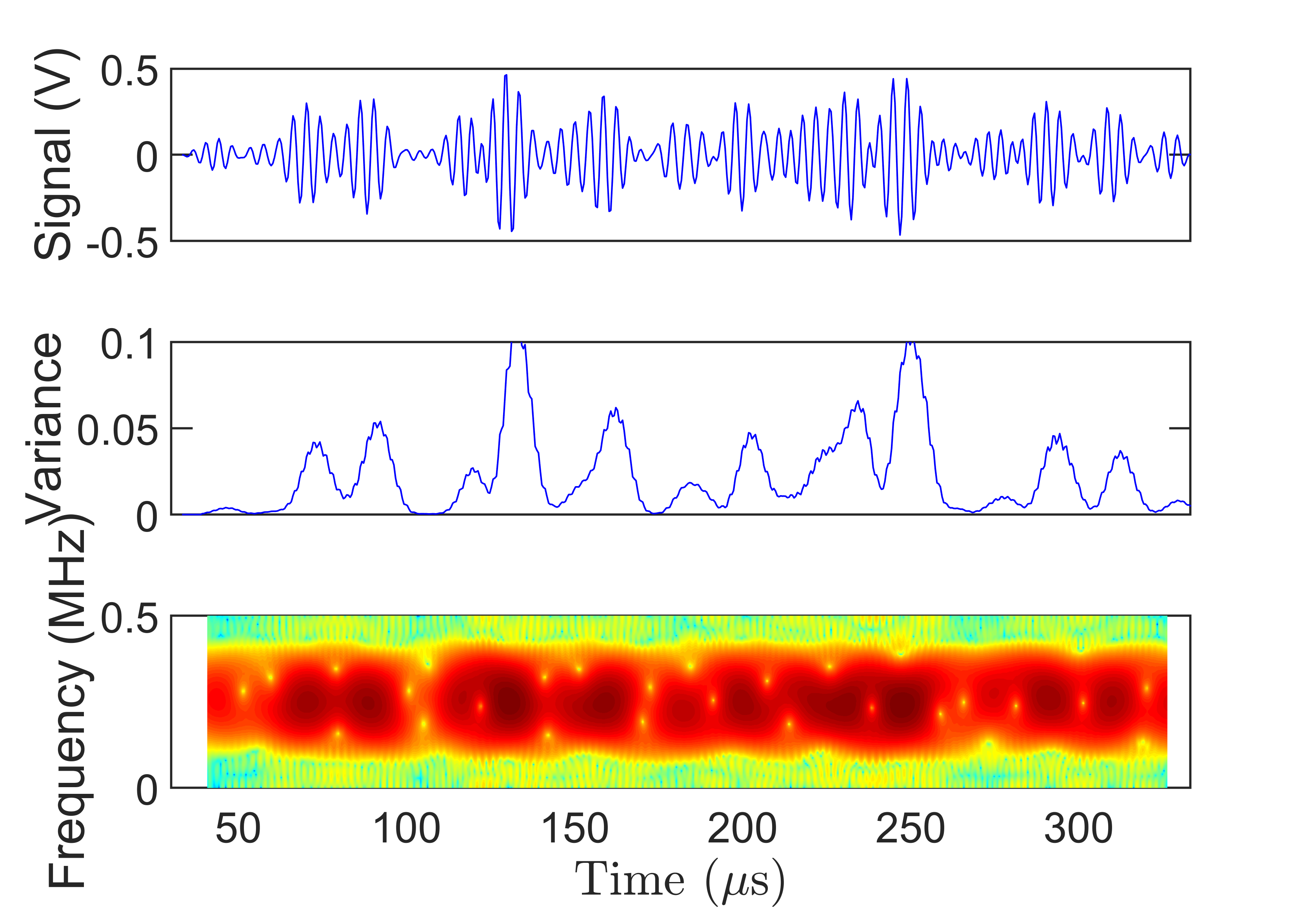}}
    \put(200,-30){\includegraphics[width=0.5\columnwidth]{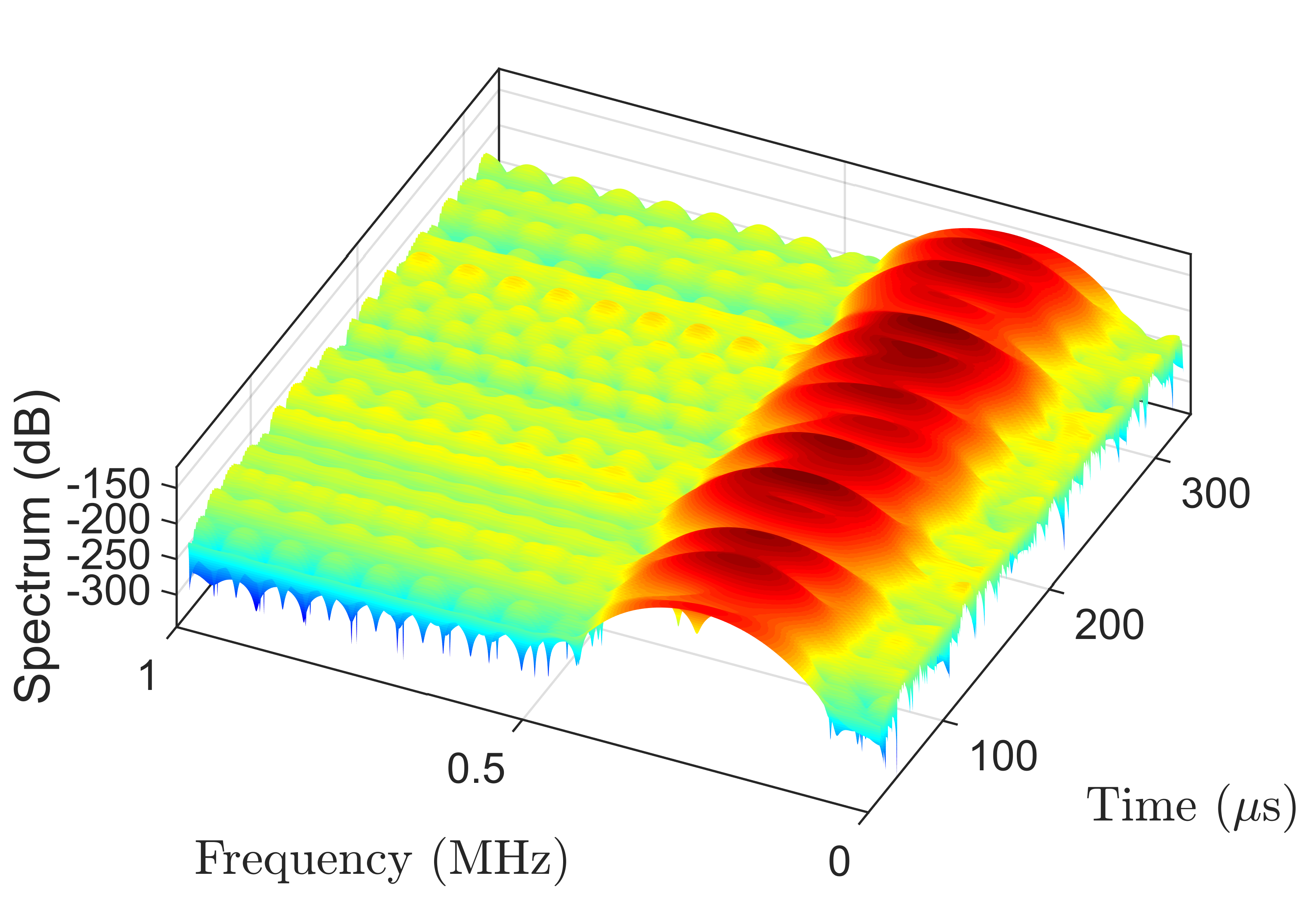}}
    \put(-50,120){ \large \textbf{(a)}}
    \put(200,120){\large \textbf{(b)}}
    \end{picture}
    \vspace{1cm}
    \caption{Non-parametric time-dependent power spectral estimates of the non-stationary experimental guided wave signal: (a) the guided wave signal at $50^o$C (top subplot), time dependent variance (middle subplot) and the 2D spectrogram (bottom subplot); (b) 3D view of the spectrogram.} 
\label{fig: spectrogram exp} 
\end{figure}

Figure \ref{fig: spectrogram exp} presents an indicative spectrogram of the non-stationary guided wave signal at $50^o$C. The top subplot of Figure \ref{fig: spectrogram exp}(a) presents the guided wave response signal from the PZT transducer. The middle subplot depicts the time-varying variance of the corresponding signal. Observe the evident non-stationarity and significant fluctuation of the estimated variance for the selected window size of 20 samples. The bottom subplot shows how the frequency spectrum of the guided wave signal changes with time during propagation within an aluminum plate (isotropic medium). It can be observed that the power of the guided wave signal is concentrated, as expected, around $250$ kHz, which is the center frequency of the actuation signal. However, the power evolves over time due to the fact that the guided waves (Lamb waves) are dispersive in nature. Figure \ref{fig: spectrogram exp}(b) shows the 3D view of the spectrogram.

\subsection{Parametric Modeling}

\subsubsection {Prediction Under Varying Temperature: RML-TAR}

Once the non-parametric spectrogram analysis was performed, the parametric RML-TAR model identification was subsequently addressed to represent the non-stationary guided wave propagation under varying temperatures, extract the time-varying spectral content, and assess how the spectral content changes with temperature.

\begin{figure}[t!]
    \centering
    \begin{picture}(400,120)
    \put(-30,-50){ \includegraphics[width=0.5\columnwidth]{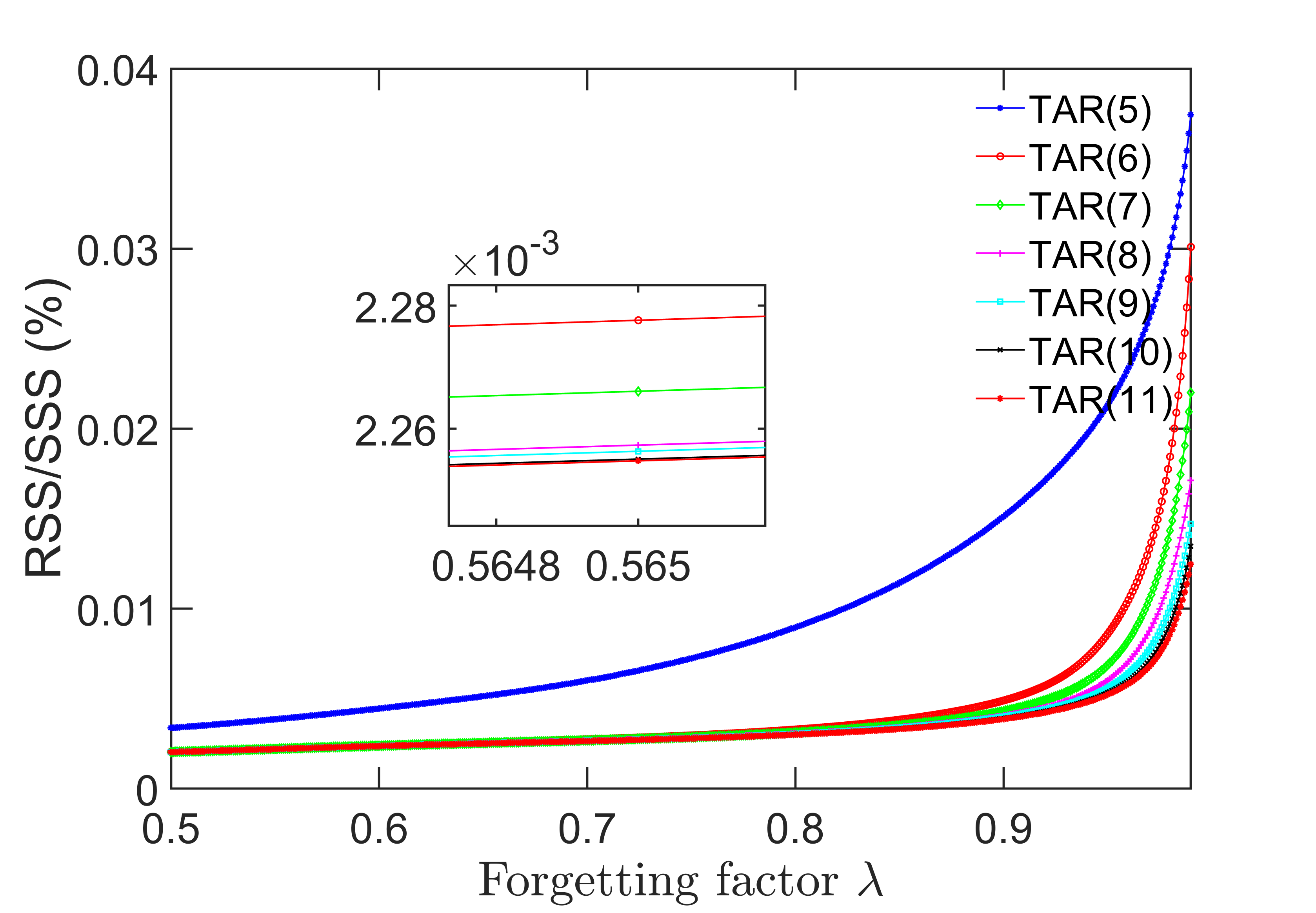}}
    \put(220,-50){\includegraphics[width=0.5\columnwidth]{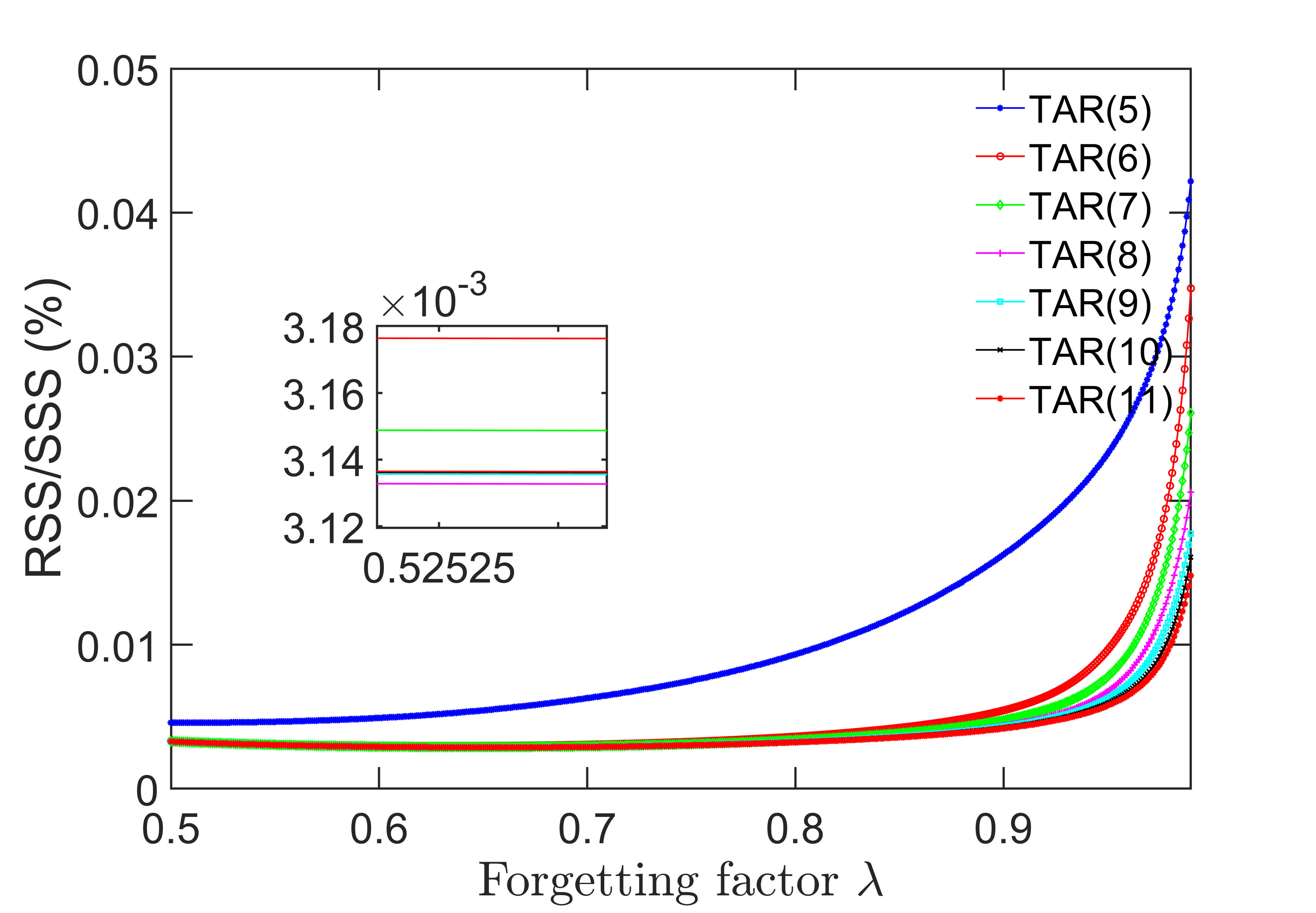}}
    
    \put(-35,110){\large \textbf{(a)}}
    \put(200,110){ \large \textbf{(b)}}
    \end{picture}
    \vspace{1.5cm}
    
    \caption{Model order selection of the RML-TAR model for one-step-ahead prediction via RSS/SSS criteria: (a) shows the plot of RSS/SSS versus forgetting factor for different model orders for a representative signal at $50^o$C; (b) similar plot for a representative signal at $95^o$C.} 
\label{fig: RSS} 
\end{figure} 
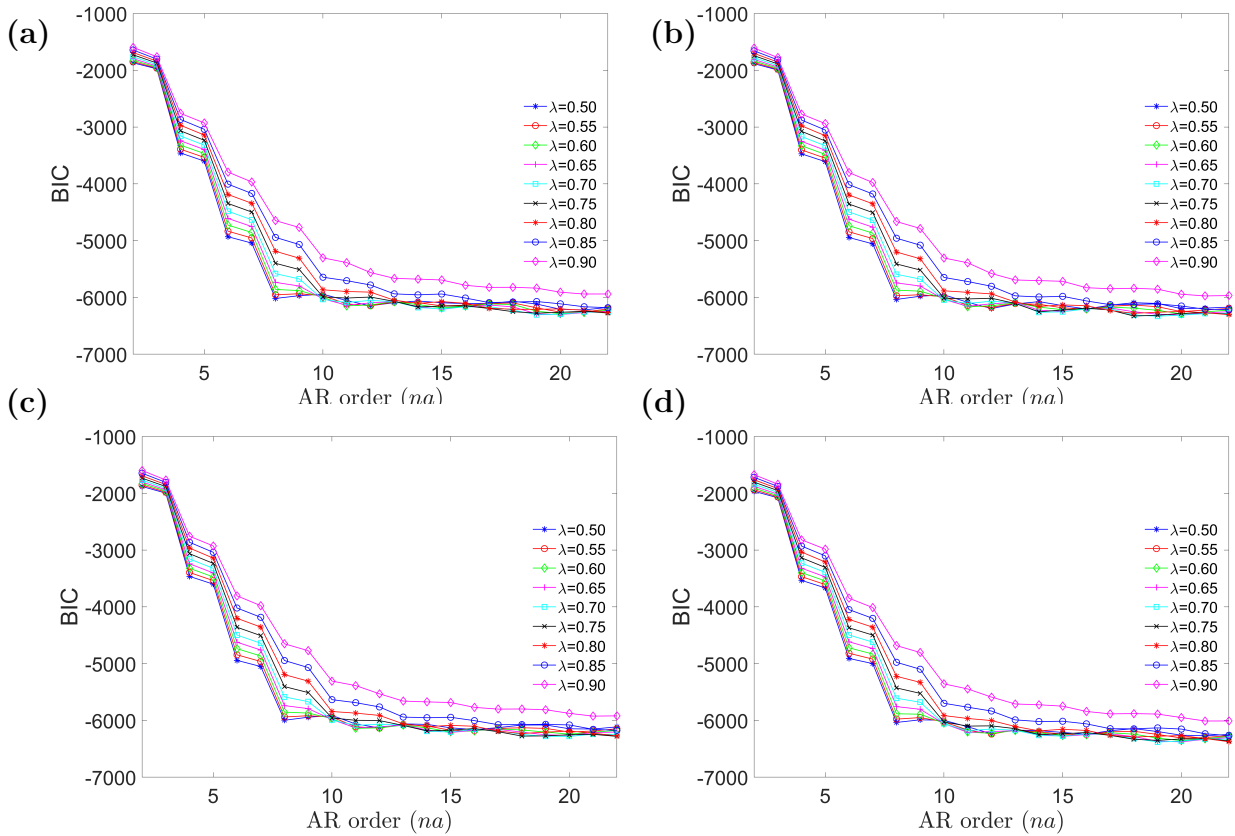
\begin{figure}[t!]
    \centering
    \begin{picture}(400,120)
    \put(-25,-40){\includegraphics[width=0.5\columnwidth]{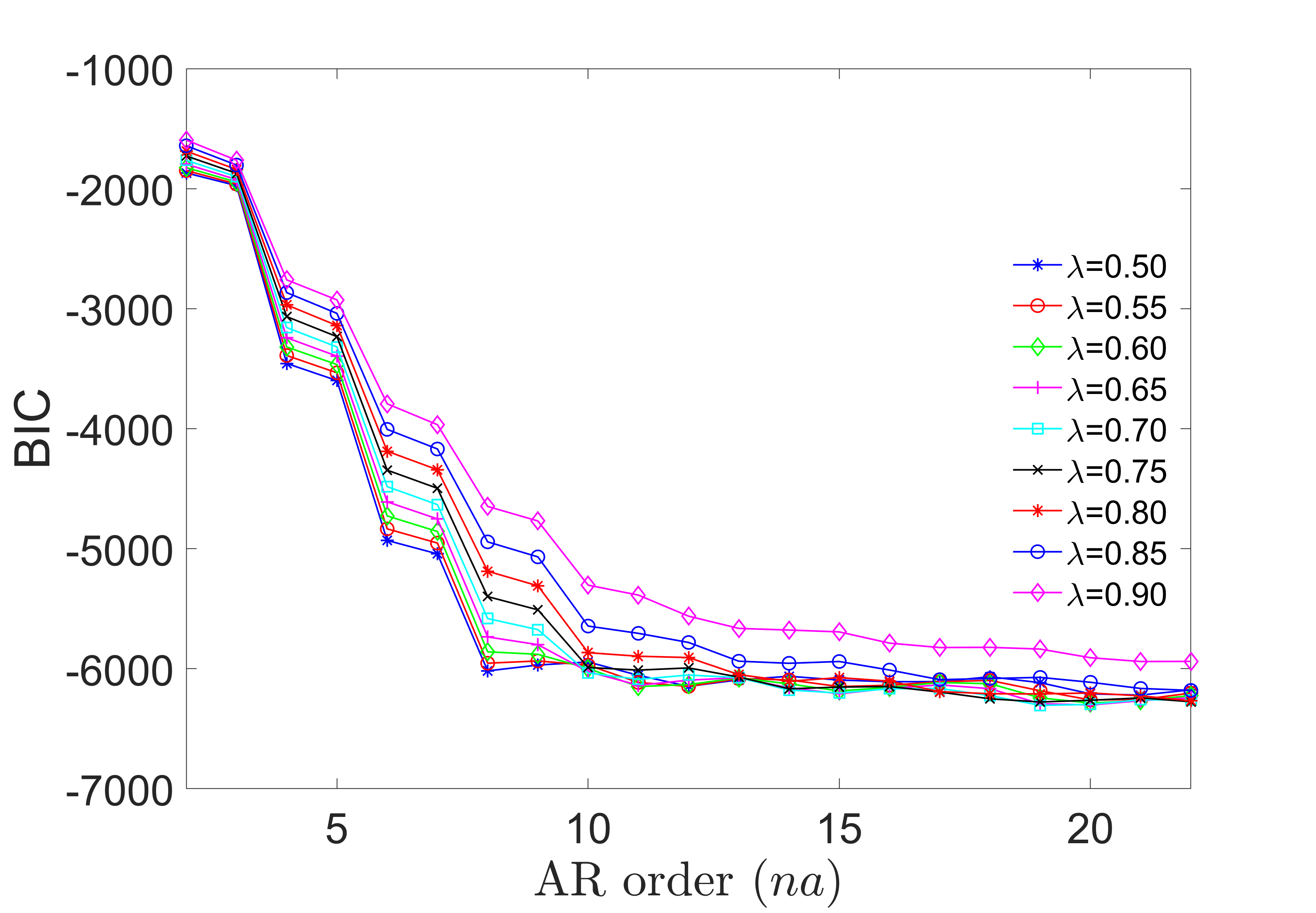}}
    \put(210,-40){\includegraphics[width=0.5\columnwidth]{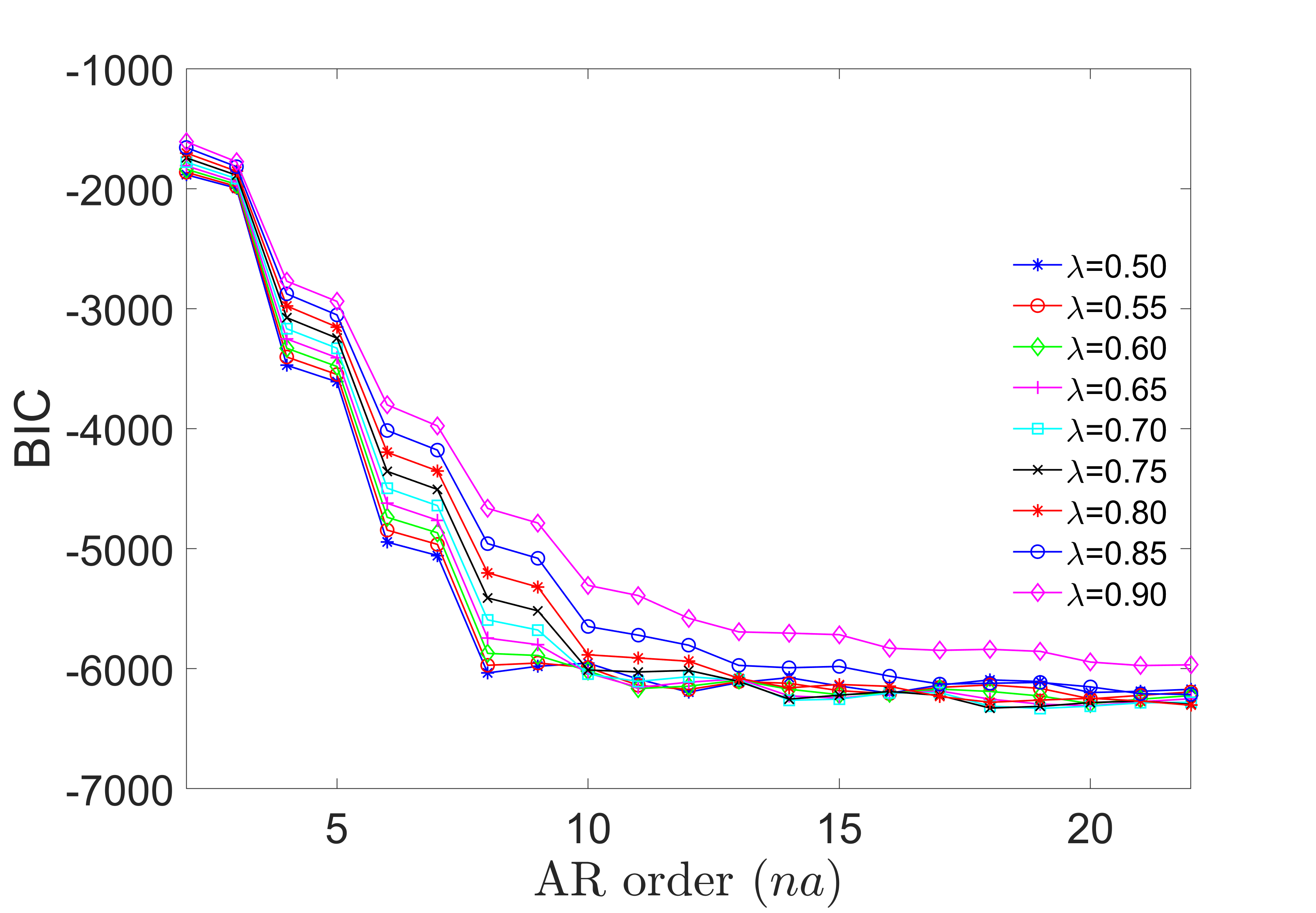}}
    
    \put(-25,-200){ \includegraphics[width=0.5\columnwidth]{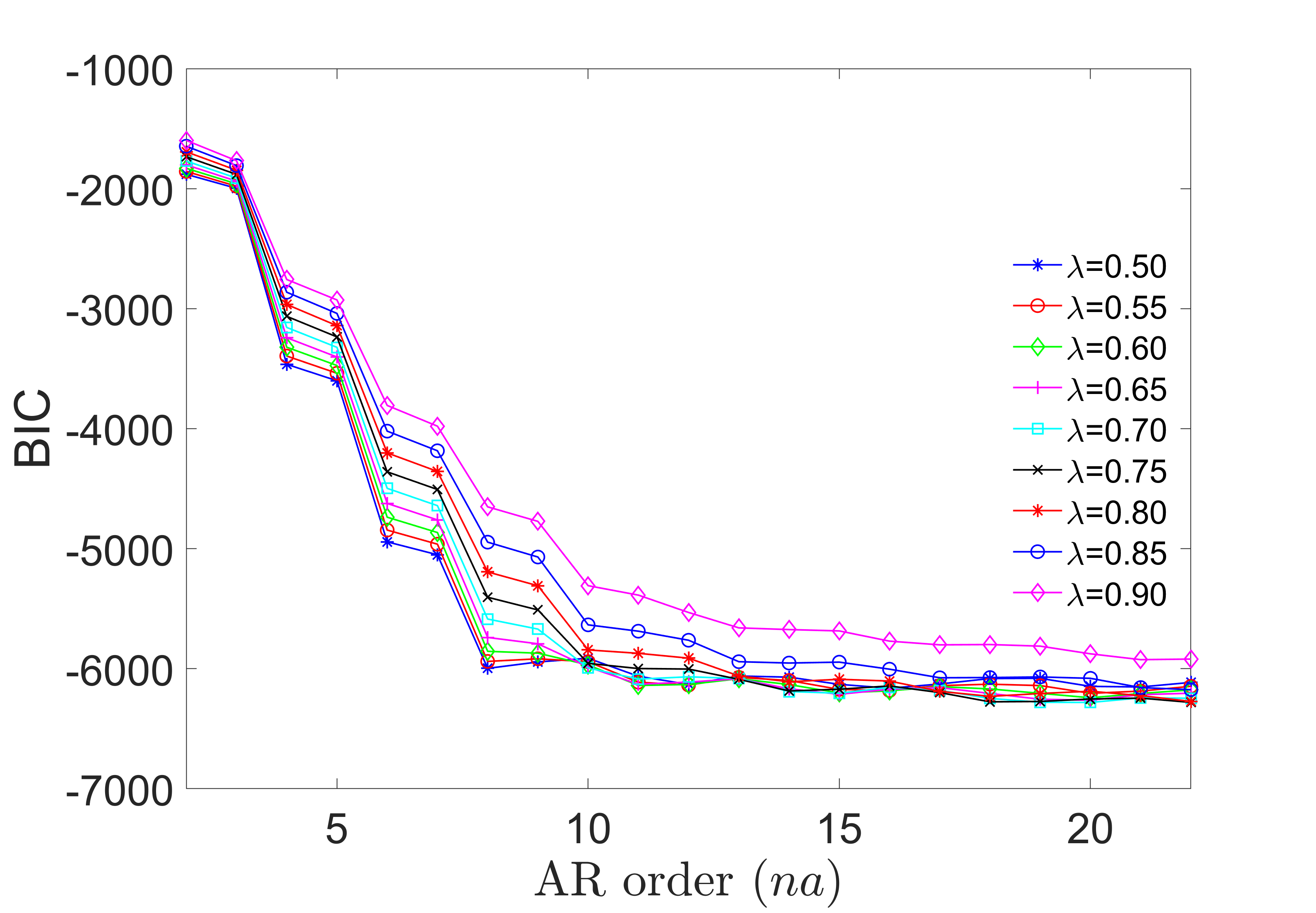}}
    \put(210,-200){\includegraphics[width=0.5\columnwidth]{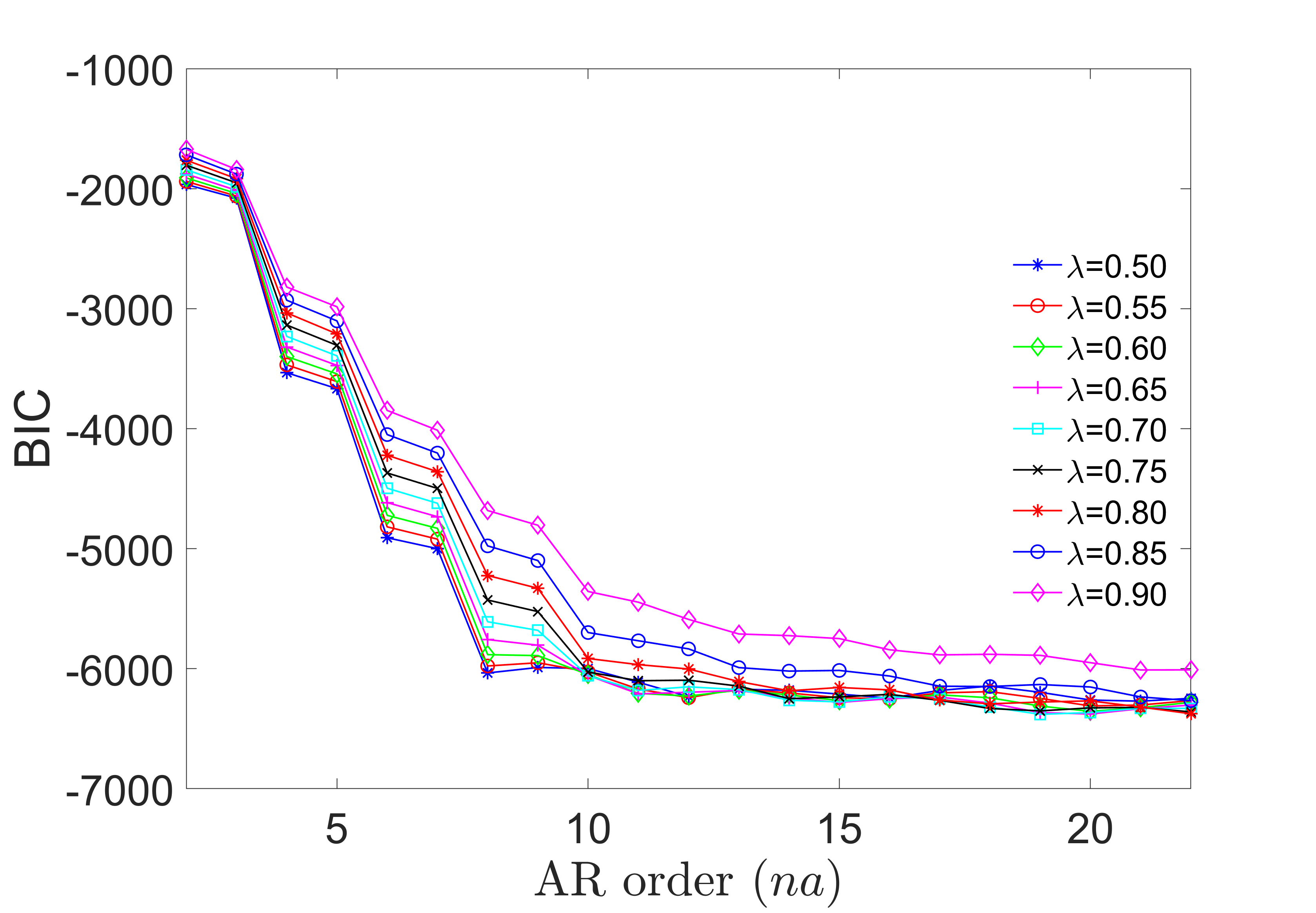}}
    \put(-40,100){\large \textbf{(a)}}
    \put(200,100){ \large \textbf{(b)}}
    \put(-40,-40){\large \textbf{(c)}}    
     \put(200,-40){\large \textbf{(d)}}
    \end{picture}
    \vspace{7cm}
    
    \caption{Model order selection of the RML-TAR model for one-step-ahead prediction using BIC criteria and shows the plot of the BIC versus the AR order for a representative signal at (a) $50^o$C; (b) $60^o$C; (c) $70^o$C; and (d) $100^o$C.} 
\label{fig: BIC} 
\end{figure} 
%
Model selection of  RML-TAR involves selecting the appropriate model order $na$ and the forgetting factor $\lambda$. The RSS/SSS (Residuals Sum of Squares/Signal Sum of Squares) criterion,  describing  the  predictive  ability of the model, was employed for the model selection process. AR  orders from $na=$ 2 to $na=$ 22 and forgetting factor $\lambda \in [0.5,0.999]$ (with an incremental step of 0.001) were considered to create a pool of candidate models. A total of 10500 models (500 $\times$ 21) were estimated, and among all these models, the best model was chosen as the one that minimizes the RSS/SSS. Following this criterion, for the signal at $50^o$C, the best model occurred at $na=$ 16 and forgetting factor $\lambda=$ 0.5. This can be compactly represented as RML-TAR$(16)_{0.5}$. However, selecting the model order $na=$ 6 and forgetting factor $\lambda = 0.5$ provides percentage RSS/SSS values that are on the same order of magnitude. In fact, the difference between the percentage RSS/SSS values between the model order $na=$ 16 and $na=$ 6 is on the order of $10^{-5}$. Hence, a model order of $na=$ 6 was chosen for the sake of model parsimony and to avoid overfitting. Again, the selection of the model order following the RSS/SSS criterion for a signal at a different temperature other than the $50^o$C results in a different model order and forgetting factor. 
In addition to the RSS/SSS criterion, Bayesian Information Criterion (BIC), which rewards the model's predictive capability while penalizing model complexity for increasing model order \cite{poulimenos2006parametric}, and ``frozen-time" natural frequencies were also taken into account. A higher model order may lead to spurious natural frequencies which are unlikely to occur in the system being considered due to the narrowband, deterministic excitation. On the other hand, a lower model order will be unable to capture some of the frequencies that are truly present in the real system.
\begin{figure}[t!]
    \centering
    \begin{picture}(400,140)
    \put(-60,-40){\includegraphics[width=0.59\columnwidth]{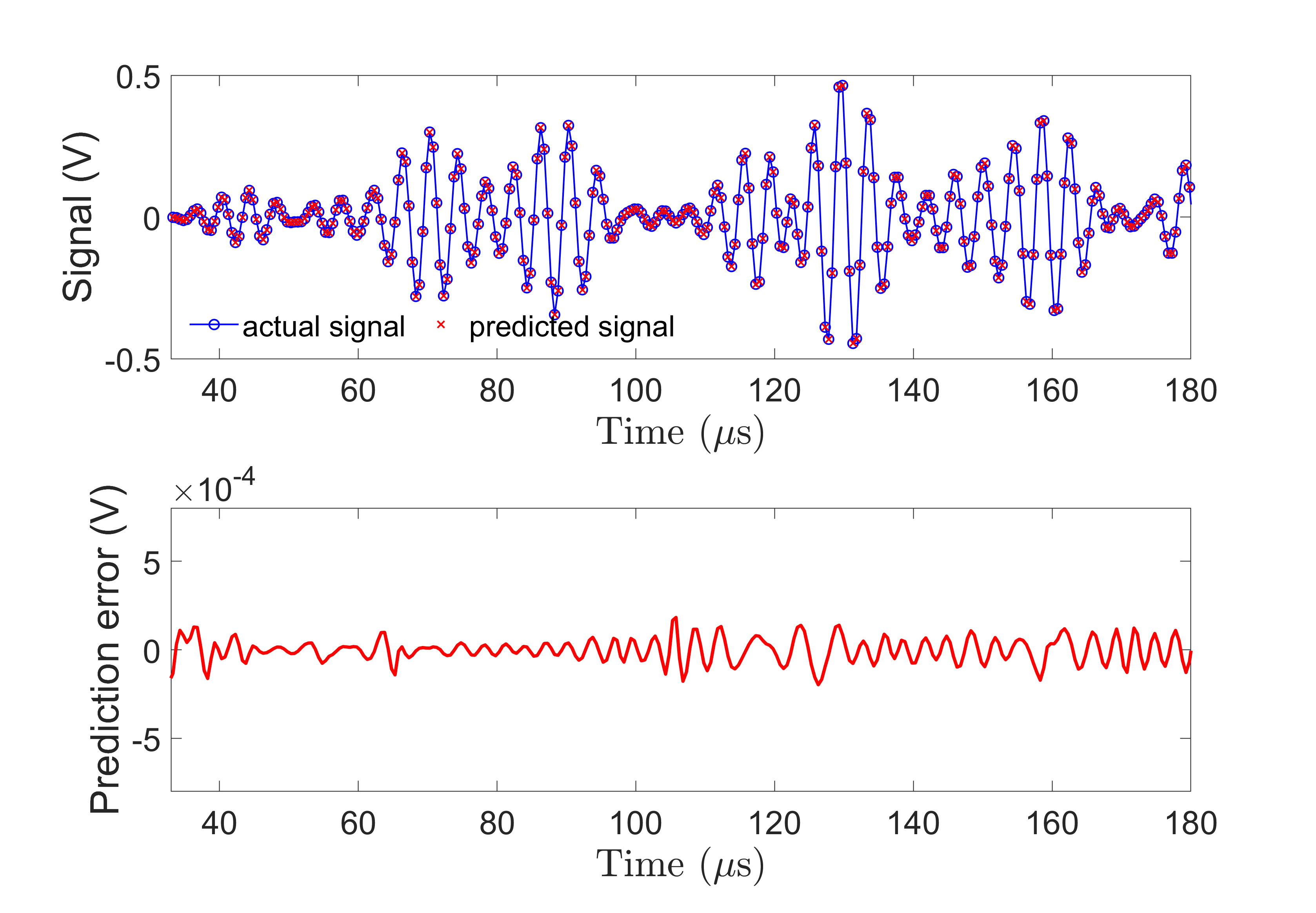}}
    \put(197,-40){\includegraphics[width=0.59\columnwidth]{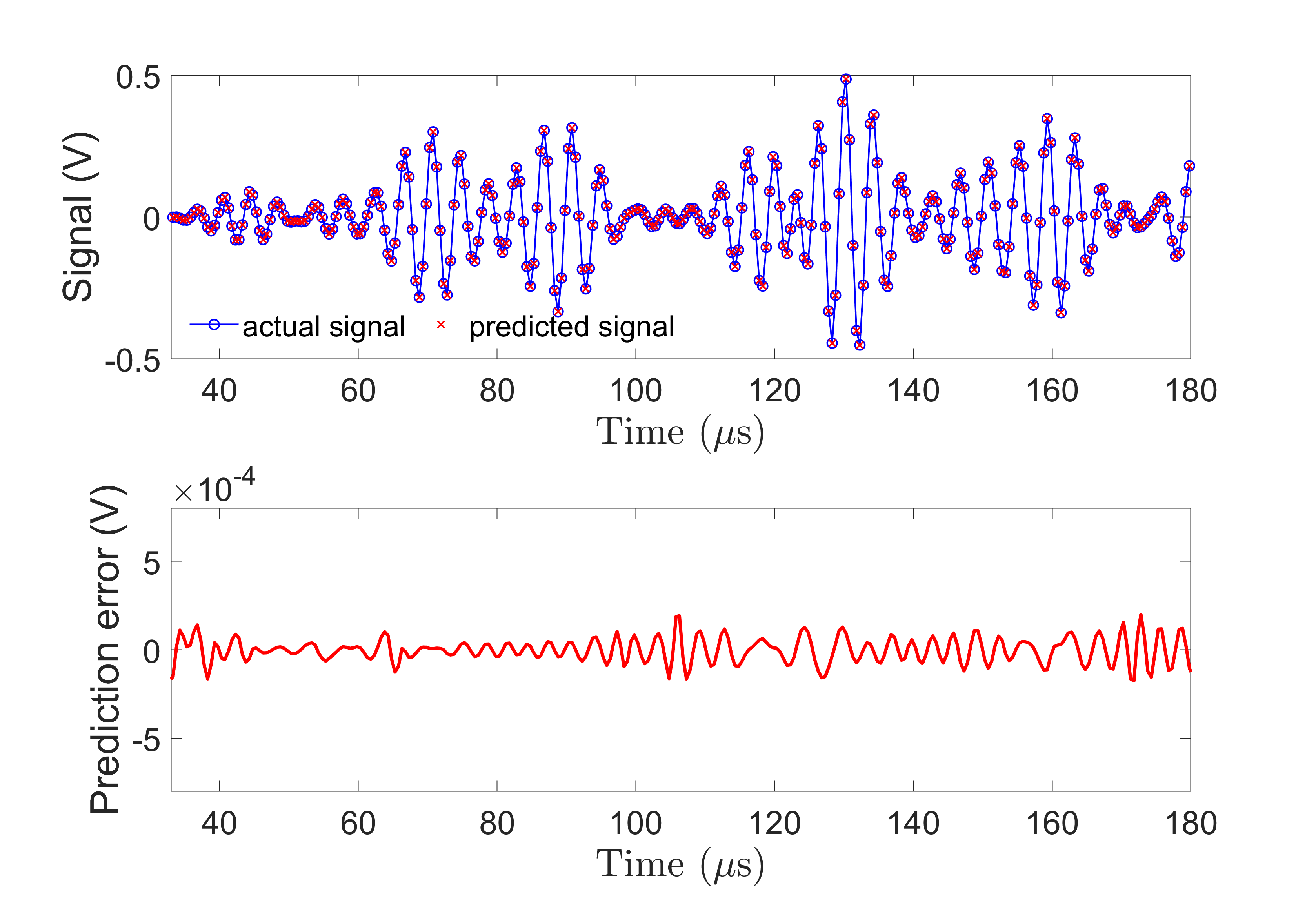}}
    \put(-60,-230){\includegraphics[width=0.59\columnwidth]{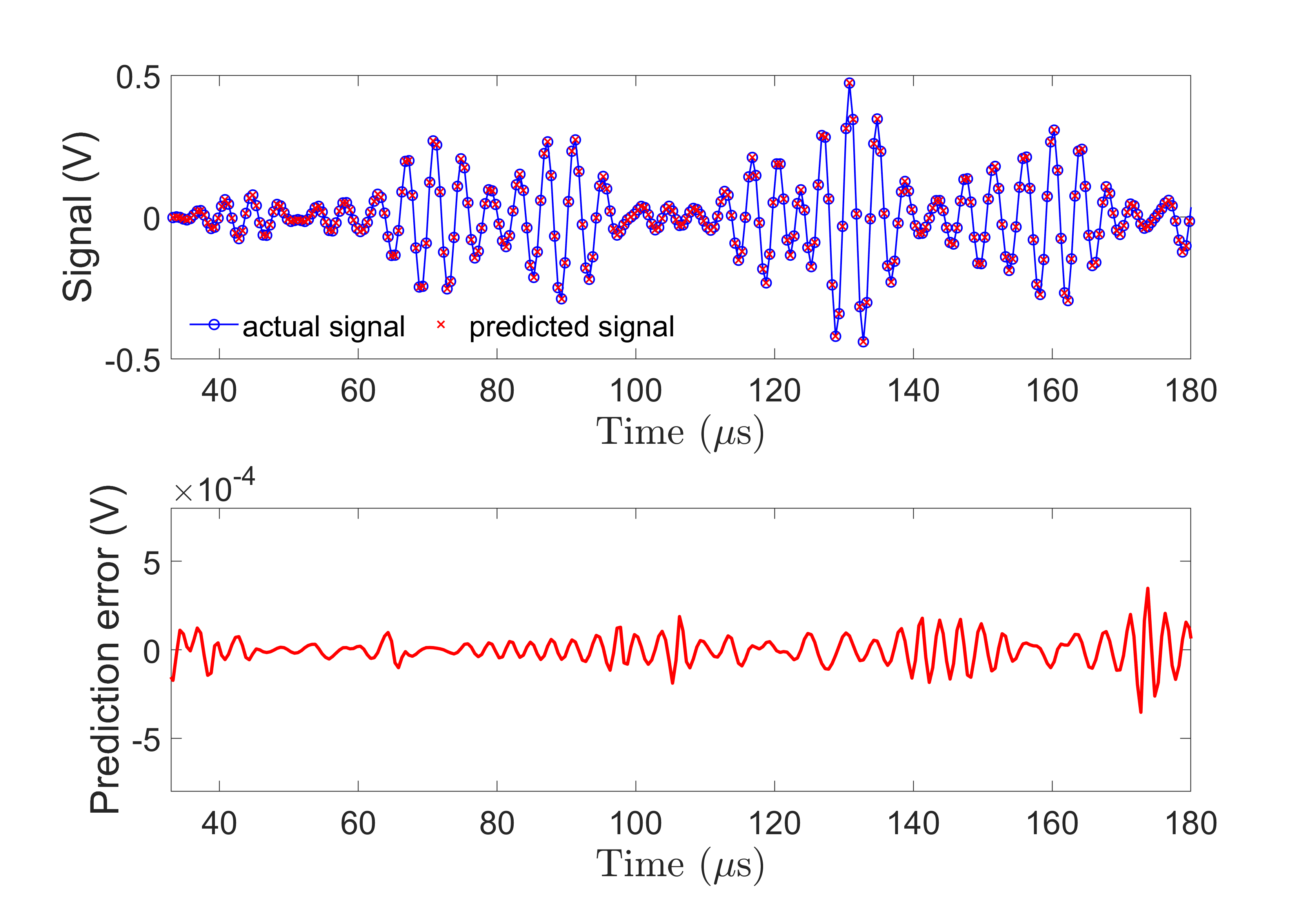}}
    \put(197,-230){\includegraphics[width=0.59\columnwidth]{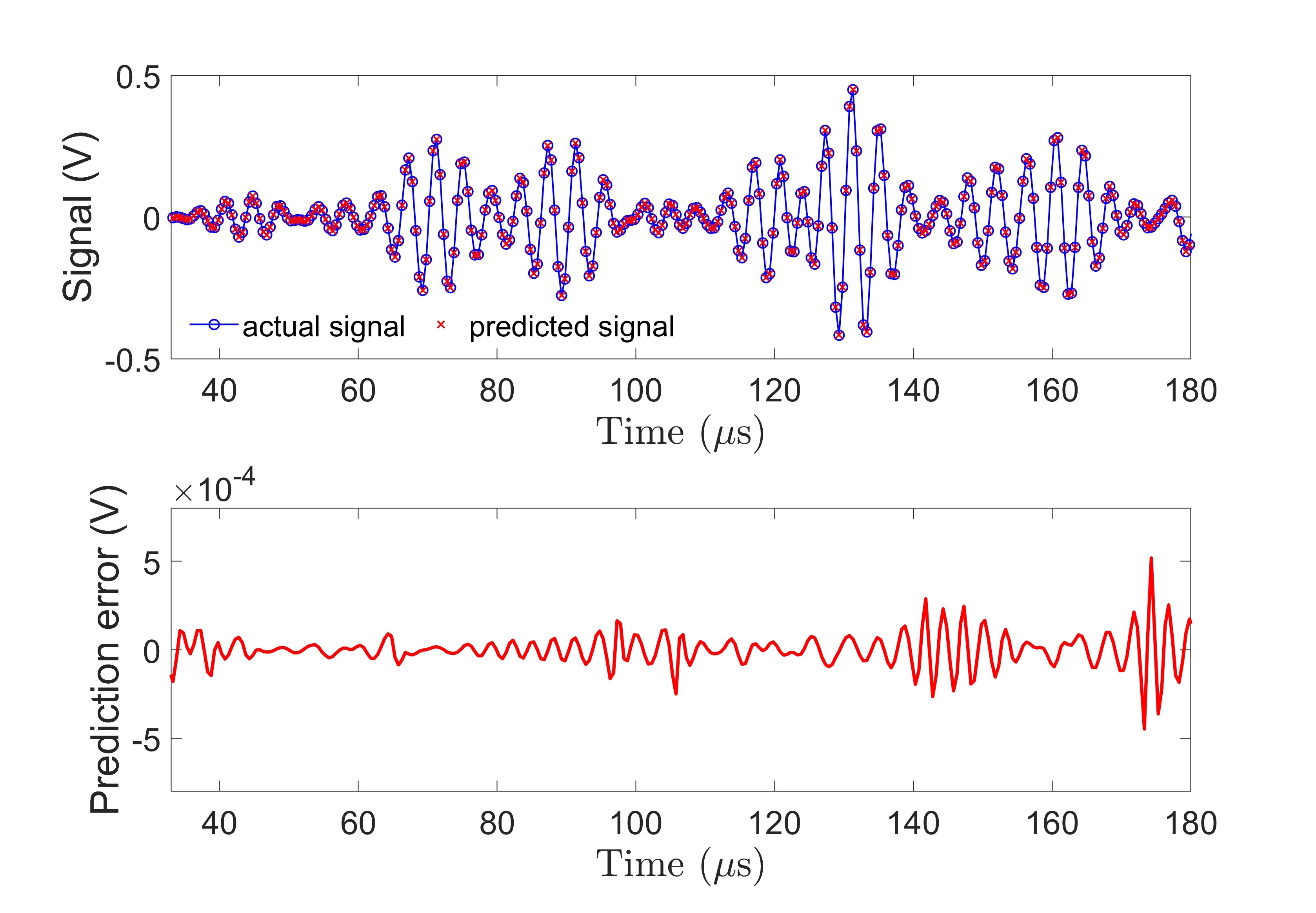}}
    
    \put(-45,145){ \large \textbf{(a)}}
    \put(210,145){ \large \textbf{(b)}}
    \put(-45,-40){\large \textbf{(c)}} 
    \put(210,-40){\large \textbf{(d)}} 
    \end{picture}
    \vspace{8cm}
    \caption{Segment of a non-stationary guided wave signal realization ($S_0$, $A_0$ mode and part of the reflected signal) superimposed with RML-TAR based one-step-ahead prediction and the corresponding residual obtained by RML-TAR$(6)_{0.6}$: (a) $50^o$C; (b) $70^o$C; (c)$90^o$C; and (d) $100^o$C. } 
\label{fig: one step ahead} \vspace{-12pt}
\end{figure}
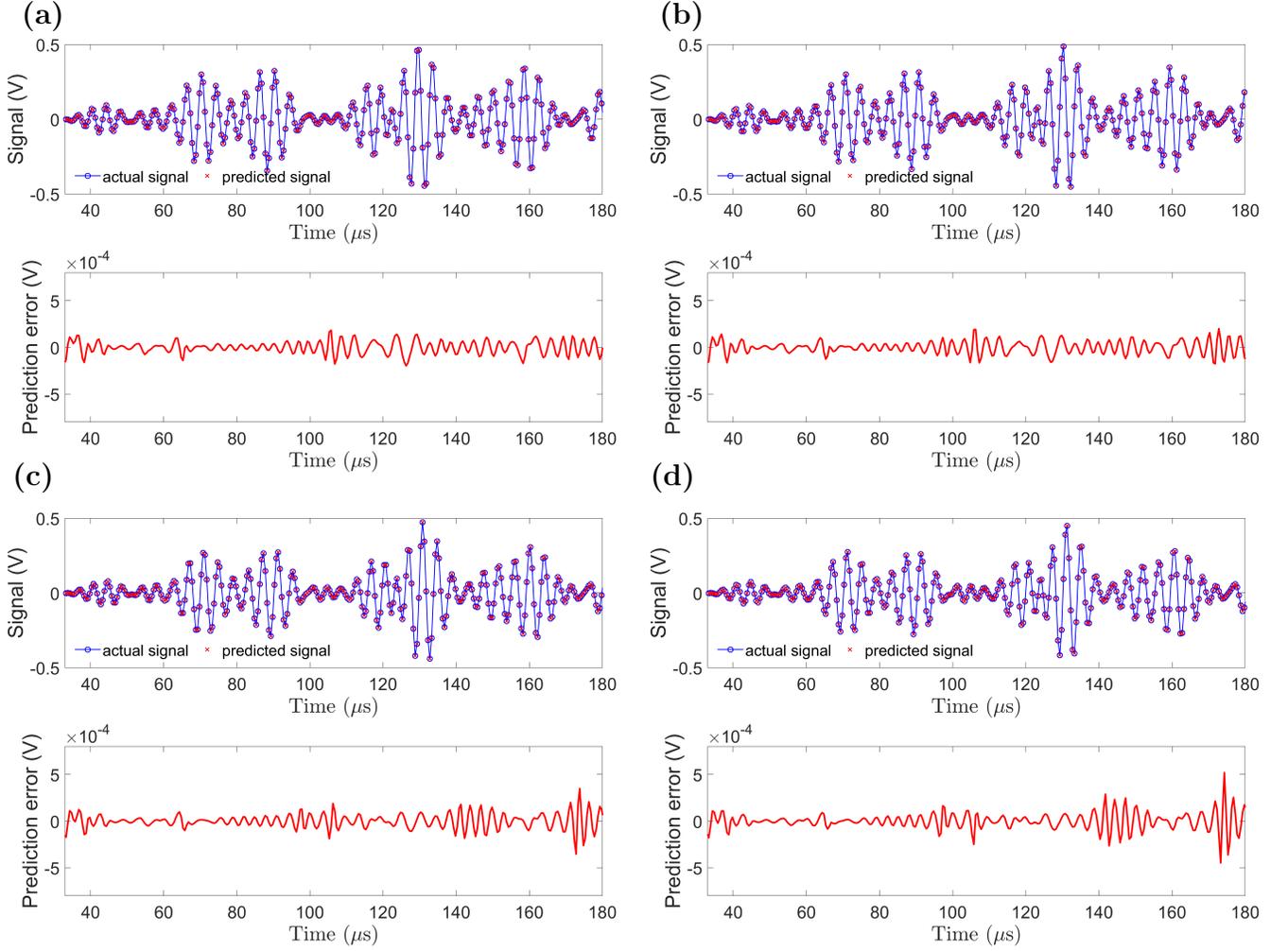 
%

Figure \ref{fig: RSS}(a) and (b) shows the RSS/SSS versus the forgetting factor plot for a representative data set at $50^o$C and $95^o$C, respectively, for a few selected model orders. It can be observed that (for both figures), for RML-TAR$(5)$, as the value of the forgetting factor increases, the RSS/SSS also increases. For RML-TAR$(6)$, RML-TAR$(7)$ ... RML-TAR$(11)$, the RSS/SSS has a plateau for a forgetting factor values between 0.5 to 0.9, and then sharply increases as it approaches 1. The overall trend for both Figures, although for two different temperatures, is similar. As a result, it is possible to identify a single model structure that will be able to model and predict the guided wave propagation for a range of different temperatures.


This fact can also be supported by looking at the plot of BIC versus model order, all of which has a similar trend for different temperatures. Figure \ref{fig: BIC}(a),(b),(c), and (d) shows the plot of BIC versus AR order for different forgetting factor values for a representative guided wave signal at $50^o$C, $60^o$C, $70^o$C and $100^o$C, respectively. It can be observed that as the model order increases, the value of the BIC sharply decreases until AR order $na=$ 8 and then reaches a plateau for higher model orders. Again, for a specific model order, the BIC values are higher for the forgetting factors closer to 1. Although selecting a higher model order provides a lower BIC value, it will also provide additional spurious ``frozen-time'' natural frequencies due to model order overdetermination.
\begin{figure}[t!]
    \centering
    \begin{picture}(400,240)
    \put(-70,50){\includegraphics[width=0.6\columnwidth]{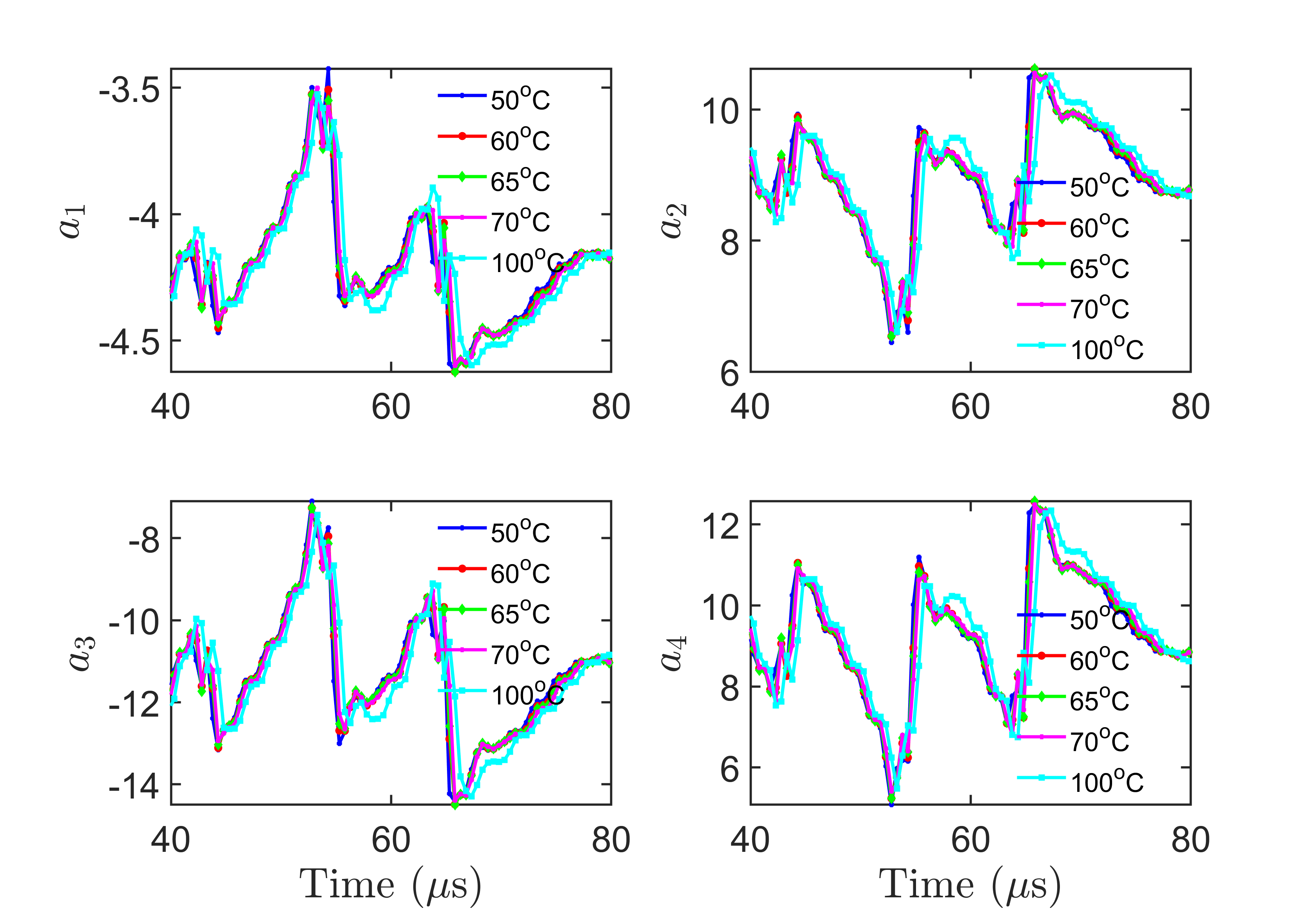}}
    \put(190,50){\includegraphics[width=0.6\columnwidth]{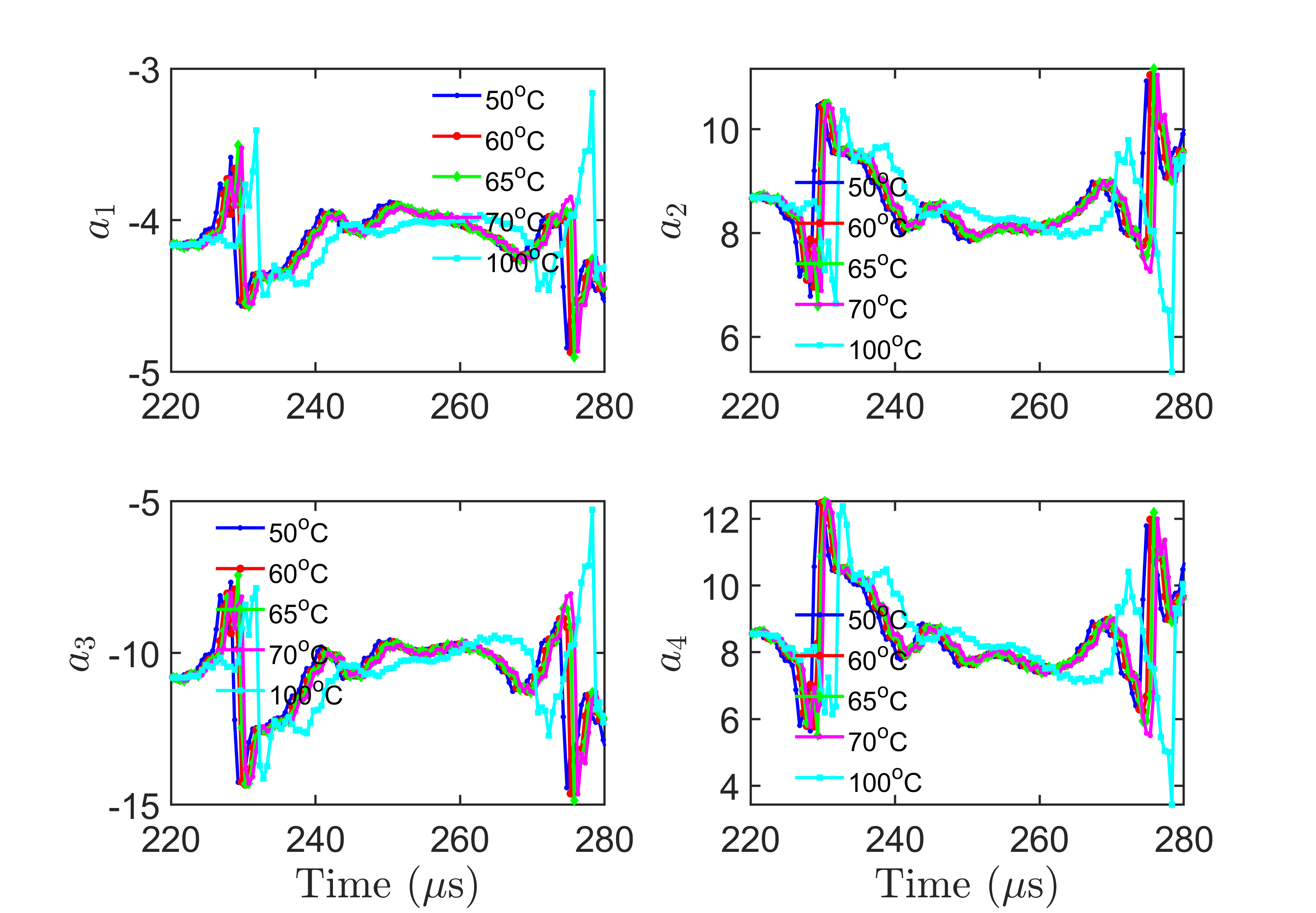}}
    \put(-70,-145){\includegraphics[width=0.6\columnwidth]{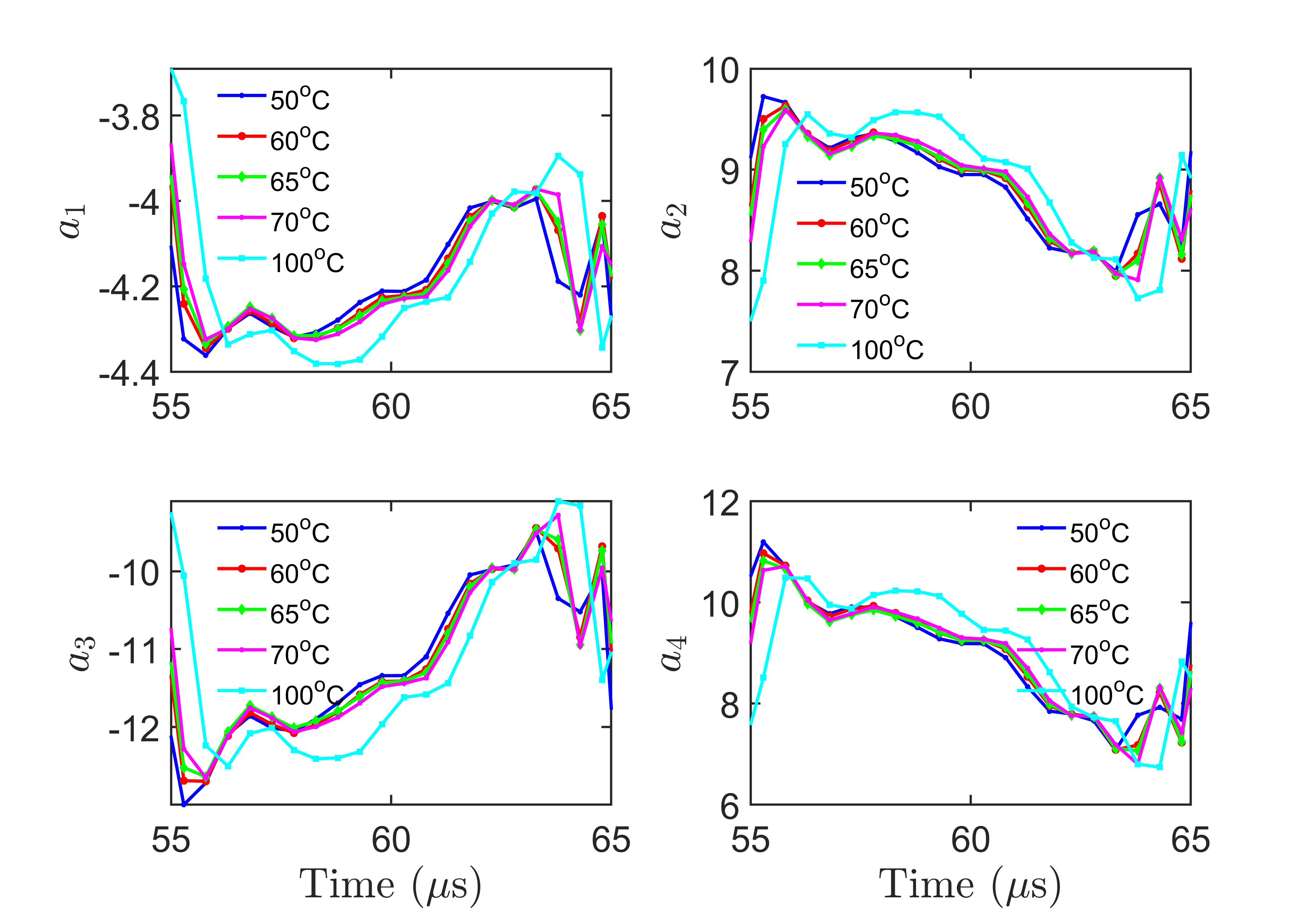}}
    \put(190,-145){\includegraphics[width=0.6\columnwidth]{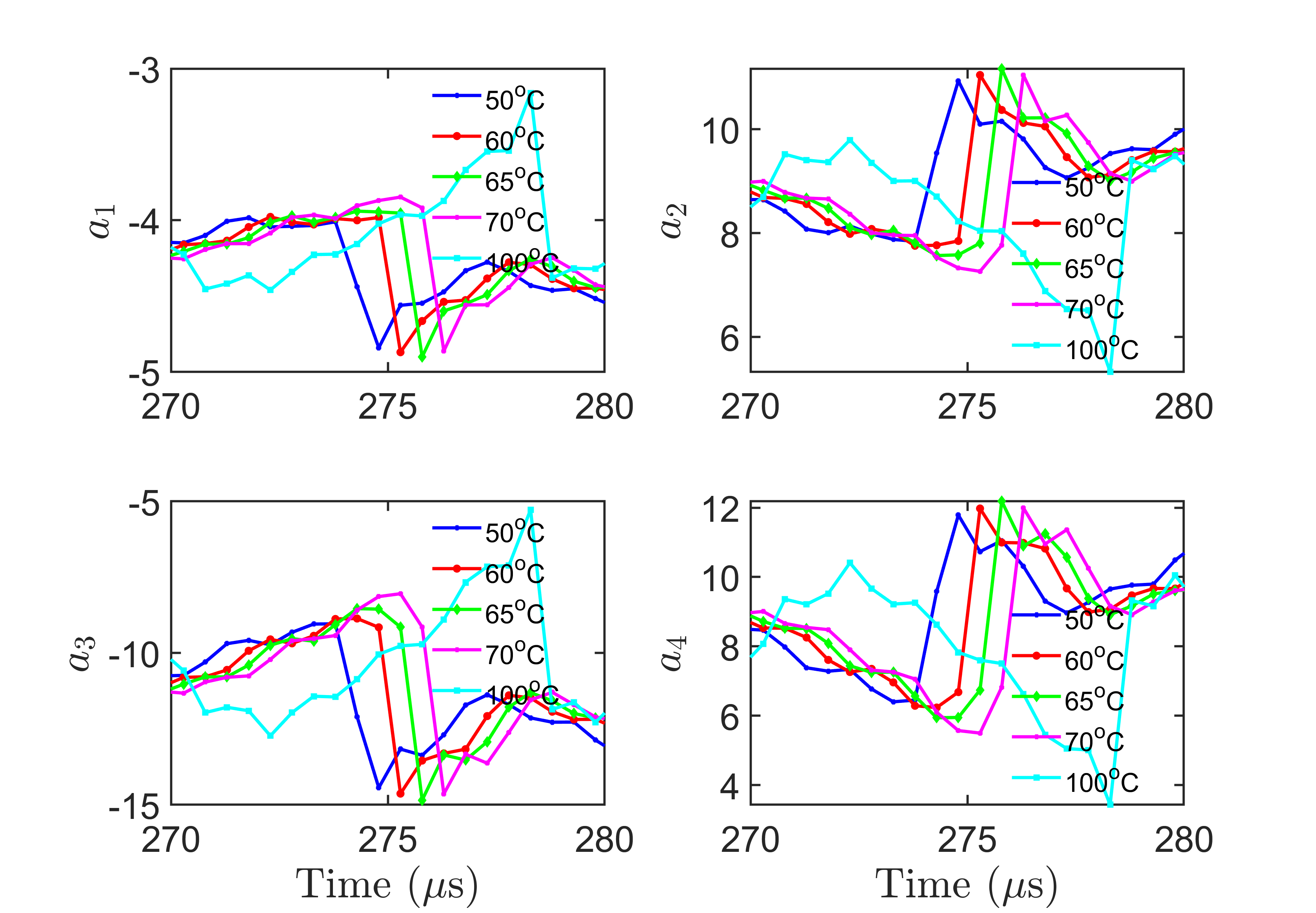}}
    \put(-60,240){ \large \textbf{(a)}}
    \put(200,240){\large \textbf{(b)}}
    \put(-60,50){\large \textbf{(c)}}    
    \put(200,50){\large \textbf{(d)}}  
    \end{picture}
    \vspace{5cm}
    \caption{First four model parameters obtained by a single model structure RML-TAR$(6)_{0.6}$ for five different temperatures: (a) model parameters in the $S_0$ and $A_0$ mode; (b) model parameters in the region of the reflected signal; (c) zoomed-in view of the model parameters in the non-reflected region; (d) zoomed-in view of the model parameters in the reflected region.} 
\label{fig: model par} 
\end{figure}
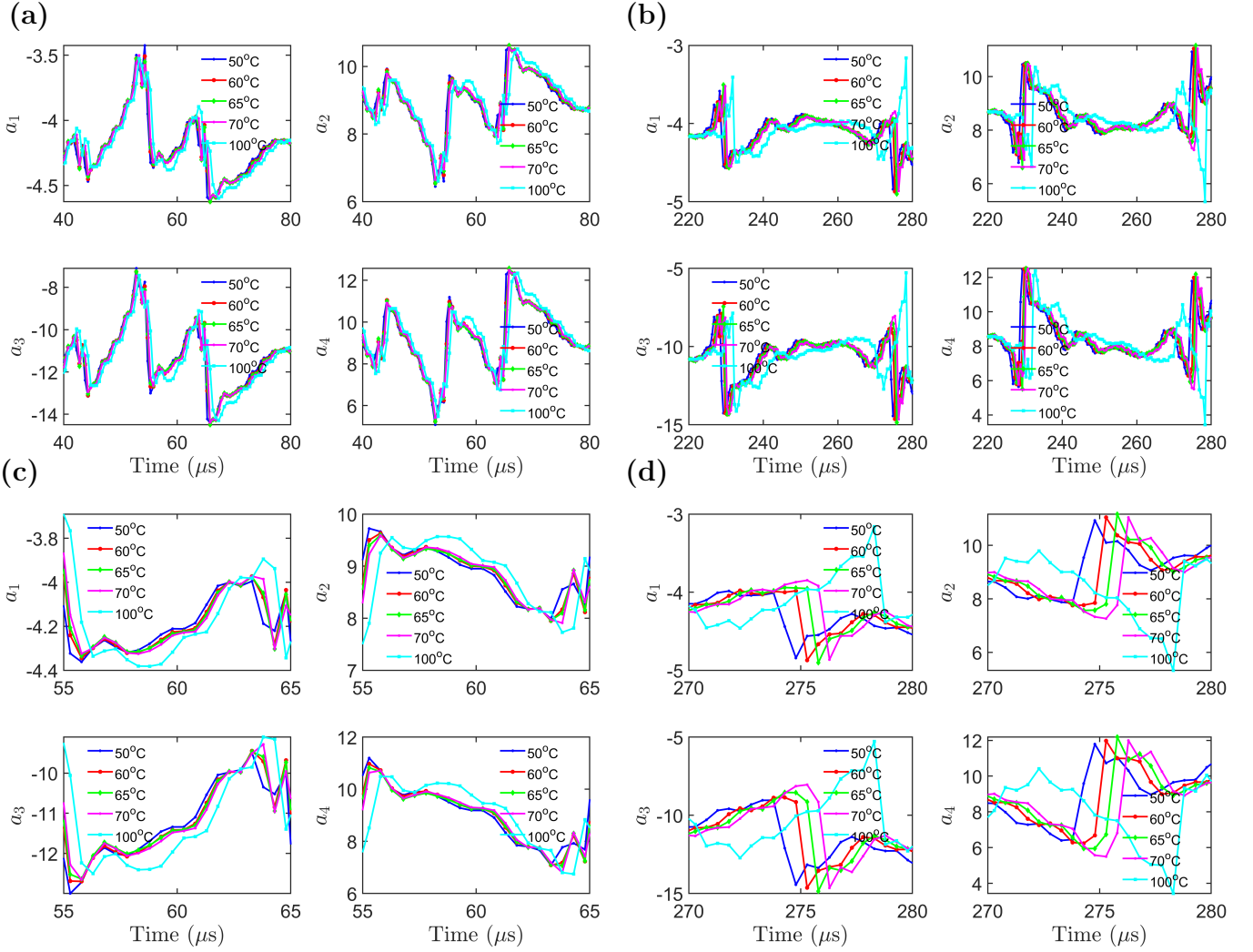 
%
Upon consideration of all these factors, the final model order for one-step-ahead prediction of the guided wave signal for modeling at a range of different temperatures was selected as RML-TAR$(6)_{0.6}$.

For the estimation of the parameters of the selected model structure, a recursive estimation scheme was used as outlined in Section \ref{sec: model par est}. At the beginning of time, the estimation accuracy of the algorithm is low as there are a limited number of data points available. There is also the effect of unknown initial conditions. In order to minimize these effects, parameter estimation was divided into three steps, namely: i) forward pass; ii) backward pass; and iii) final forward pass. In the forward pass, initial conditions for the parameters were assumed zero, and a larger value of the covariance matrix was assumed ($P = \alpha I$). A larger value of the covariance matrix at the beginning reflects that there is lower confidence in the assumed initial conditions. After several iterations, as more data becomes available, the parameter estimation converges to its true parameter values. In the backward pass, the final values of the signal are used as the initial values and the final estimated parameters and covariance matrix from the forward pass are used as the initial conditions. In the final forward pass, the final estimated parameters and covariance from the backward pass are used as the initial conditions. This helps achieve better parameter estimation at the beginning of time as informed initial conditions are being used in the estimation algorithm. All the results presented in this section are for the final forward pass.

Figure \ref{fig: one step ahead}(a), (b), (c), and (d) shows the  RML-TAR$(6)_{0.6}$ based one-step-ahead prediction of the segment of the guided wave signal ($S_0$, $A_0$ mode and part of reflection) and their corresponding prediction error for $50^o$C, $70^o$C, $90^o$C, and $100^o$C. It can be observed that an excellent match exists between the experimental guided wave signal and the one-step-ahead prediction using a single model structure for different temperatures. The prediction error is also very low (on the order of $10^{-4}$).

Figure \ref{fig: model par} shows the first four model parameters of the RML-TAR$(6)_{0.6}$ model for five different temperatures, namely: $50^o$C, $60^o$C, $65^o$C, $70^o$C, and $100^o$C. Figure \ref{fig: model par}(a) shows the segment of the model parameter from 40 $\mu$s to 80 $\mu$s that corresponds to the major portion of the $S_0$ and $A_0$ mode of the guided wave signal (non-reflected part). Figure \ref{fig: model par}(b) shows the  segment of the model parameters from 220 $\mu$s to 280 $\mu$s that corresponds to the portion of the reflected signal. It can be observed that the model parameters corresponding to the reflection part of the signal shows higher variation than the non-reflection part of the signal ($S_0$ and $A_0$ mode). 

Figure \ref{fig: model par}(c) and (d) shows the zoomed-in view of the model parameters corresponding to the non-reflection and reflection part of the signal for $50^o$C, $60^o$C, $65^o$C, $70^o$C and $100^o$C, respectively. It can be observed that the parameters follow a certain trend with increasing temperature. As for example, from Figure \ref{fig: model par}(c), parameter $\alpha_4$, the notch around 56 $\mu$s shifts downward and moves to the right with the increase in temperature. Again, in Figure \ref{fig: model par}(d) which is zoomed-in view of the segment of the reflection part of the signal, it can be observed that the notch around 275 $\mu$s for $\alpha_1$, $\alpha_2$, $\alpha_3$, and $\alpha_4$ shifts to the right with the increase in temperature. This change is more pronounced in the reflection part of the signal rather than the non-reflection part. This observation of parameter change under varying temperature is commensurate with the change in the guided wave signals that shift to the right with the increase in temperature.

%
\begin{figure}[t!]
    \centering
    \begin{picture}(400,240)
    \put(-70,50){\includegraphics[width=0.59\columnwidth]{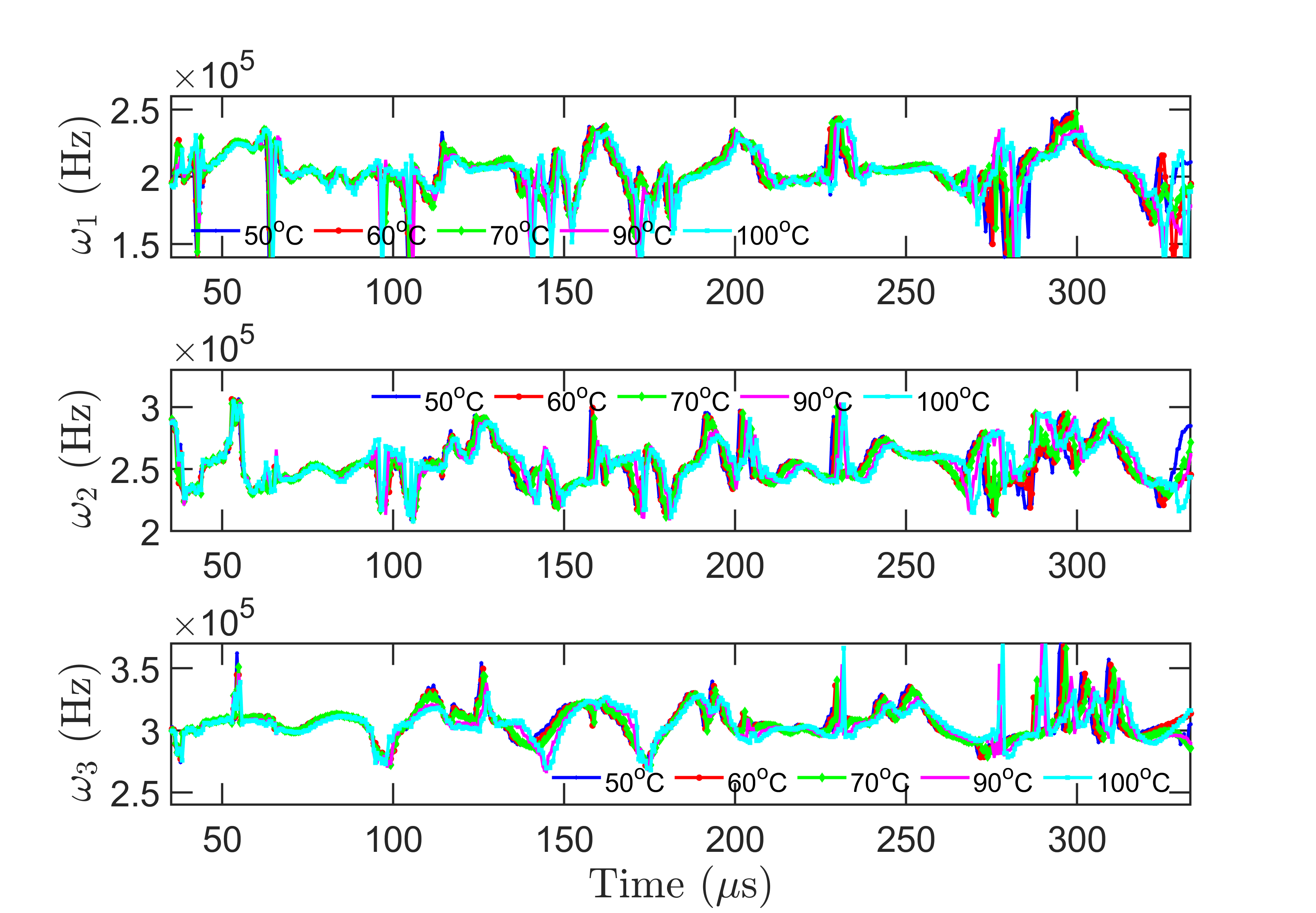}}
    \put(190,50){\includegraphics[width=0.59\columnwidth]{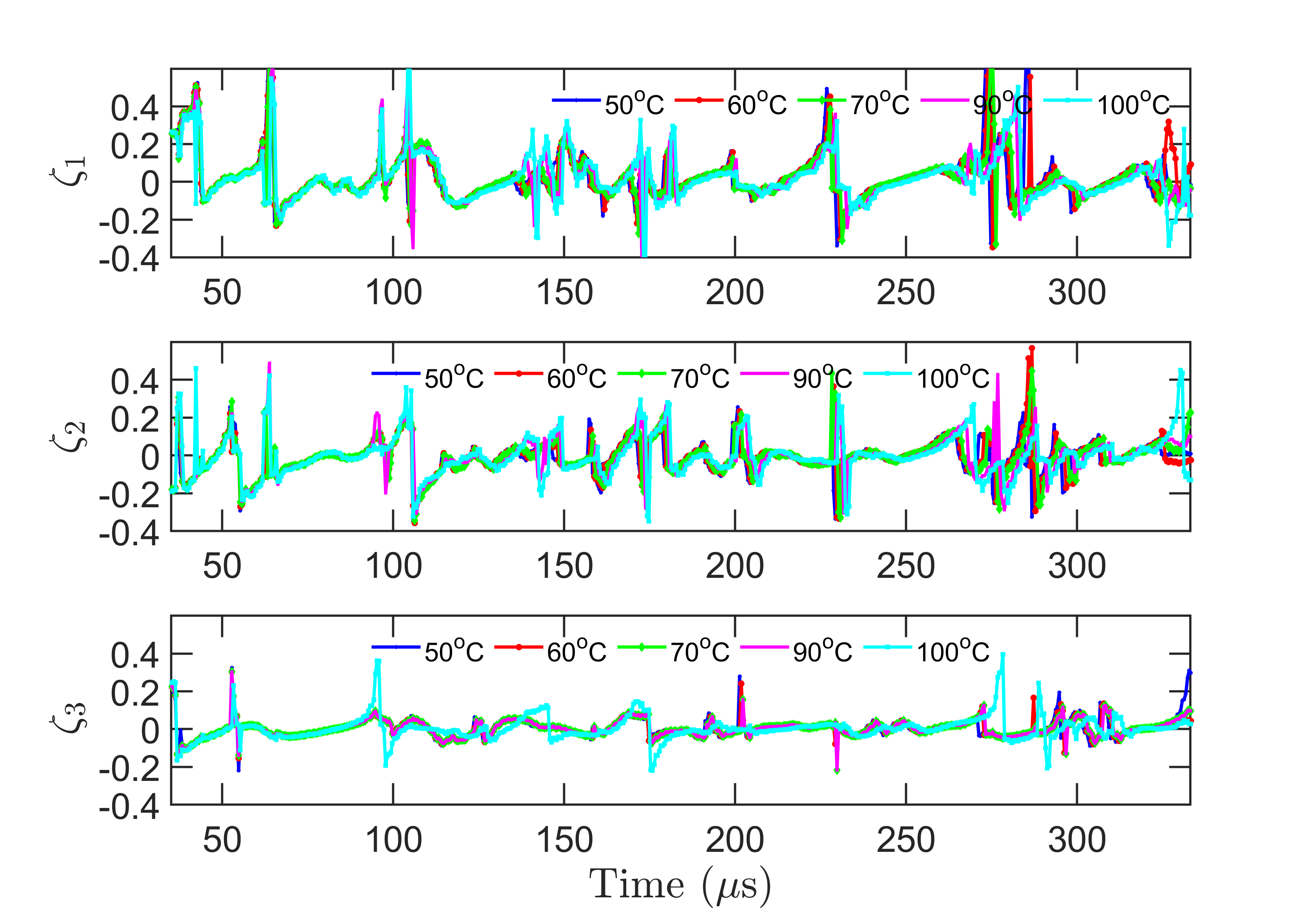}}
    \put(-70,-140){\includegraphics[width=0.59\columnwidth]{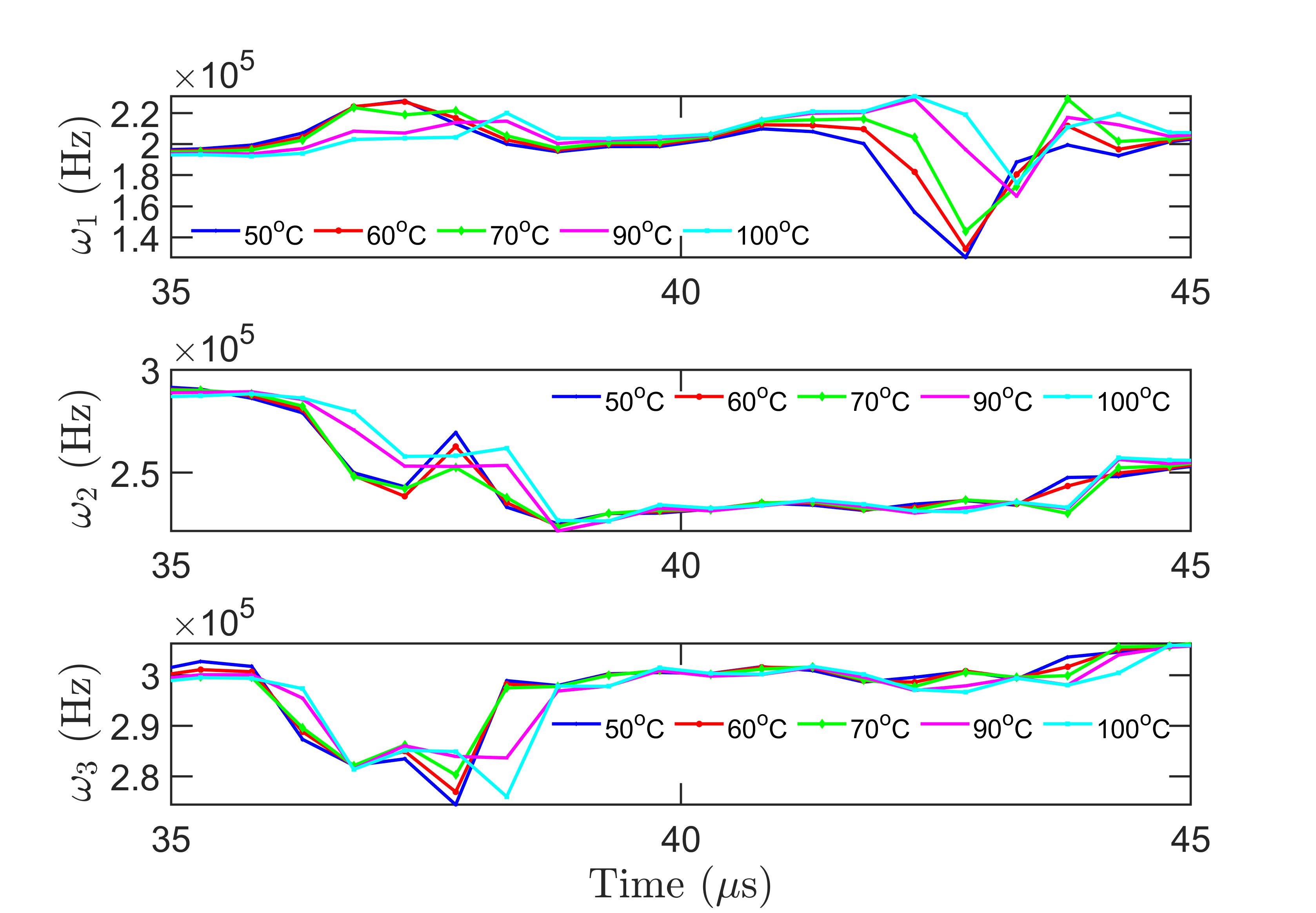}}
    \put(190,-140){\includegraphics[width=0.59\columnwidth]{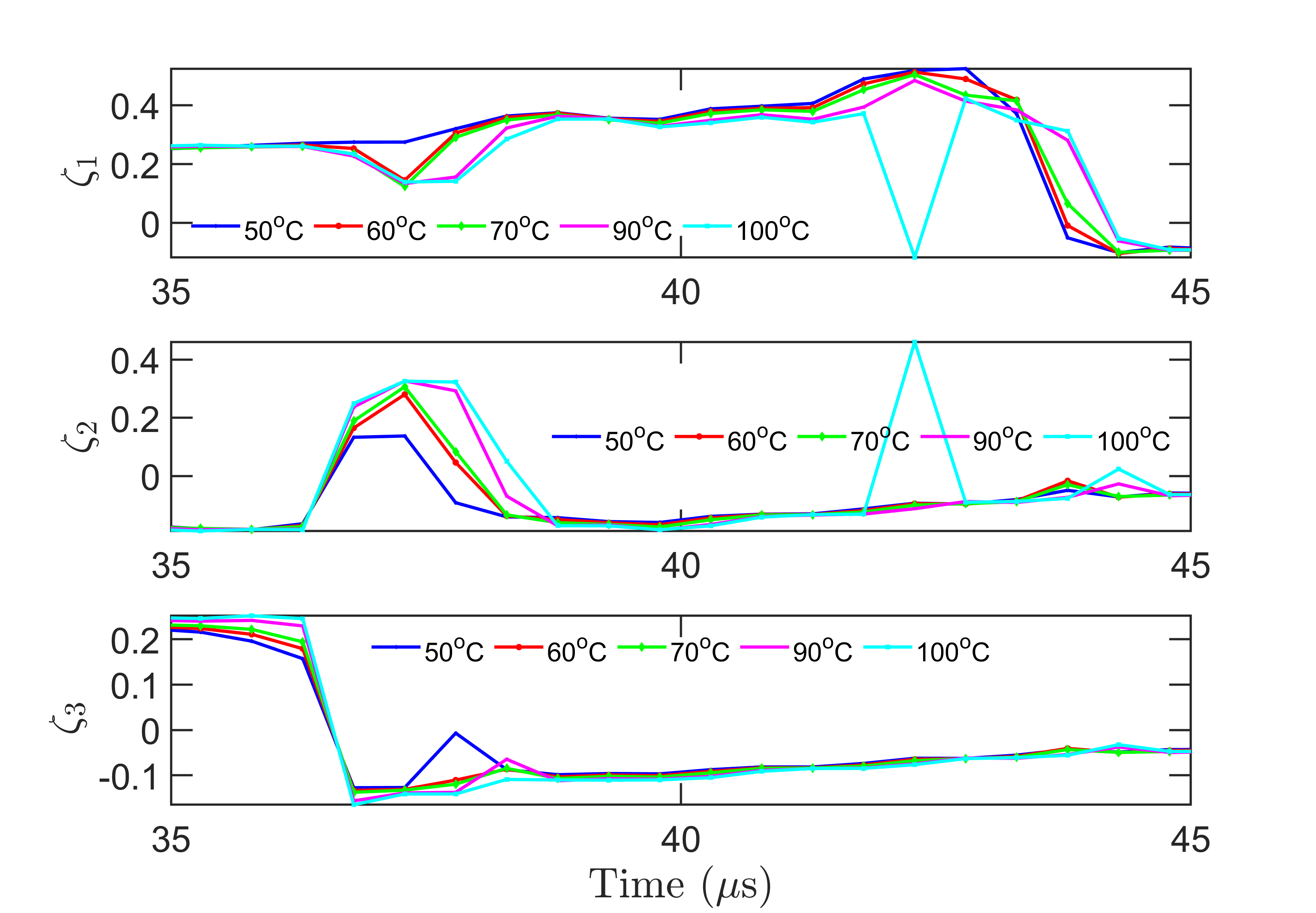}}
     \put(-70,-340){\includegraphics[width=0.59\columnwidth]{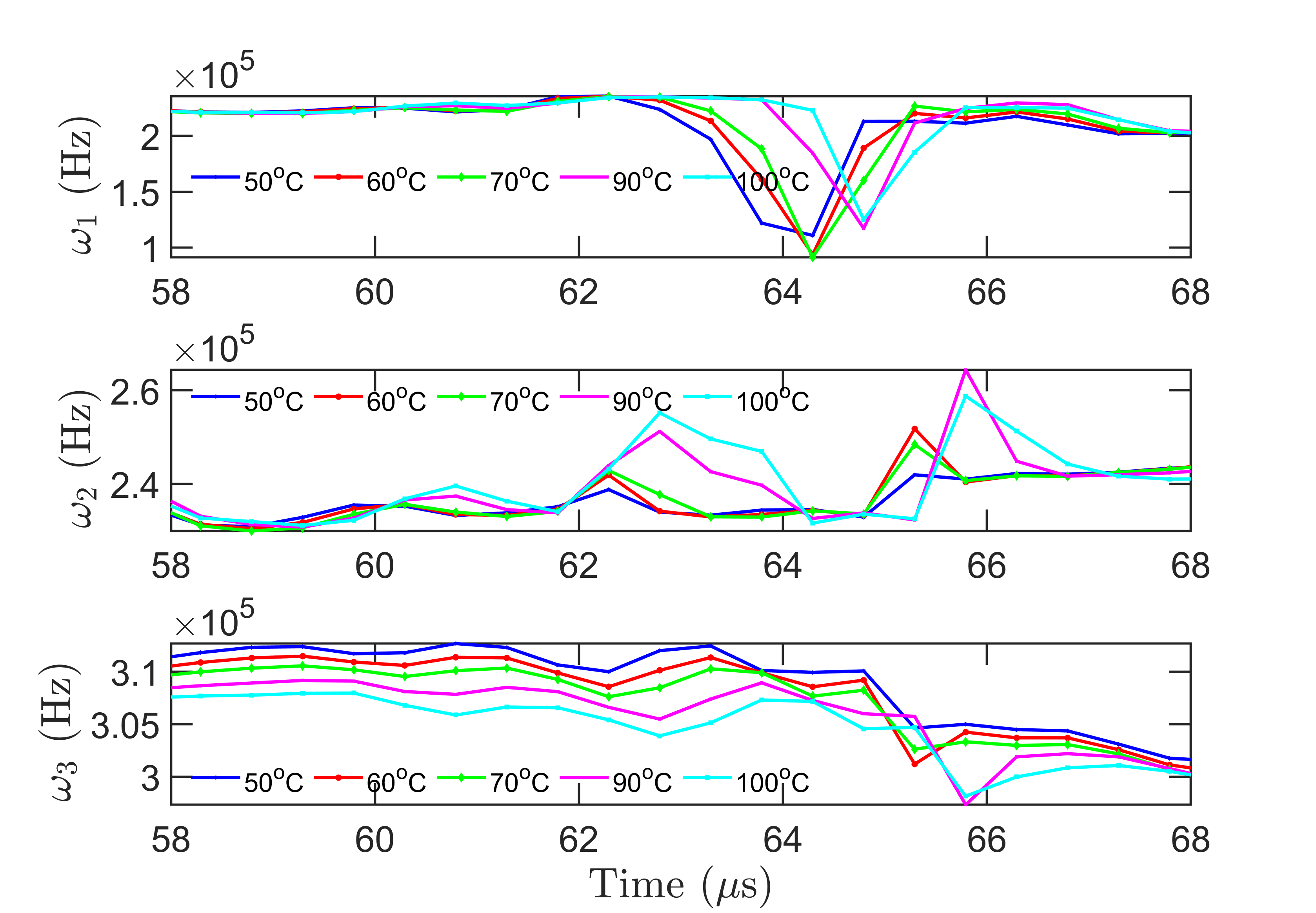}}
    \put(190,-340){\includegraphics[width=0.59\columnwidth]{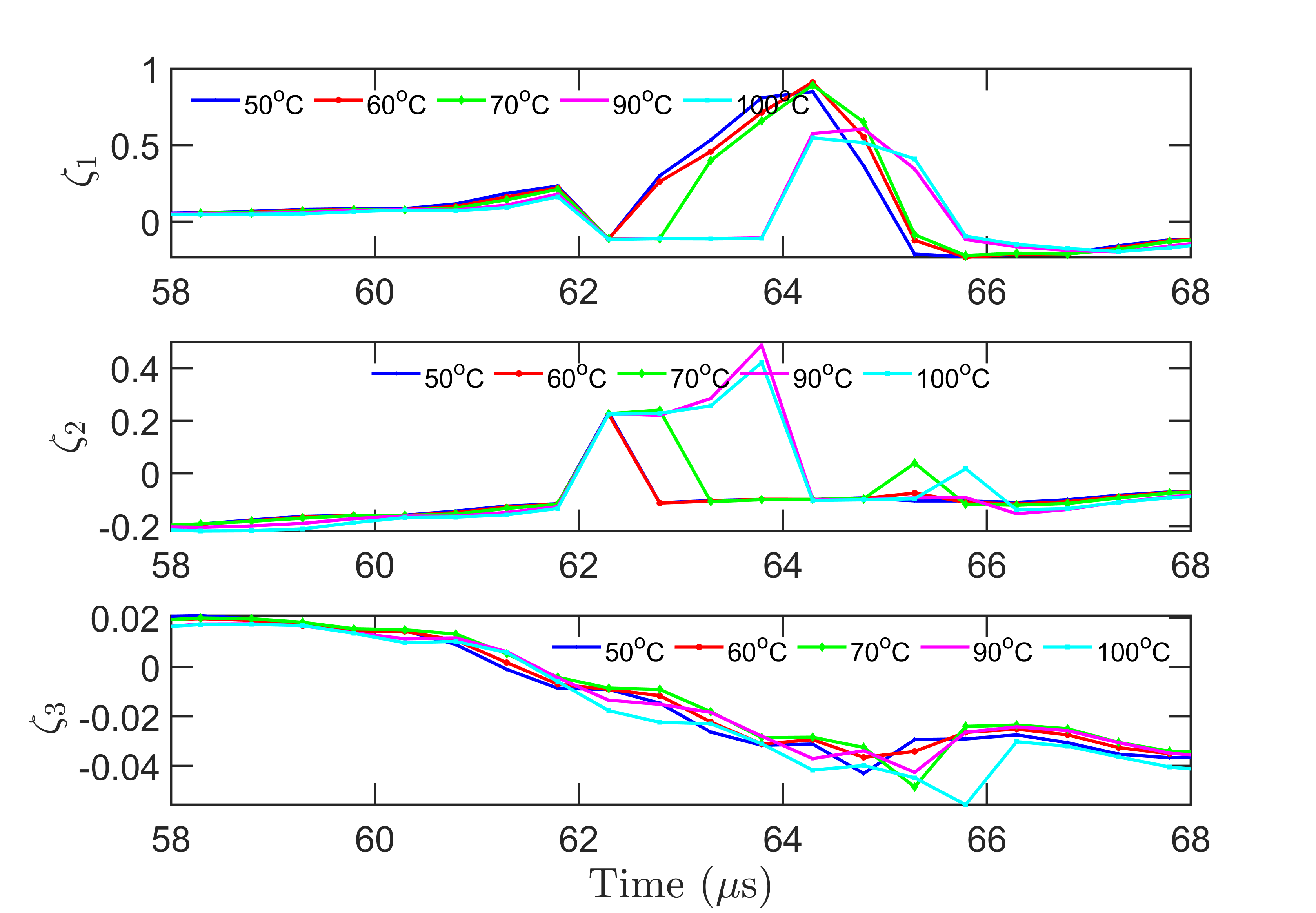}}
    \put(-60,230){ \large \textbf{(a)}}
    \put(190,230){\large \textbf{(b)}}
    \put(-60,40){\large \textbf{(c)}}    
    \put(190,40){\large \textbf{(d)}}  
     \put(-60,-150){\large \textbf{(e)}}    
    \put(190,-150){\large \textbf{(f)}}
    \end{picture}
    \vspace{11.6cm}
    \caption{Natural frequencies and damping ratios of the system obtained from a single model structure RML-TAR$(6)_{0.6}$ for five different temperatures: (a) natural frequencies and (b) damping ratios for the whole signal length; (c) zoomed-in view of the natural frequencies and (d) damping ratios of the $S_0$ mode; (e) zoomed-in view of the natural frequencies and (f) damping ratios of the $A_0$ mode.} 
\label{fig: freq} 
\end{figure}
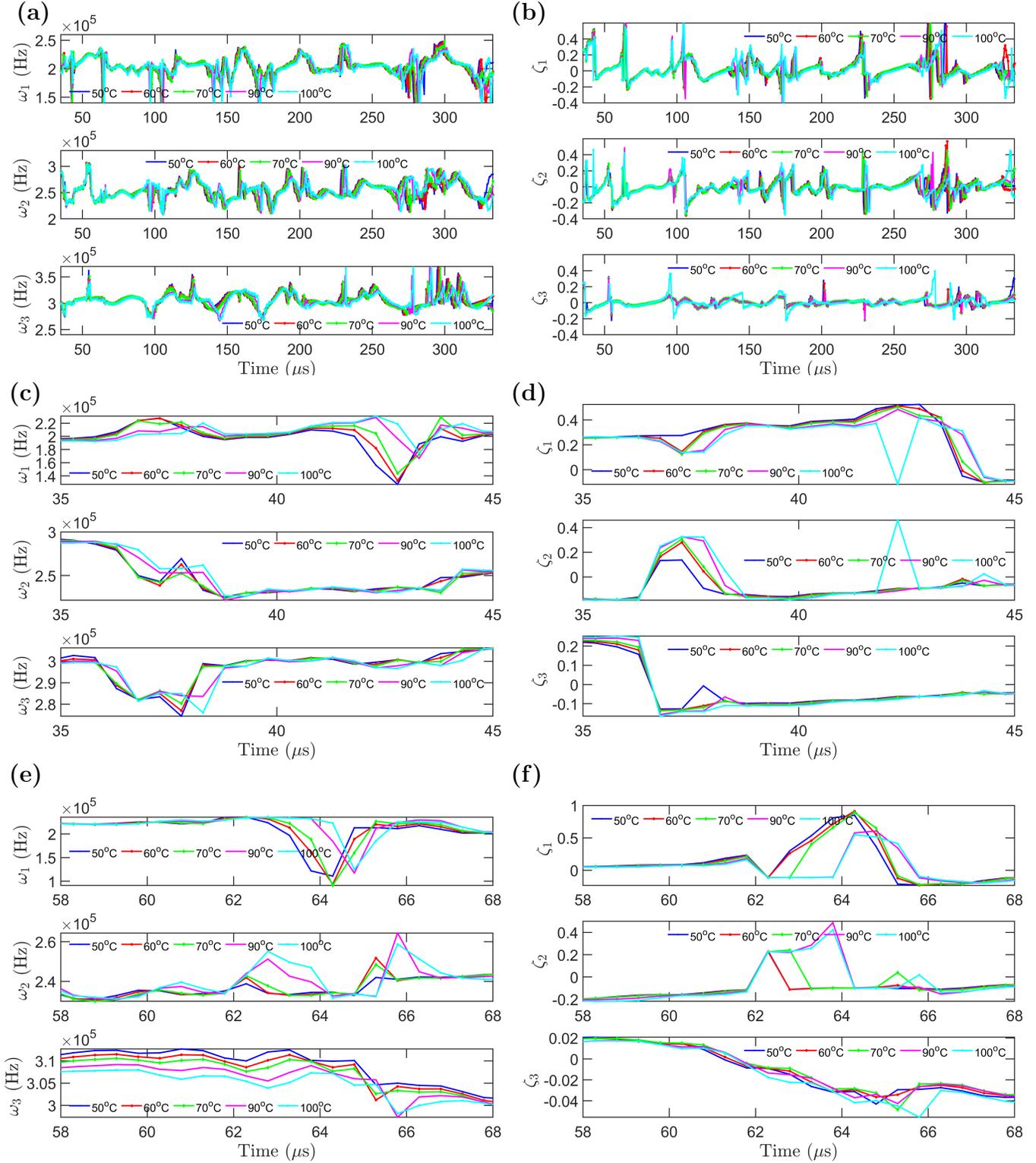 
\FloatBarrier

Figure \ref{fig: freq}(a) and (b) shows the ``frozen-time" natural frequencies and damping ratios, respectively, estimated by RML-TAR$(6)_{0.6}$ model for the five different temperatures, namely: $50^o$C, $60^o$C, $70^o$C, $90^o$C, and $100^o$C. As the model order $na=$ 6, so there are three natural frequencies and damping ratios. The three natural frequencies are seen to exist near 200 kHz, 250 kHz and 300 kHz. A closer observation reveals that with the increase in temperature, different peaks and valleys of the time-varying natural frequencies slightly shifts to the right. This is also the case for the three damping ratios.

Figure \ref{fig: freq}(c) and (d) shows a zoomed-in view of the ``frozen-time" natural frequencies and damping ratios, respectively, corresponding to the $S_0$ mode. It can be observed that, from 35 $\mu$s to 45 $\mu$s, the three natural frequencies $\omega_1$, $\omega_2$, and $\omega_3$ shifts to the right with the increase in temperature. For $\omega_1$, at 43 $\mu$s, the natural frequencies gradually move upward with the increase in temperature. The notch remains in the same position from $50^o$C to $70^o$C. However, for $90^o$C and $100^o$C, the notch shifts to the right. The three damping ratios $\zeta_1$, $\zeta_2$, and $\zeta_3$, from 35 $\mu$s to 45 $\mu$s, shifts to the right with the increase in temperature. For $\zeta_2$, at 37 $\mu$s, the notch goes upward with the increase in temperature and shifts to the right. 

Similarly, Figure \ref{fig: freq}(e) and (f) shows a zoomed-in view of the ``frozen-time" natural frequencies and damping ratios, respectively, corresponding to the $A_0$ mode for different temperatures. In this case, for $\omega_1$, at 64 $\mu$s, the notch goes down from $50^o$C to $70^o$C, and then goes up for $90^o$C and $100^o$C. Again, the notch shifts to the right with the increase in temperature. The damping ratios, similarly, shift to the right with the increase in temperature. 

\begin{figure}[t!]
    \centering
    \begin{picture}(400,140)
    \put(-60,-40){\includegraphics[width=0.59\columnwidth]{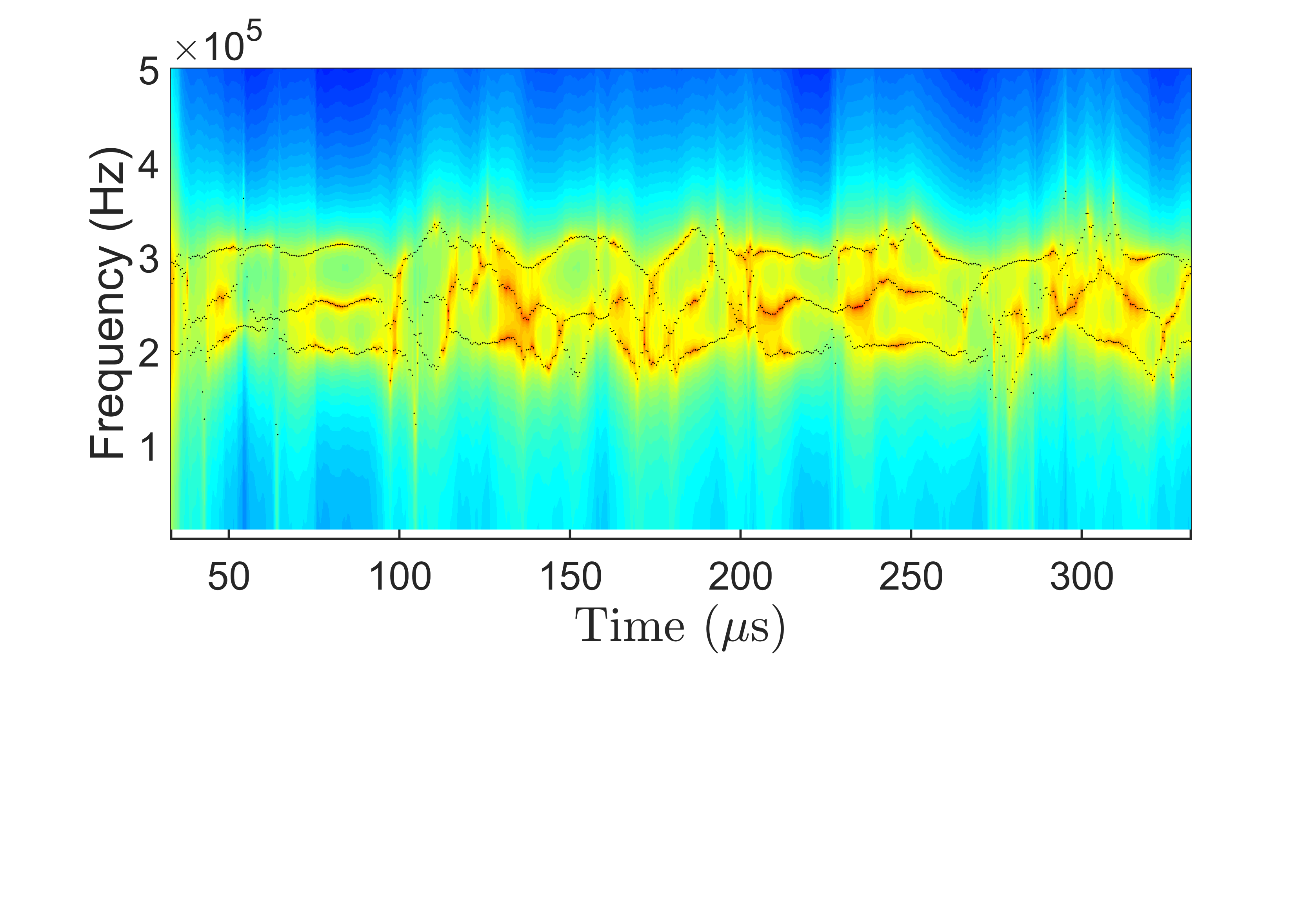}}
    \put(197,-40){\includegraphics[width=0.59\columnwidth]{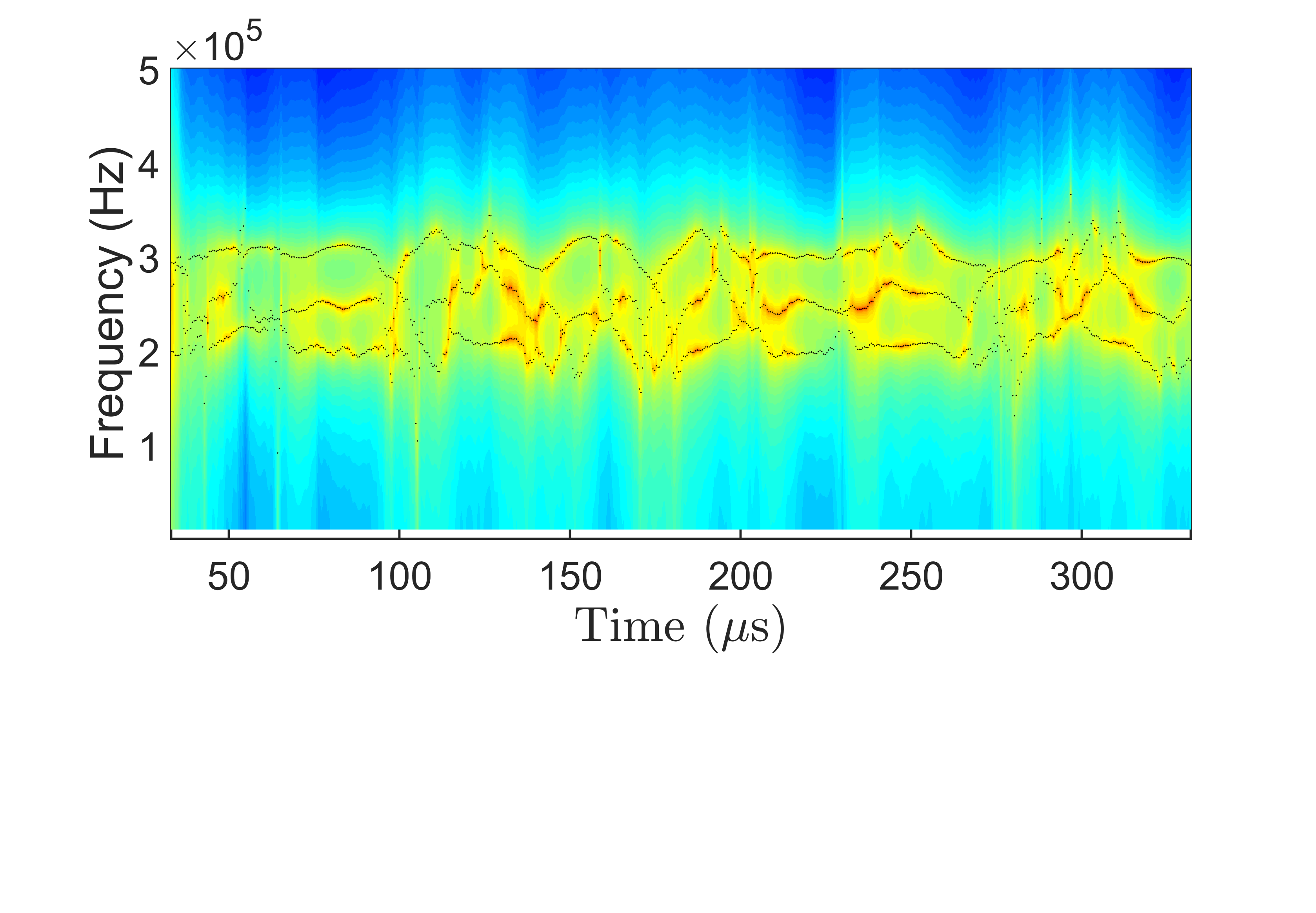}}
    
    \put(-60,-180){ \includegraphics[width=0.59\columnwidth]{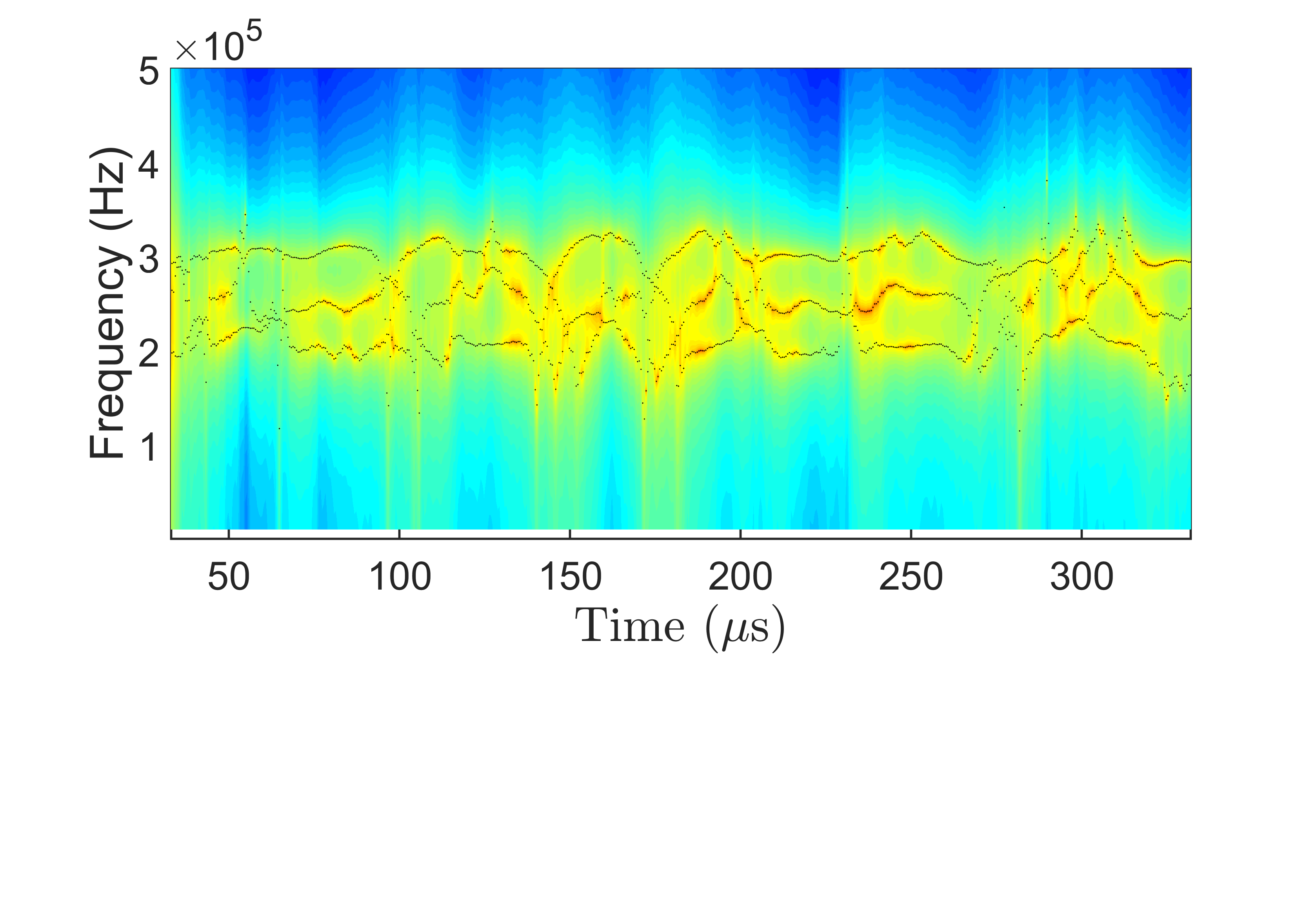}}
    \put(197,-180){\includegraphics[width=0.59\columnwidth]{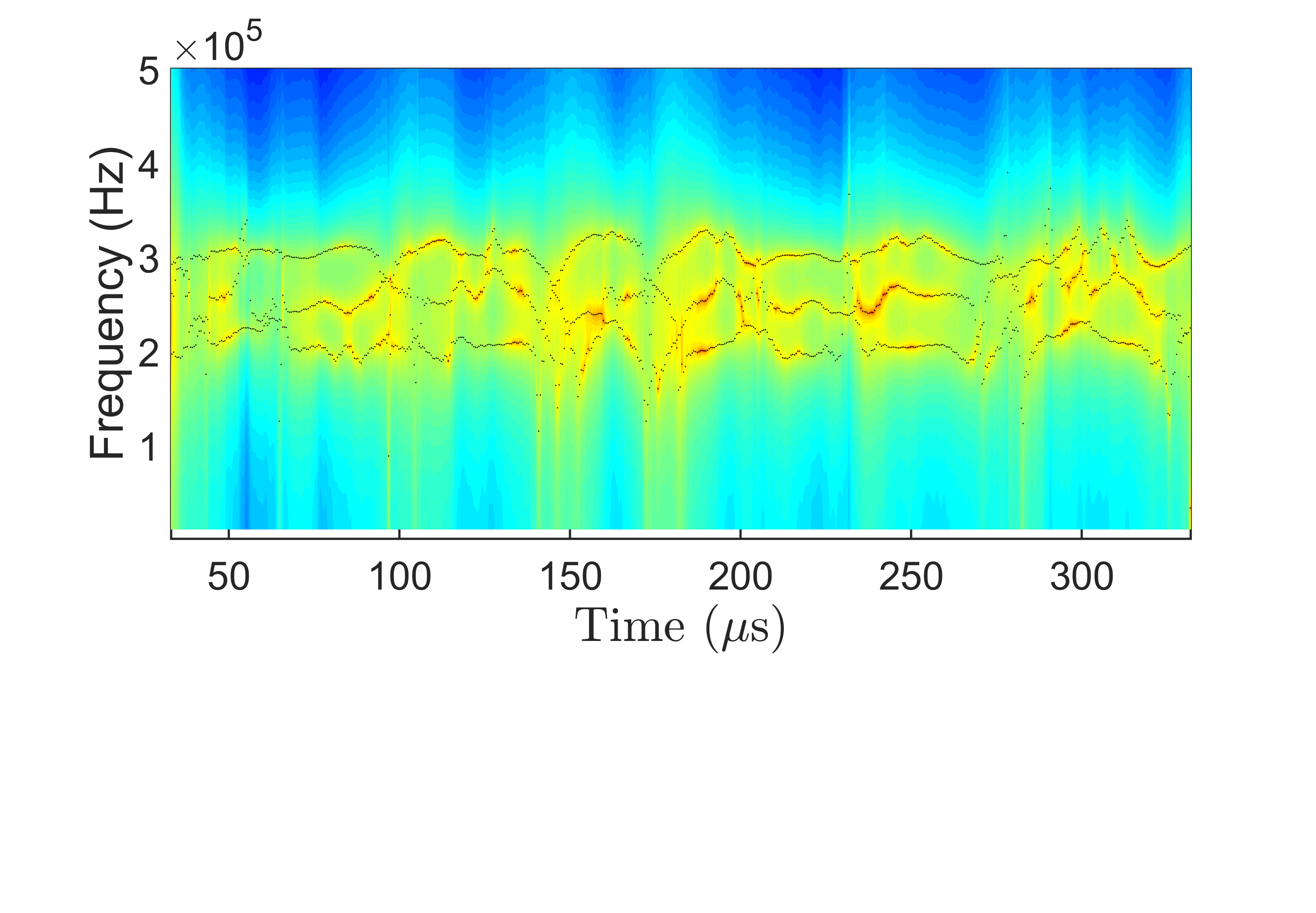}}
    \put(-50,150){\large \textbf{(a)}}
    \put(200,150){ \large \textbf{(b)}}
    \put(-50,10){\large \textbf{(c)}}    
     \put(200,10){\large \textbf{(d)}}
    \end{picture}
    \vspace{4.5cm}
    
    \caption{ Top view of the 3D frequency response function (FRF) along with ``frozen-time" natural frequencies estimated by RML-TAR$(6)_{0.6}$ model for different temperatures: (a) $50^0$C; (b) $70^0$C; (c) $90^0$C; and (d) $100^0$C.} 
\label{fig: Exp FRF} 
\end{figure}
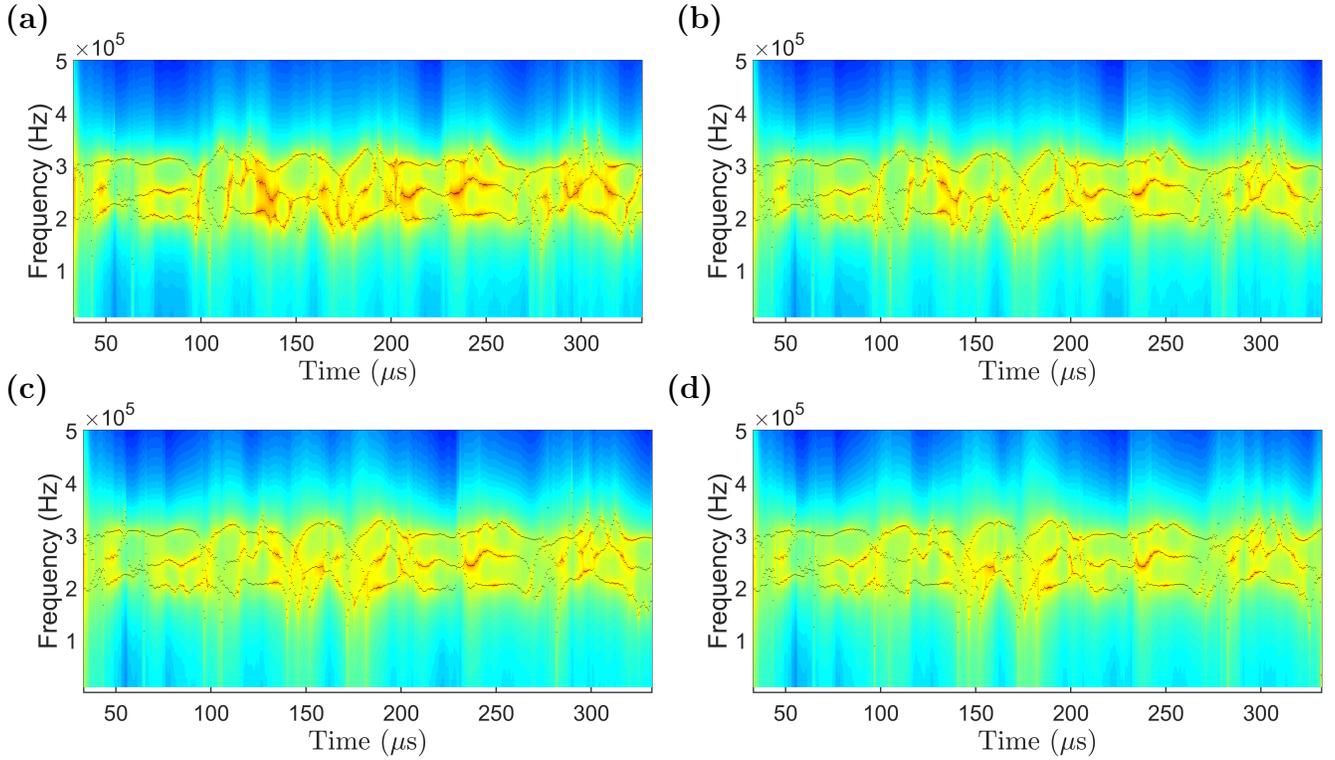

Figure \ref{fig: Exp FRF} shows the plot of the top view of the 3D frequency response function (FRF) superimposed with the three ``frozen-time" natural frequencies based on RML-TAR$(6)_{0.6}$ model for four different temperatures, namely: $50^o$C, $70^o$C, $90^o$C, and $100^o$C. It can be observed that excellent agreement exists between the ``frozen-time" natural frequencies and the 3D FRF.   

%

If non-parametric spectrogram (Figure \ref{fig: spectrogram exp}) is compared with the underlying system's global modal parameters and parametric FRFs (Figure \ref{fig: freq} and Figure \ref{fig: Exp FRF}, respectively), it can be observed that the parametric representation achieves better accuracy, resolution and tracking. The spectrogram shows that the frequency spectrum is centered around 250 kHz, as expected, and the power is evolving over time. However, the parametric identification reveals the presence of the three distinct frequencies and their time-varying characteristics. In addition, they also reveal how these frequencies change with varying temperatures. Regarding the terminology used, it should be clarified that the term accuracy refers to potential bias errors and
variability (scatter) in the estimates. Resolution refers to the discrimination of the characteristics of interest achieved in the time and frequency domains (this is mainly of interest for comparisons with non-parametric methods). Tracking refers to the ability of the estimates to precisely follow (track) the time-varying system dynamics. In order to use parametric models, it requires some familiarity and expertise on behalf of the user. It is, for instance, well known that the model structure selection is a task requiring caution and expertise. Although it may be automated to a certain degree, some user familiarity is still necessary. Also, the interpretation of the method’s behavior or results may require attention.

\subsubsection {Simulation Under Varying Temperature: RML-TARX}

Although FEM models have enabled the investigation of phenomena that are difficult to investigate experimentally due to the money and cost involved and complex instrumentation needed, the computational power and time needed is sometimes prohibitively large if every details of the system is modeled. As a result, surrogate models are formed of the corresponding system to simulate the system's response which are not as high fidelity as the FEM model but still can provide the system's response or output with reasonable accuracy and in a short amount of time.

In this section, TARX models are used to simulate both the experimental as well as the FEM guided wave signals under varying temperatures. For the TARX model, the output response as well as the input excitation signals are taken into account. 

\paragraph{Experimental Signal Simulation}

In order to simulate the experimental guided wave propagation signals under varying temperature, it is first necessary to identify the proper model structure, and the system transfer function.

Model selection of RML-TARX involves selecting appropriate model order $na$, $nb$ and the  forgetting factor $\lambda$. The ESS/SSS (Error Sum of Squares/Signal Sum of Squares)  criterion, describing how well the model simulates the given signal, was employed for the model selection process.  AR orders from $na=$ 2 to $na=$ 22, X orders from $na=$ 2 to $na=$ 22, and forgetting factor $\lambda \in [0.5,0.999]$ (with an incremental  step  of 0.001) were considered to create a pool of candidate models. A total of 220,500 models (500 $\times$ 21 $\times$ 21) were estimated and among all these models, the best model was chosen as the one that minimizes the ESS/SSS. Following this criterion, for the signal at $50^o$C, the best model occured at $na=$ 7, $nb=$ 6, and forgetting factor $\lambda = 0.516$. This can be represented compactly as RML-TARX$(7,6)_{0.516}$. Again, the selection of the model order following the ESS/SSS criterion for a signal at a different temperature other than $50^o$C results in a different model order and forgetting factor. As for example, in order to simulate the experimental signal at $70^o$C and $90^o$C, the best model occurred for RML-TARX$(6,7)_{0.533}$ and RML-TARX$(6,7)_{0.596}$, respectively, by employing the ESS/SSS criterion.
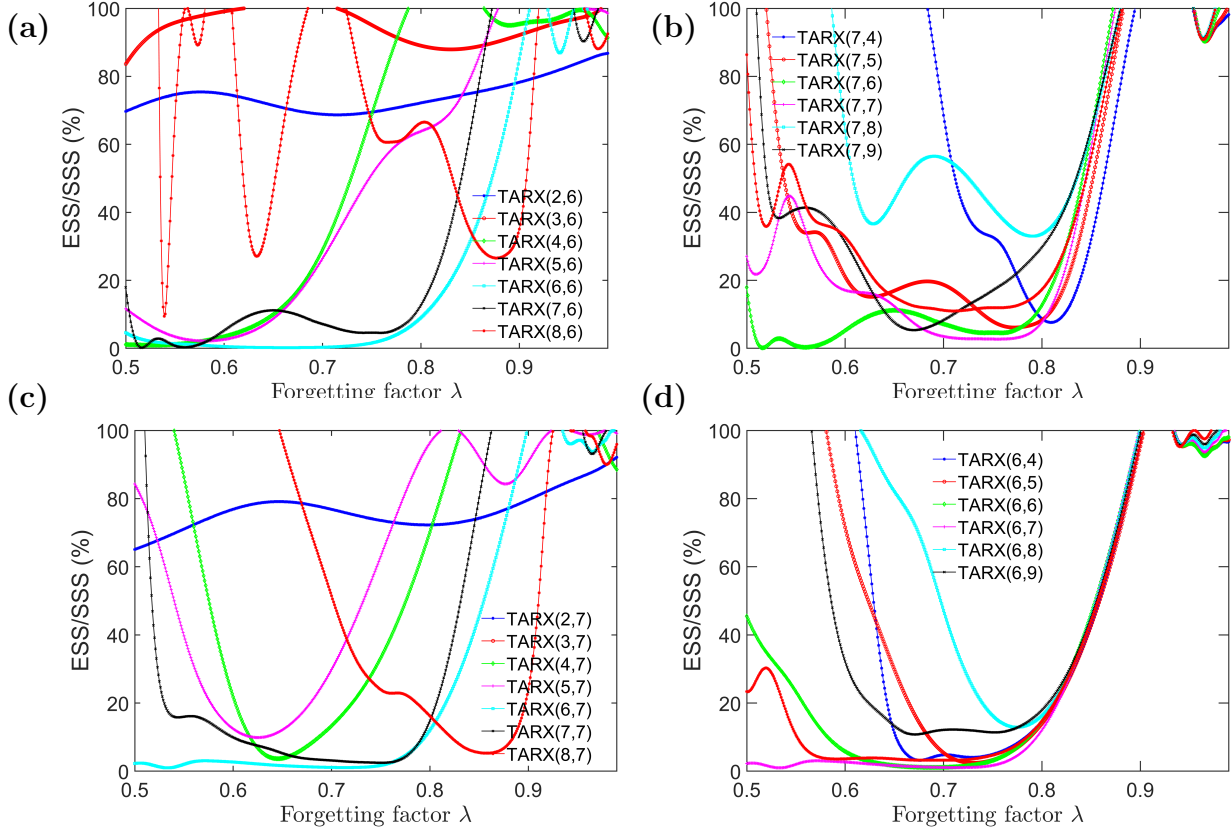
\begin{figure}[t!]
    \centering
    \begin{picture}(400,120)
    \put(-25,-40){\includegraphics[width=0.5\columnwidth]{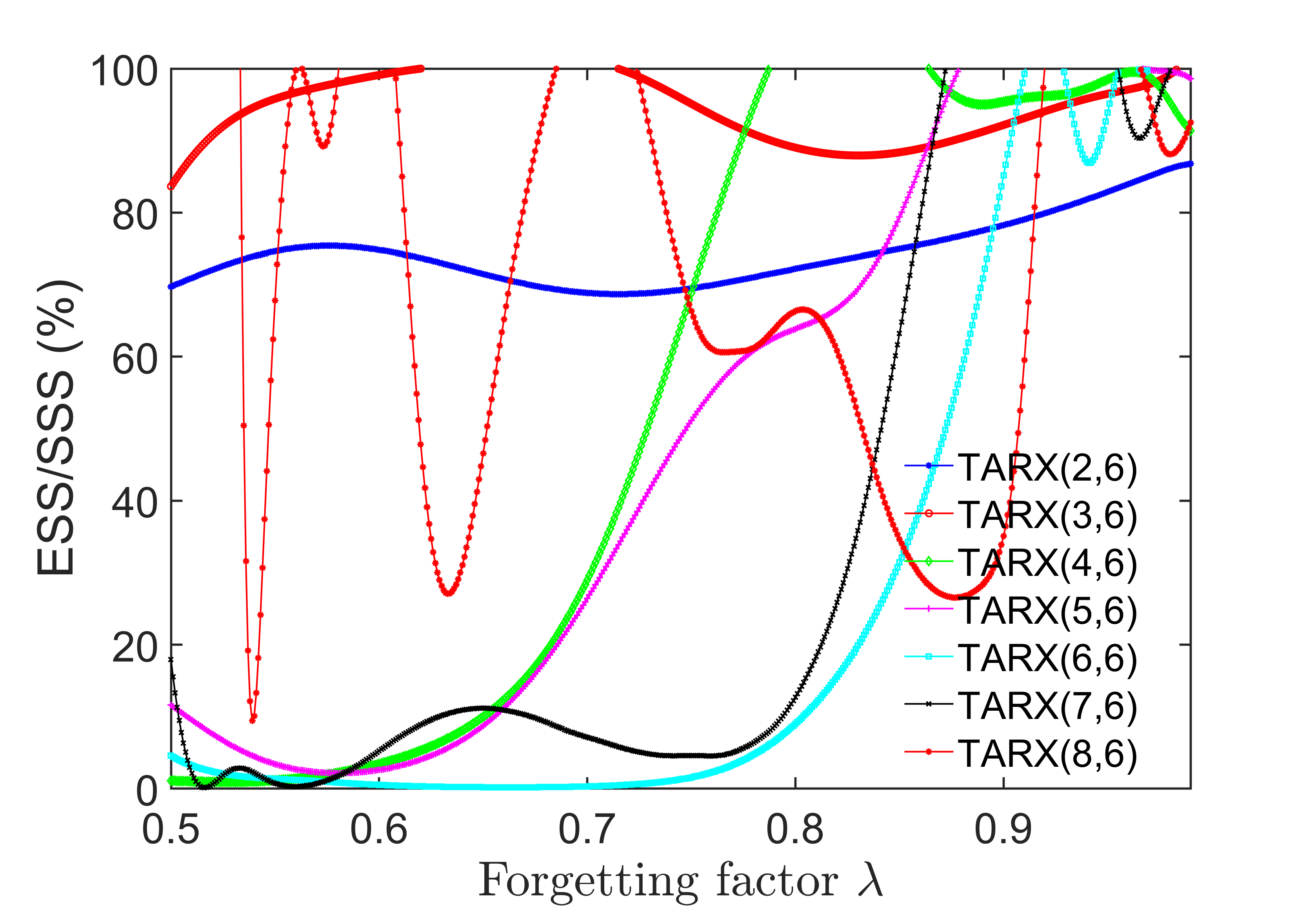}}
    \put(210,-40){\includegraphics[width=0.5\columnwidth]{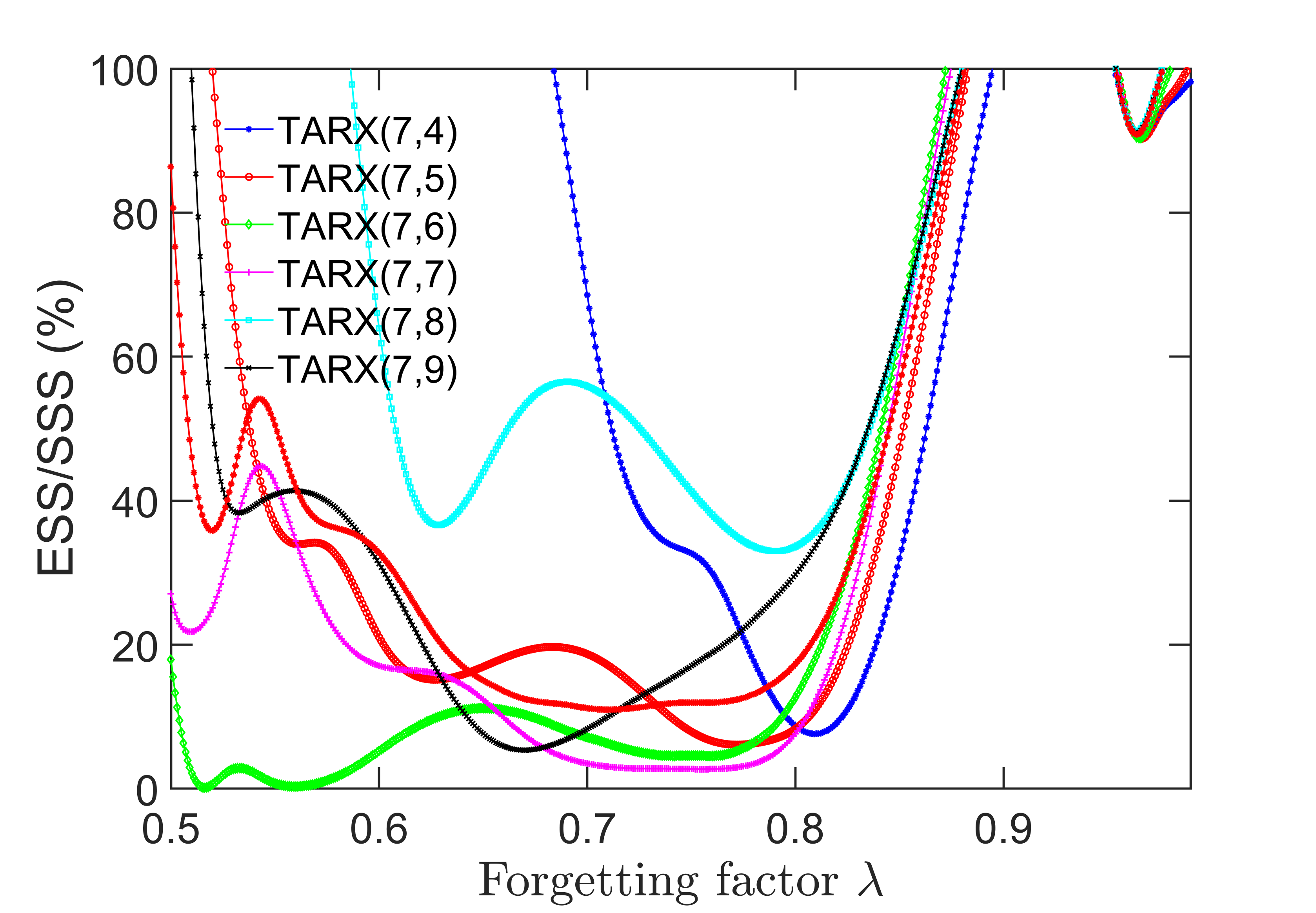}}
    
    \put(-25,-200){ \includegraphics[width=0.5\columnwidth]{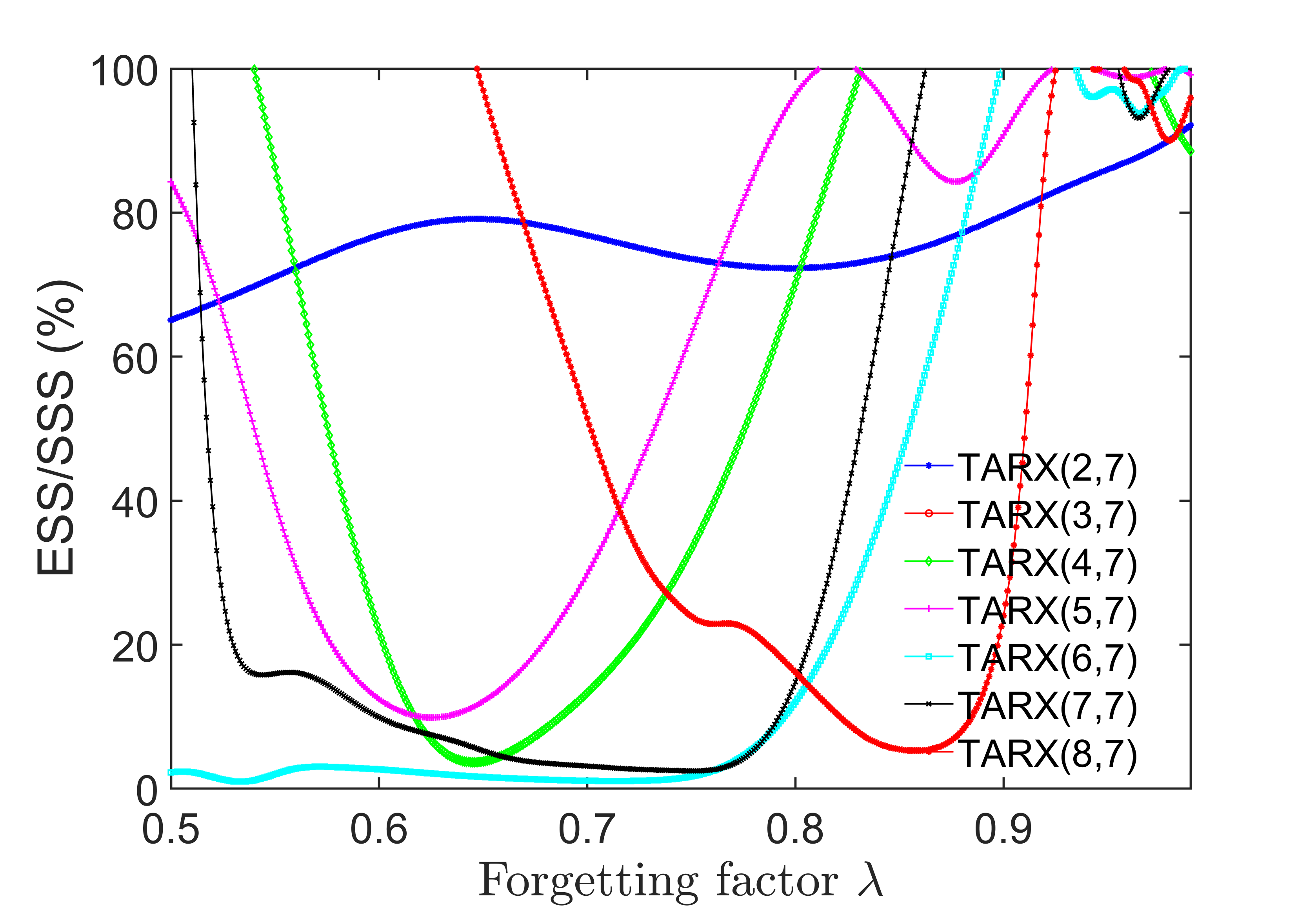}}
    \put(210,-200){\includegraphics[width=0.5\columnwidth]{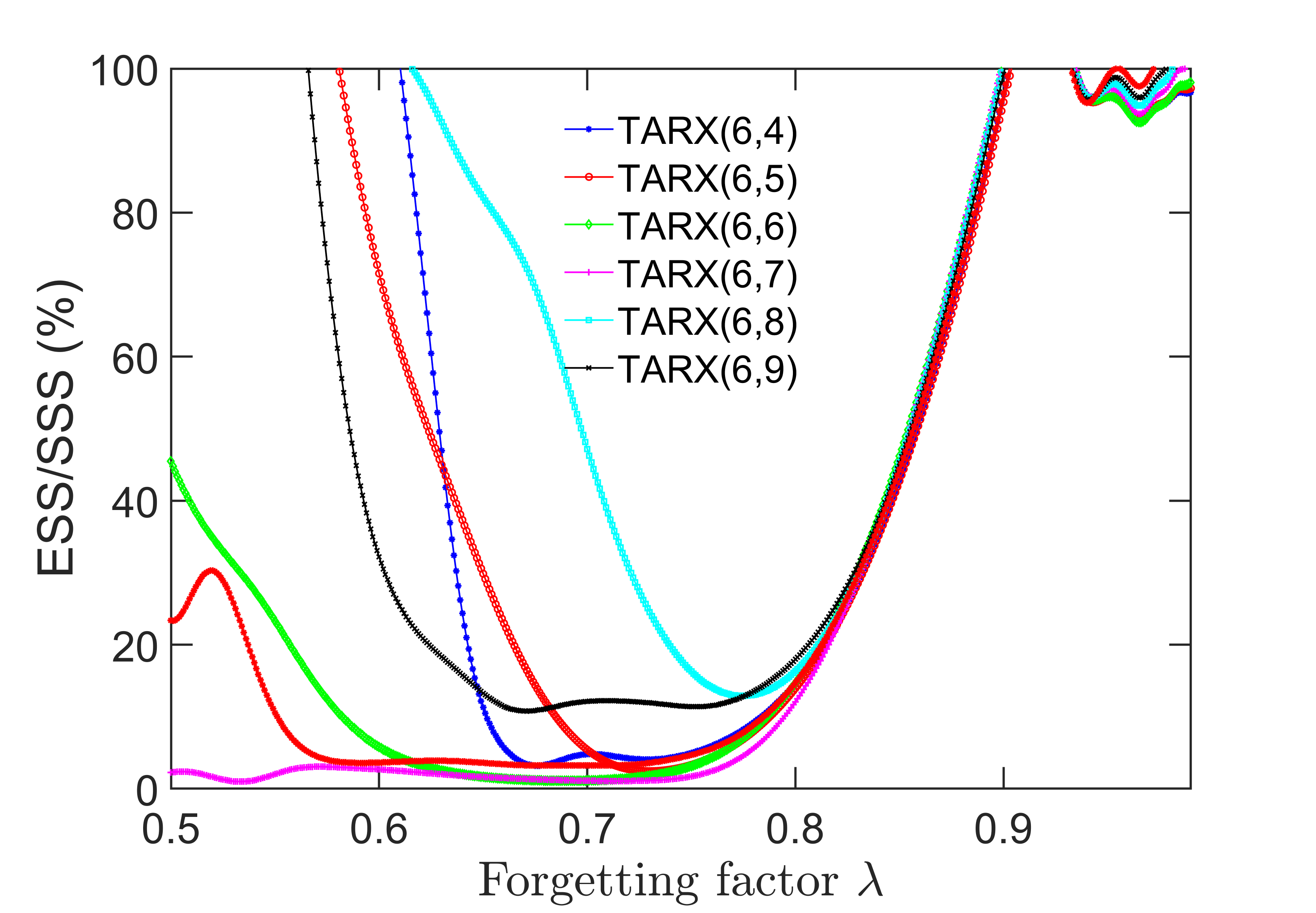}}
    \put(-40,100){\large \textbf{(a)}}
    \put(200,100){ \large \textbf{(b)}}
    \put(-40,-40){\large \textbf{(c)}}    
     \put(200,-40){\large \textbf{(d)}}
    \end{picture}
    \vspace{7cm}
    
    \caption{Model order selection of the RML-TARX model for the simulation of the experimental guided wave signal using the ESS/SSS criteria: (a) plot of the ESS/SSS versus the forgetting factor for different AR order $na$ keeping the X order $nb=$ 6 fixed for a representative signal at $50^o$C; (b) similar plot for different $nb$ order keeping $na=$ 7 fixed; (c) plot of the ESS/SSS versus the forgetting factor for different $na$ order keeping $nb=$ 7 fixed for $70^o$C; (d) similar plot for varying $nb$ order keeping $na=$ 6 fixed. } 
\label{fig: ESS} 
\end{figure} 
%

Figure \ref{fig: ESS} shows the RML-TARX model selection process for a signal at $50^o$C and $70^o$C. Figure \ref{fig: ESS}(a) shows the plot of the ESS/SSS versus the forgetting factor for different $na$ order keeping a fixed $nb$ order. It can be observed that the ESS/SSS changes very rapidly with the change in the model order $na$ and forgetting factor. The minimum ESS/SSS occurs at the model order $na=$ 7 with forgetting factor $\lambda=$ 0.516 for a fixed $nb$ order. Figure \ref{fig: ESS}(b) shows the ESS/SSS values for different $nb$ order keeping a fixed $na$ order and varying forgetting factors. It can be observed that the minimum value occurs at $nb=$ 6 for a fixed $na$ order with forgetting factor $\lambda=$ 0.516. Figure \ref{fig: ESS}(c) and (d) shows the RML-TARX model selection process for a representative signal for $70^o$C for a fixed $nb$ and $na$ order, respectively. The best model occurred for RML-TARX$(6,7)_{0.533}$ model for $70^o$C.   

%
    
%

%
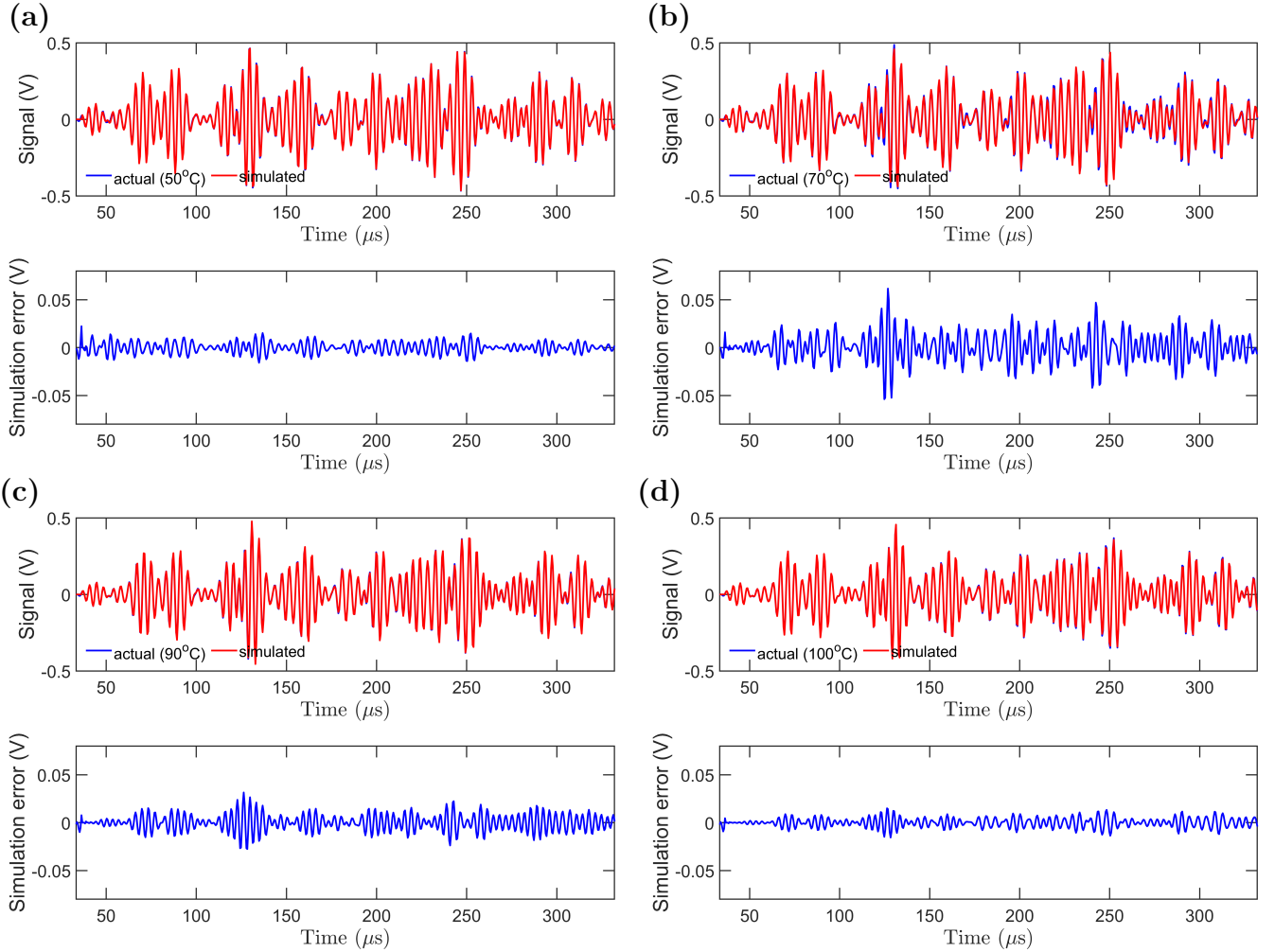
\begin{figure}[t!]
    \centering
    \begin{picture}(400,120)
    \put(-60,-40){\includegraphics[width=0.59\columnwidth]{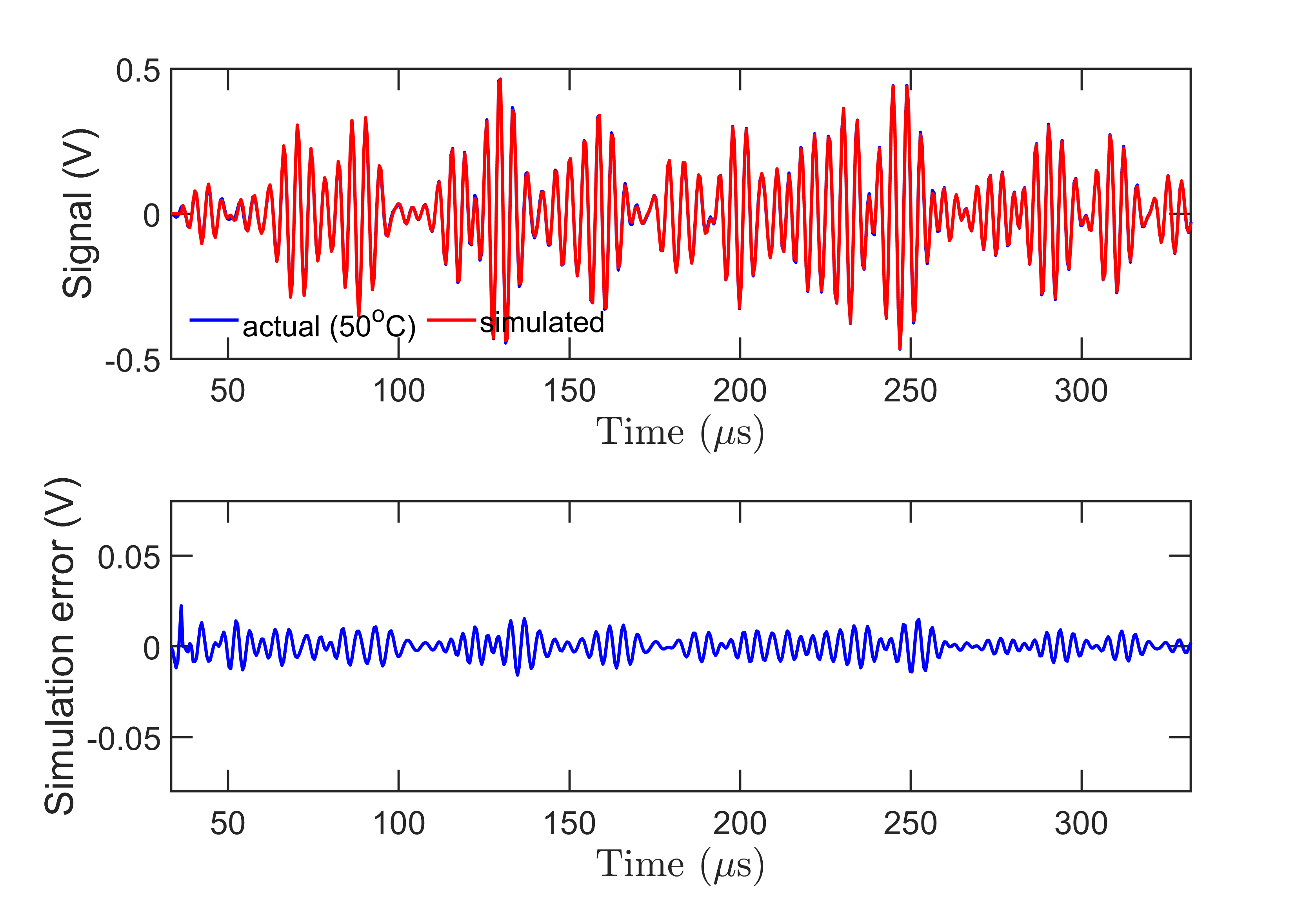}}
    \put(197,-40){\includegraphics[width=0.59\columnwidth]{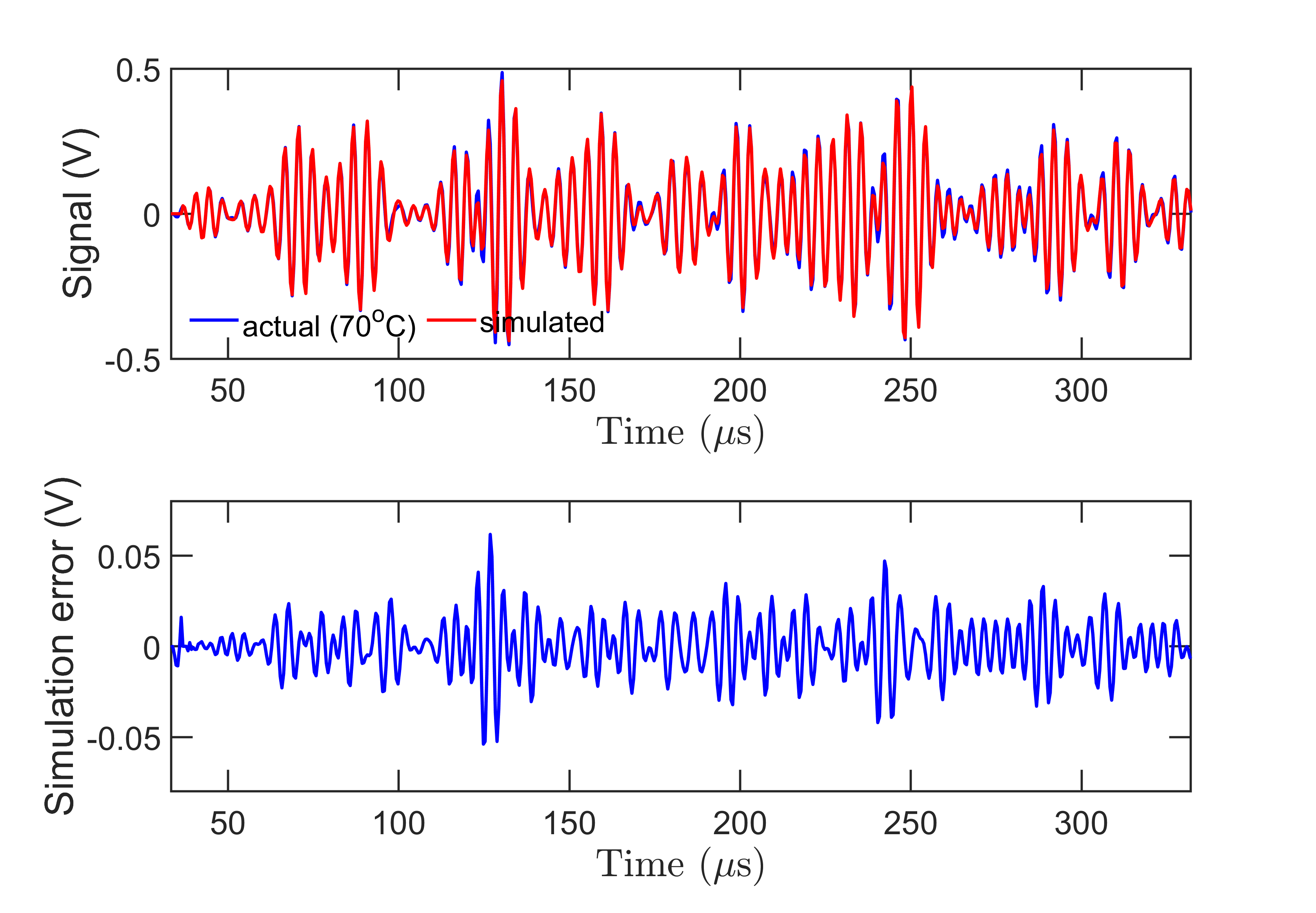}}
    \put(-60,-230){\includegraphics[width=0.59\columnwidth]{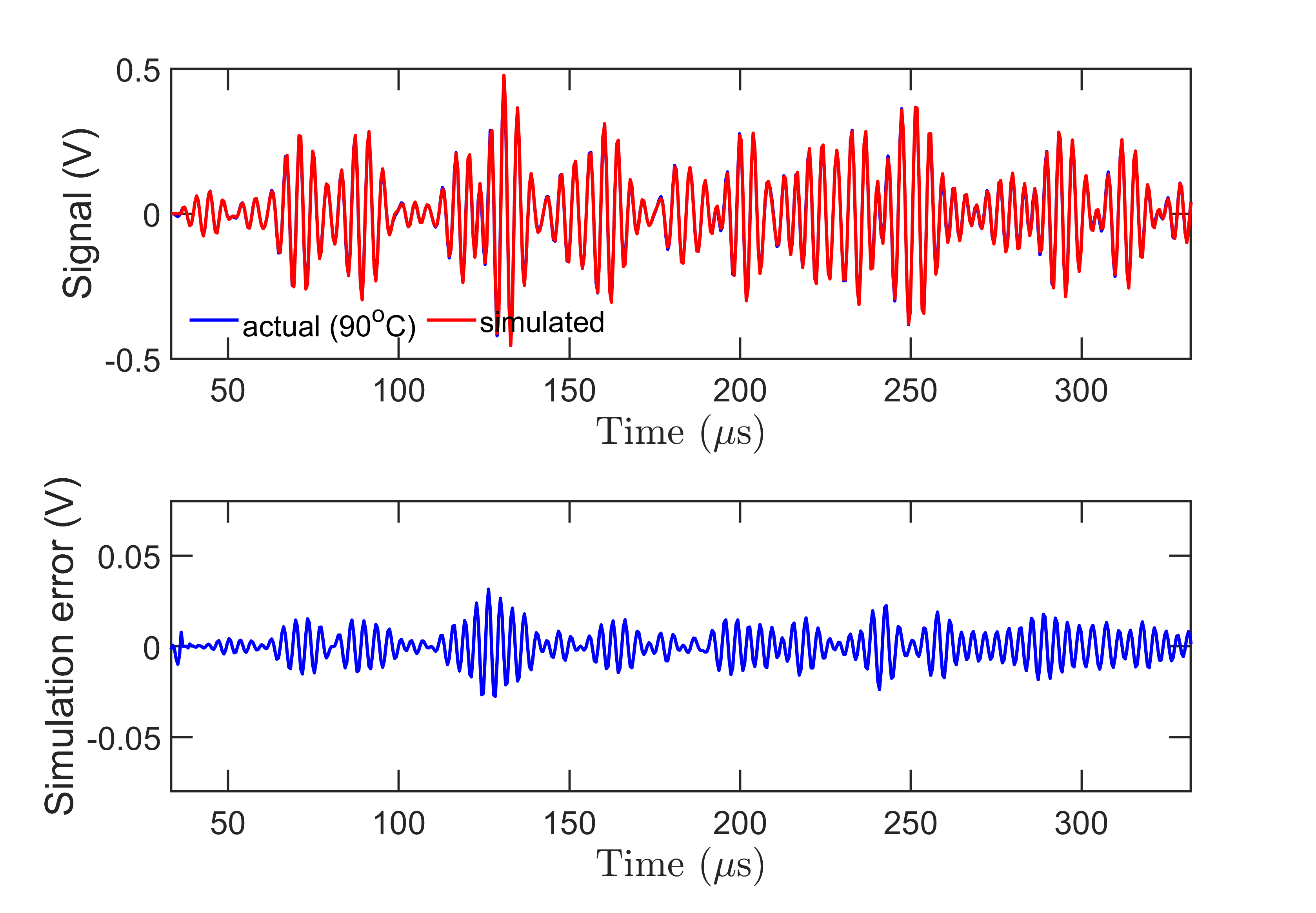}}
    \put(197,-230){\includegraphics[width=0.59\columnwidth]{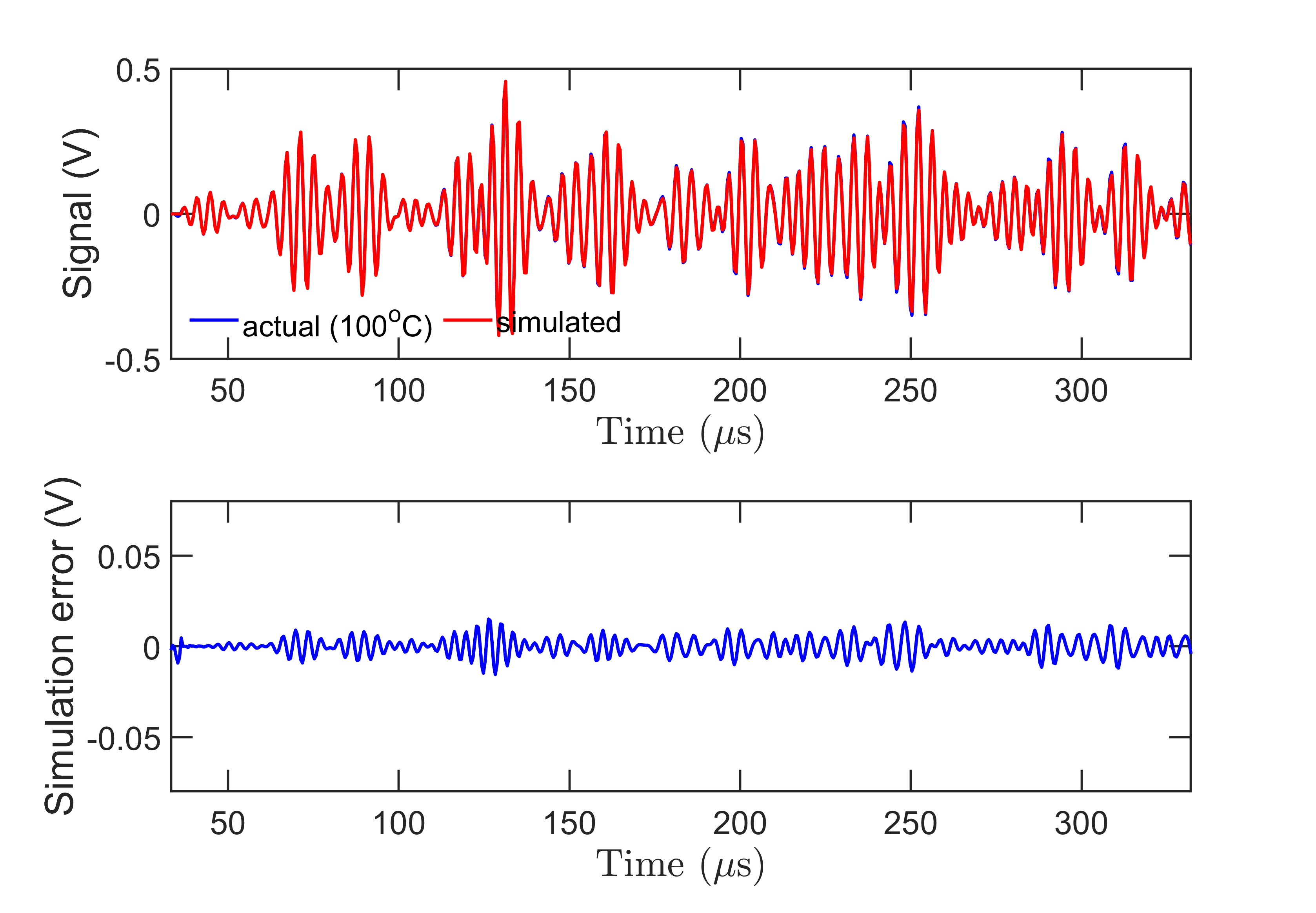}}
    
    \put(-55,145){ \large \textbf{(a)}}
    \put(200,145){ \large \textbf{(b)}}
    \put(-55,-45){\large \textbf{(c)}}   
    \put(200,-45){\large \textbf{(d)}}
    \end{picture}
    \vspace{8cm}
    \caption{ Actual guided wave signal and simulated experimental signal: (a) simulation performed by RML-TARX$(7,6)_{0.516}$ for $50^o$C; (b) simulation performed by RML-TARX$(6,7)_{0.533}$ for $70^o$C; (c) simulation performed by RML-TARX$(6,7)_{0.596}$ for $90^o$C; (d) simulation performed by RML-TARX$(6,7)_{0.606}$ for $100^o$C; } 
\label{fig: simulation} \vspace{-12pt}
\end{figure} 

Figure \ref{fig: simulation}(a), (b), (c), and (d) shows the simulation of the guided wave signal with the help of the RML-TARX model after selecting an appropriate model structure for a representative signal at $50^o$C, $70^o$C, $90^o$C, and $100^o$C. The signal at $50^o$C was simulated with the help of RML-TARX$(7,6)_{0.516}$ model. The blue line represents the actual signal and the red line represents the simulated signal. An excellent match exists between the actual signal and the simulated signal during the total duration of the time. That is, the model is able to simulate the $S_0$ mode, $A_0$ mode as well as the reflected part of the signal. The simulation error is also low for the total duration of the time. The ESS/SSS value for the signal at $50^o$C and $70^o$C is 0.1355\% and 0.9775\%, respectively (goodness of fit values using the best model structure). The computational time required for this simulation process is also very low (usually about the fraction of a second).

In addition to performing simulation of guided wave propagation at different temperatures after identifying the best model at that specified temperature, it is also possible to identify a single model structure that can simulate guided wave propagation signals at a range of different temperatures. This approach provides a convenient way of simulating guided wave signal at any required temperature using a single model structure, given the single model structure is identified properly. The single model structure for simulating experimental signal was identified as RML-TARX$(6,7)_{0.6}$. This model structure was used to simulate guided wave signal at any temperature after identifying the proper transfer function from the available data.

It is also possible to simulate the guided wave signal at any in-between temperature where no data is available, after having only a few guided wave data set at some specific temperature. This can be accomplished by interpolating the known model parameters for unknown temperatures. As for example, if guided wave signal data is available at $50^o$C, $70^o$C and $90^o$C, then after estimating the model parameters at those temperatures, it is possible to get the model parameters at any in-between temperature by interpolation, such as, at $55^o$C or $80^o$C. Then using those interpolated model parameters, it is possible to simulate the guided wave signal at those in-between temperatures. 

\begin{figure}[t!]
    \centering
    \begin{picture}(400,160)
    \put(-70,-10){\includegraphics[width=0.59\columnwidth]{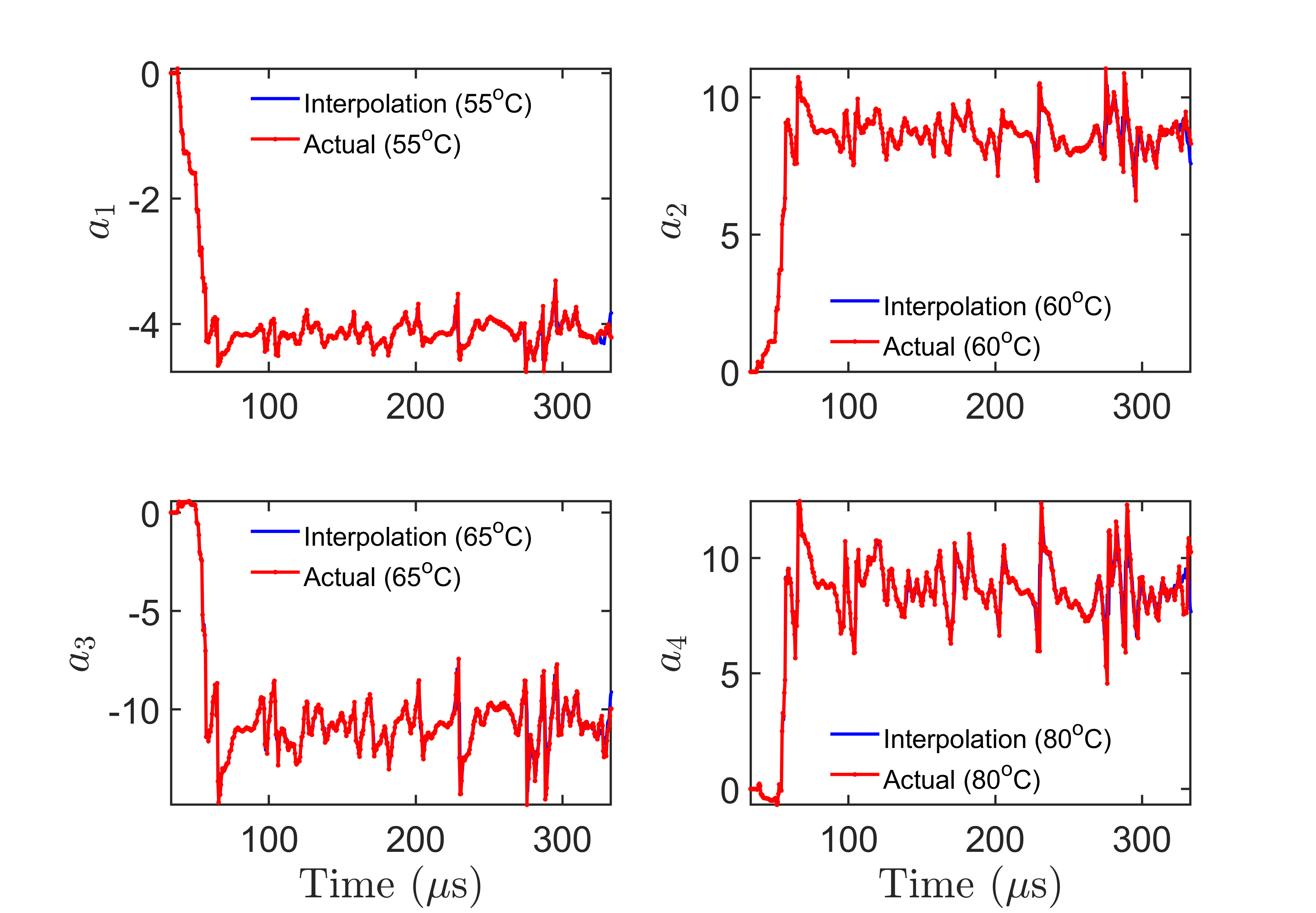}}
    \put(190,-10){\includegraphics[width=0.59\columnwidth]{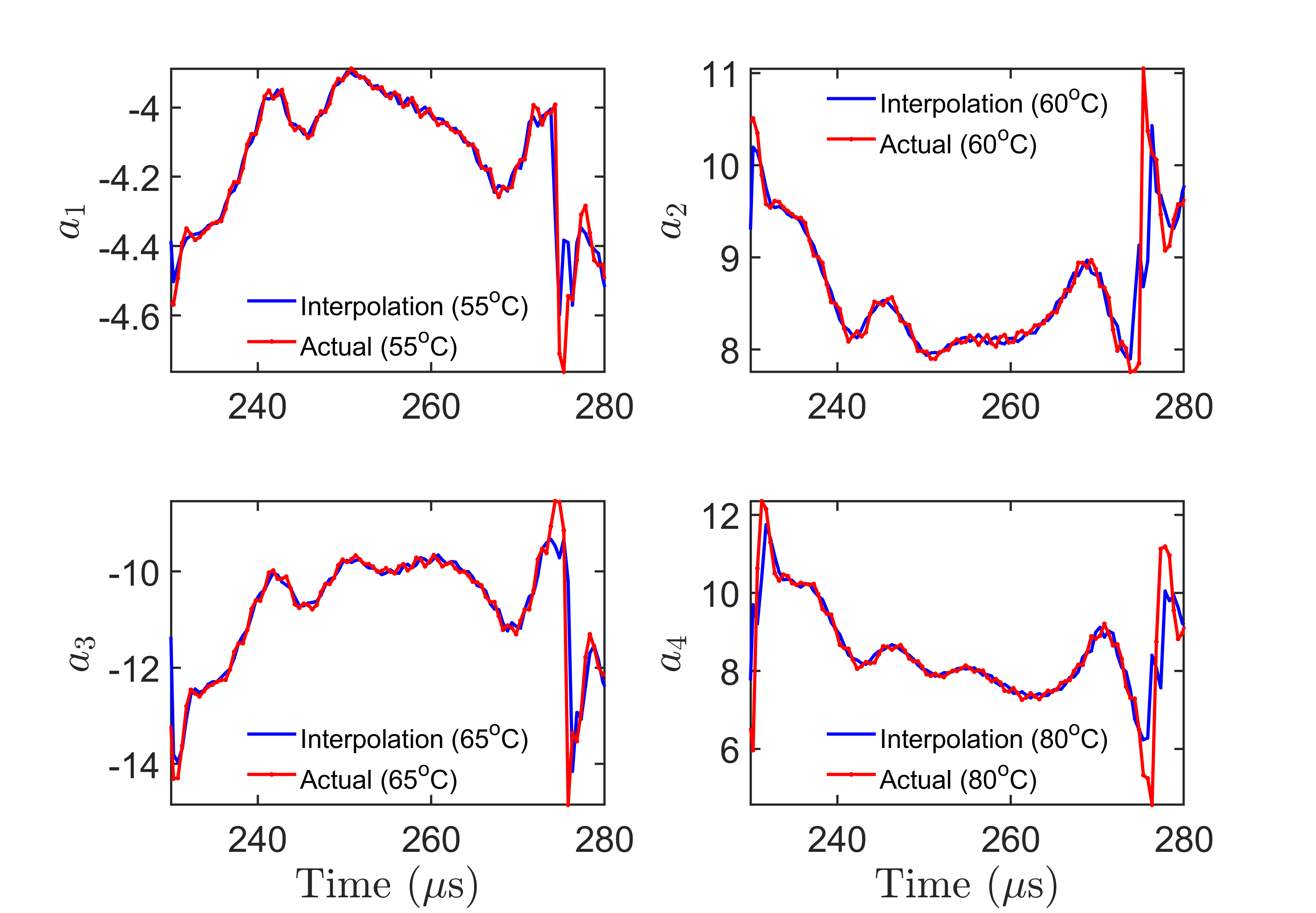}}
    \put(-70,-210){\includegraphics[width=0.59\columnwidth]{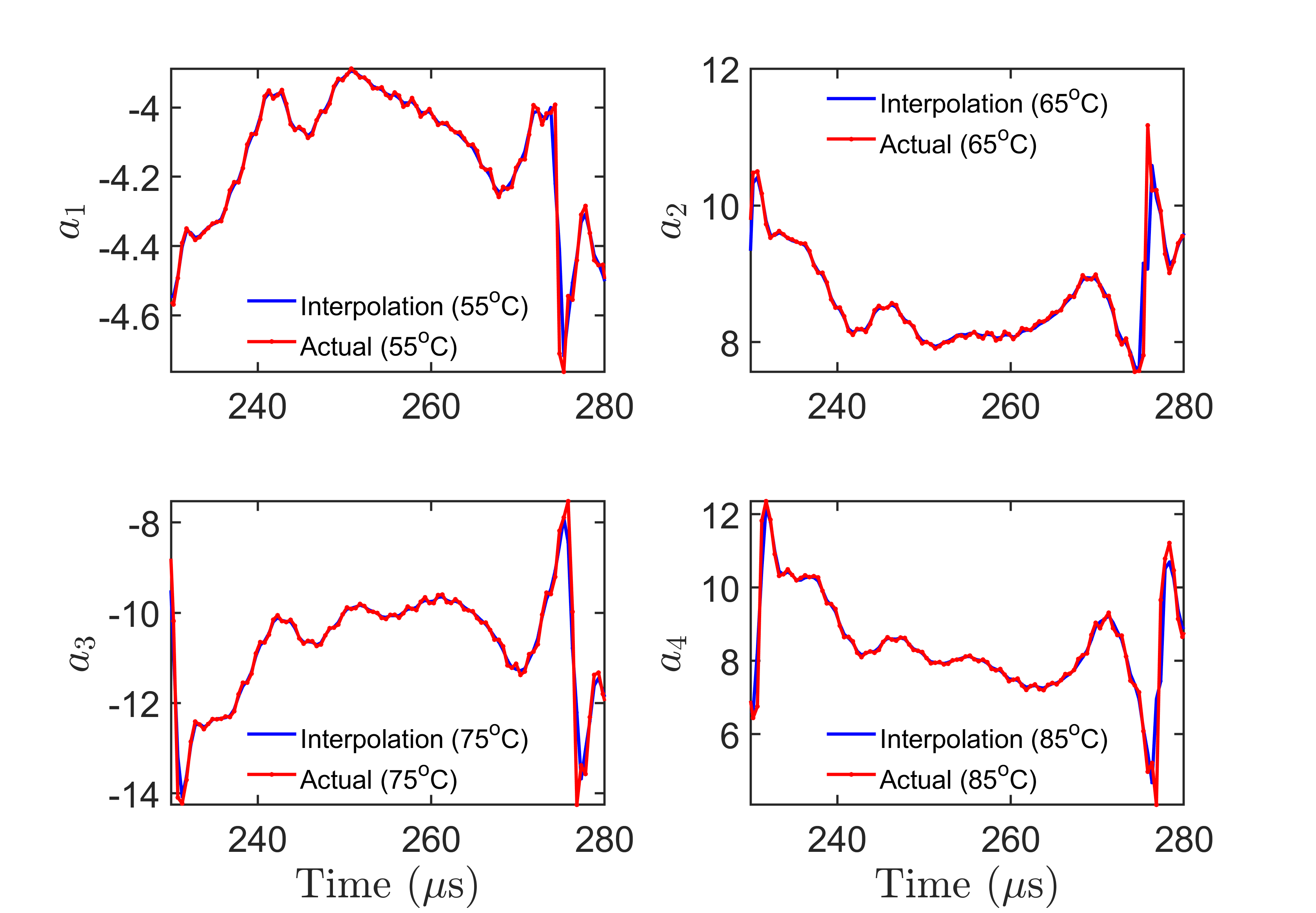}}
    \put(190,-210){\includegraphics[width=0.59\columnwidth]{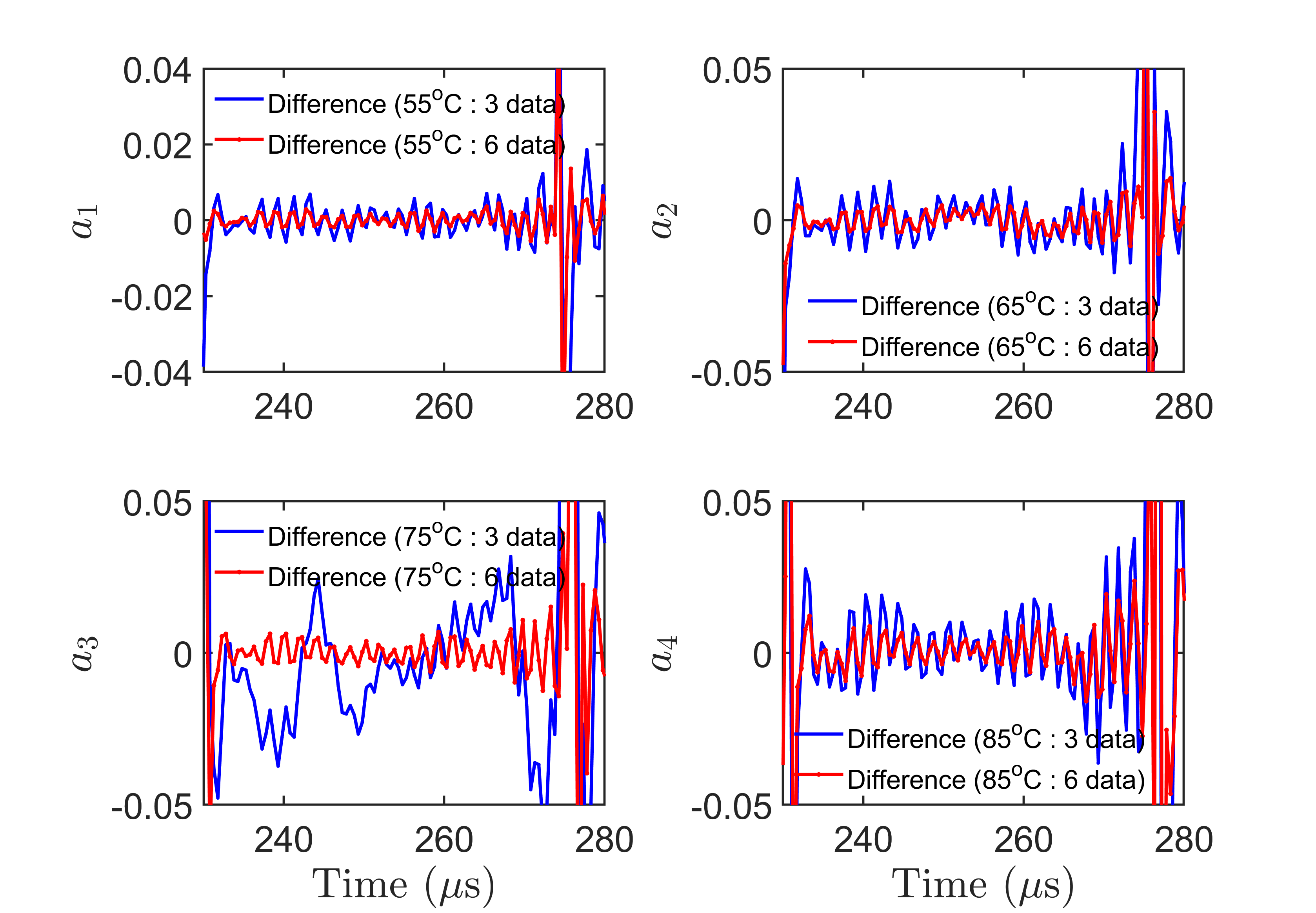}}
    \put(-50,180){ \large \textbf{(a)}}
    \put(190,180){\large \textbf{(b)}}
    \put(-50,-20){\large \textbf{(c)}}    
    \put(190,-20){\large \textbf{(d)}}  
    \end{picture}
    \vspace{7cm}
    \caption{First four model parameters identified by RML-TARX$(6,7)_{0.6}$ model: (a) actual and interpolated parameters for different temperatures using 3 data sets; (b) zoomed in view of the model parameters for 3 data sets; (c) zoomed-in view of the model parameters using 6 data sets at different temperatures; (d) normalized difference between the actual and interpolated model parameters at different temperatures.} 
\label{fig: interp} 
\end{figure}
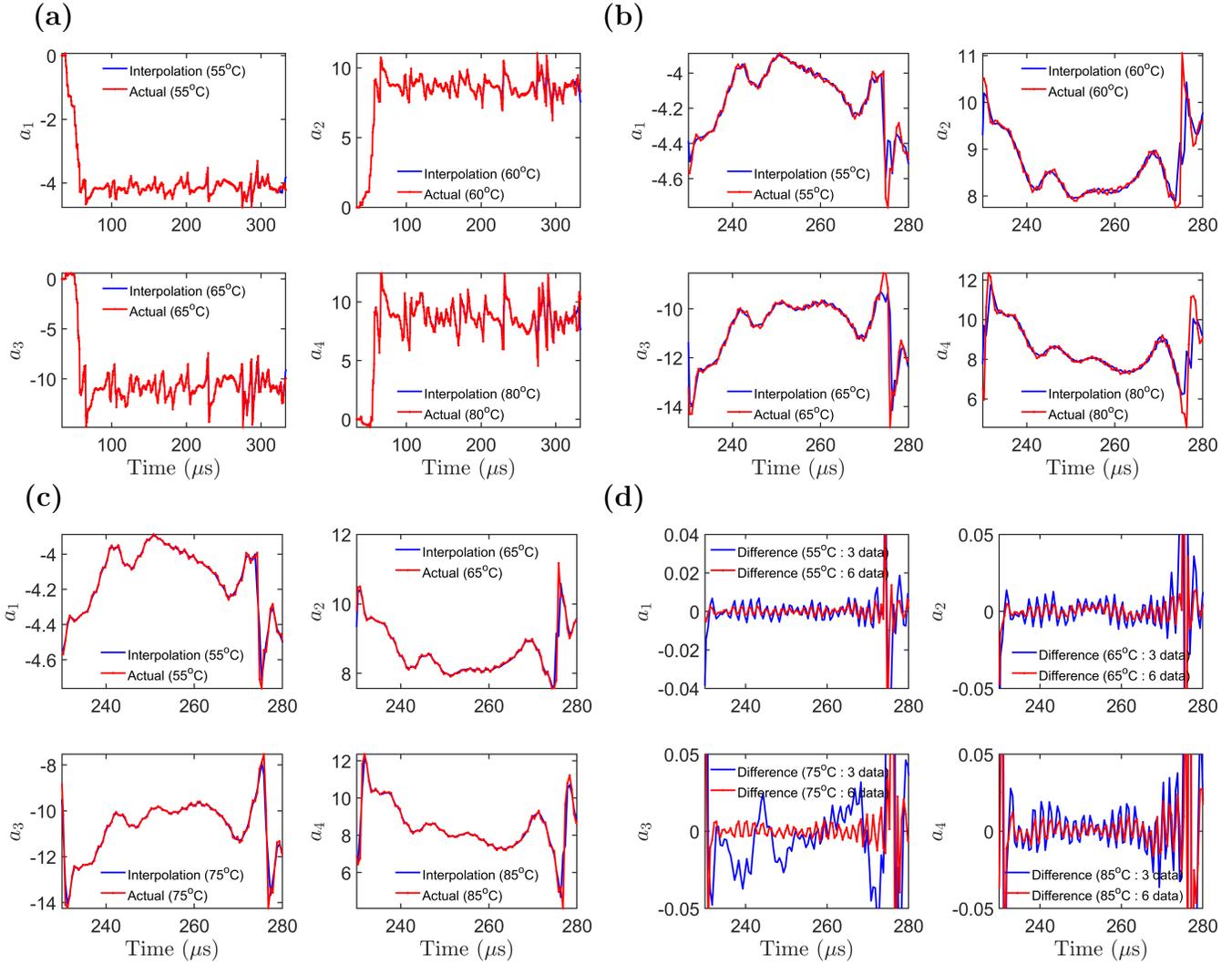 

Figure \ref{fig: interp}(a) shows the comparison between the actual parameter estimated by RML-TARX$(6,7)_{0.6}$ model and the interpolated model parameters using only three guided wave data sets. Figure \ref{fig: interp}(b) shows a zoomed-in view of Figure \ref{fig: interp}(a). It can be observed that a close match exists between the actual parameters and the interpolated parameters at the representative temperatures. Figure \ref{fig: interp}(c) shows a similar comparison between the actual and interpolated parameters using 6 data sets. Observe that when using 6 data sets, the interpolation gets more accurate. Figure \ref{fig: interp}(d) shows the normalized difference between the actual and interpolated parameters using 3 and 6 data sets for different temperatures as well as for different parameters. It can be observed that the magnitude of the difference for 3 data sets (blue lines) is higher than the 6 data sets (red lines) for all different temperatures and parameters. As a result, using more data sets increase the accuracy of the simulation for in-between temperatures.  
Table \ref{tbl: exp} compactly summarizes the results of the simulation of the guided wave signals for in-between temperatures using 3 data sets and 6 data sets. ESS/SSS values provide a single quantitative value of how well the simulation matches with the actual guided wave signal. Three kinds of interpolation schemes namely: piecewise cubic spline (spline), linear, and cubic convolutional (v5cubic) were used using the ``interp1" function in MATLAB. It can be observed that v5cubic interpolation provides better results than the linear interpolation for three data sets. However, for 6 data sets, both v5cubic and spline interpolation provides better results than the linear interpolation.  

It should be noted that piecewise cubic spline interpolation cannot be used for three data sets. It requires at least 4 data points to perform cubic spline interpolation as it approximates a cubic polynomial which has four constants \cite{burden2011numerical}. Being a cubic polynomial, it allows twice continuous differentiability which allows smoother approximation between nodal points (``knots"). For cubic splines, the end conditions can be formulated in several different ways. A very common approach is to assume that the second derivatives at the first and last knots are equal to zero. That is, the function becomes a straight line at the end nodes. This condition is usually referred to as ``natural" splines. In addition to this, there are ``clamped" condition and the ``not-a-knot" conditions. Depending on the end conditions, a cubic spline may assume different values \cite{chapra2008applied}.  

On the other hand, cubic convolutional interpolation (v5cubic) approximates a second-degree polynomial and as such, it is singly differentiable \cite{keys1981cubic}. It requires at least three data points to perform cubic convolutional interpolation. The convergence rate for cubic convolutional interpolation function is $\mathcal{O}(h^3)$. The computational cost for cubic convolutional interpolation is lower than the cubic spline interpolation. The linear interpolation scheme is the simplest of all and requires at least 2 data points. The convergence rate for linear interpolation function is $\mathcal{O}(h^2)$ \cite{keys1981cubic}. For the extrapolation of model parameters, only linear and cubic spline interpolation techniques can be used. The cubic convolutional interpolation method cannot be used for the extrapolation of the model parameters.

\paragraph{Surrogate Models for FEM}

The simulation approach described for the case of experimental guided wave signals is equally applicable for the guided wave signals obtained from FE simulations. However, for the sake of brevity, those results have not been included here. For the simulation of the FEM guided wave signals, a single model structure was also identified. The identified model was RML-TARX$(4,4)_{0.5}$.

Table \ref{tbl: simulation} compactly presents the simulation results using the RML-TARX model for the guided wave signal obtained by FEM simulations using 3, 6, and 7 data sets. For three data sets, guided wave signals at $30^o$C, $60^o$C, and $90^o$C were used, and model parameters were estimated at those temperatures. Then using model parameters at these three temperatures, unknown model parameters at required temperatures, where data are not available, were obtained by interpolation. Using these interpolated model parameters, guided wave signals were simulated and compared with the FEM guided wave signals. For six data sets, guided wave signals at temperatures $\{30, 40, 50 \cdots , 80^o\}$C were used. For 7 data sets, guided wave signals at temperatures $\{30, 40, 50 \cdots , 90^o\}$C were used. For obtaining the interpolated model parameters, three interpolation schemes were used namely: spline, linear and v5cubic interpolation using the ``interp1" function from MATLAB. It can be observed from Table \ref{tbl: simulation} that even using only 3 data sets, FEM guided wave signal can be simulated using RML-TARX model at in-between temperatures, such as, $35^o$C, $40^o$C, $80^o$C, etc. with high accuracy. With the use of more data sets, accuracy of the interpolation of model parameters increases, and, as such, the simulation becomes more accurate. 

\begin{table}[b!]
\centering
\caption{ESS/SSS(\%) values obtained for experimental signals}
\begin{tabular}[t]{lcccccc}
\toprule
\multicolumn{1}{| c |}{Temperature} & \multicolumn{3}{| c |}{3 data set} & \multicolumn{3}{| c |}{6 data set}\\
\hline
&spline & linear & v5cubic & spline & linear & v5cubic\\
\midrule
50 & --& --&-- & -- & -- & -- \\
55 & --& 5.1376 & 1.1585 & 0.9474 & 1.3120 & 0.6530 \\
60 & --& 8.5507 & 1.1931 & -- & -- & -- \\
65 &--&6.4540 &0.9038 & 2.0833 &3.2990 & 1.9140 \\
70 & --& -- & -- & -- & -- & -- \\
75 &--& 19.8499 & 9.2157& 1.8159 & 4.6519 & 1.9652 \\
80 & -- & 19.6836 & 13.0091 & -- & -- & -- \\
85 &  & 6.4024 & 6.6210 & 0.8245 & 2.2054 & 0.9370 \\
90 & --& -- & -- & -- & -- & -- \\
95 & --& 3.8904 & -- & 1.6159 & 1.1599 & 0.8960 \\
100 & --& 24.2597 & -- & -- & -- & -- \\

\bottomrule
\end{tabular}
\label{tbl: exp}
\end{table}%

In addition to performing interpolation of the model parameters, extrapolation can also be performed (if the guided wave signals required are outside the temperature range ). As for example, for 6 data sets ($\{30, 40, 50 \cdots , 80^o\}$C), if guided wave signals are required at $25^o$C or $29^o$C, then extrapolation of the model parameters have to be performed. From Table \ref{tbl: simulation}, it can be observed that for $25^o$C, $29^o$, or $95^o$C, where extrapolation of the model parameters was performed, the accuracy is also good. However, if extrapolation temperature is far away from the temperature where data is available, then the accuracy of the TARX simulation diminishes, as the interpolated model parameters deviate from the actual model parameters. This situation can be observed for the case of three data sets at temperatures $98.5^o$C and $100^o$C where the ESS/SSS is very high (62.87\% and 254.23\%, respectively). A similar situation occurs for 6 data sets and 7 sets. However, the accuracy of the TARX simulation increases at temperature $98.5^o$C and $100^o$C for the case of 7 data sets as additional set of data have been used (for the linear case, ESS/SSS are 1.83\% and 3.22\%, respectively).

\begin{figure}[t!]
    \centering
    \begin{picture}(400,120)
    \put(-60,-40){\includegraphics[width=0.59\columnwidth]{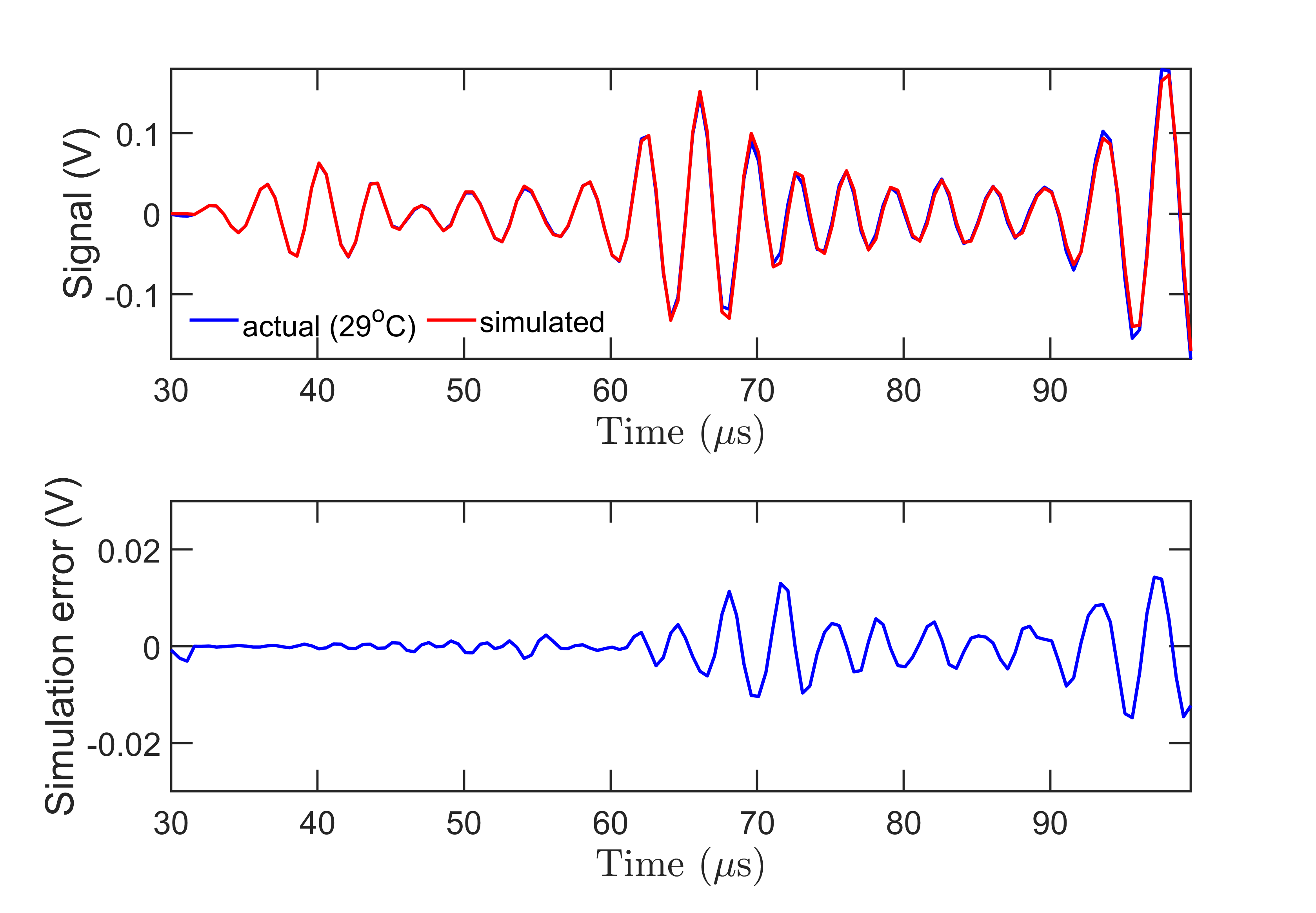}}
    \put(197,-40){\includegraphics[width=0.59\columnwidth]{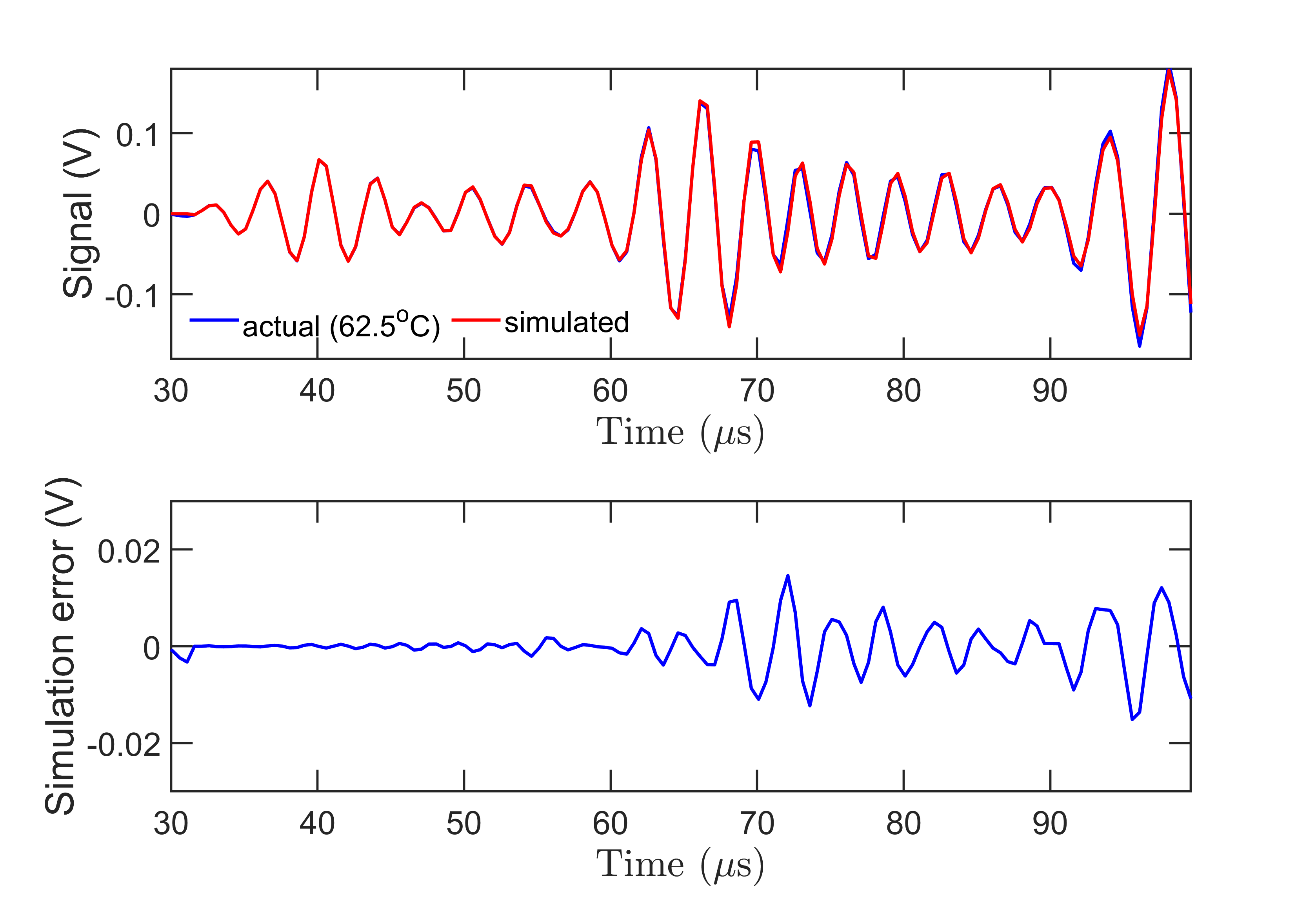}}
    \put(-60,-230){\includegraphics[width=0.59\columnwidth]{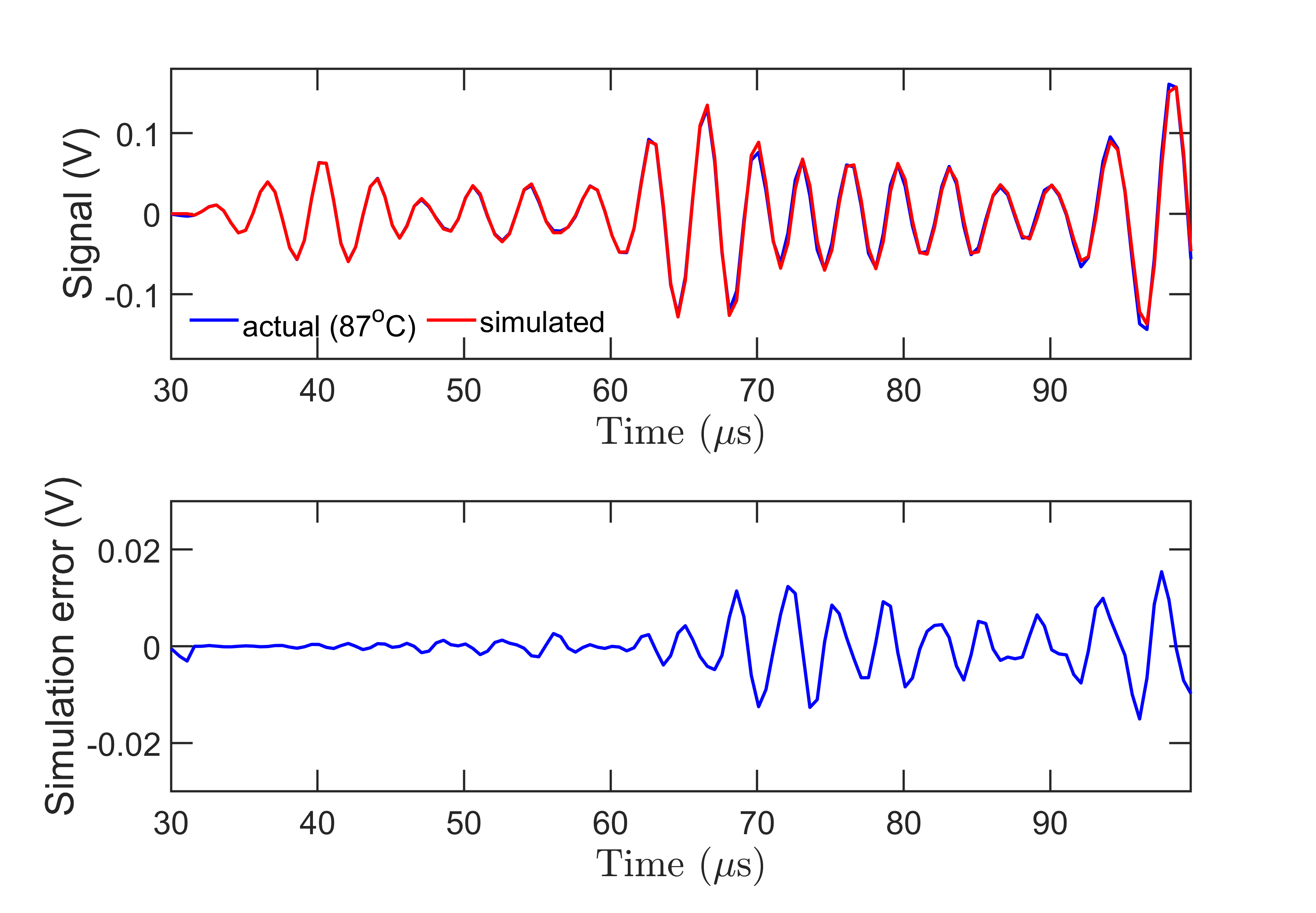}}
    \put(197,-230){\includegraphics[width=0.59\columnwidth]{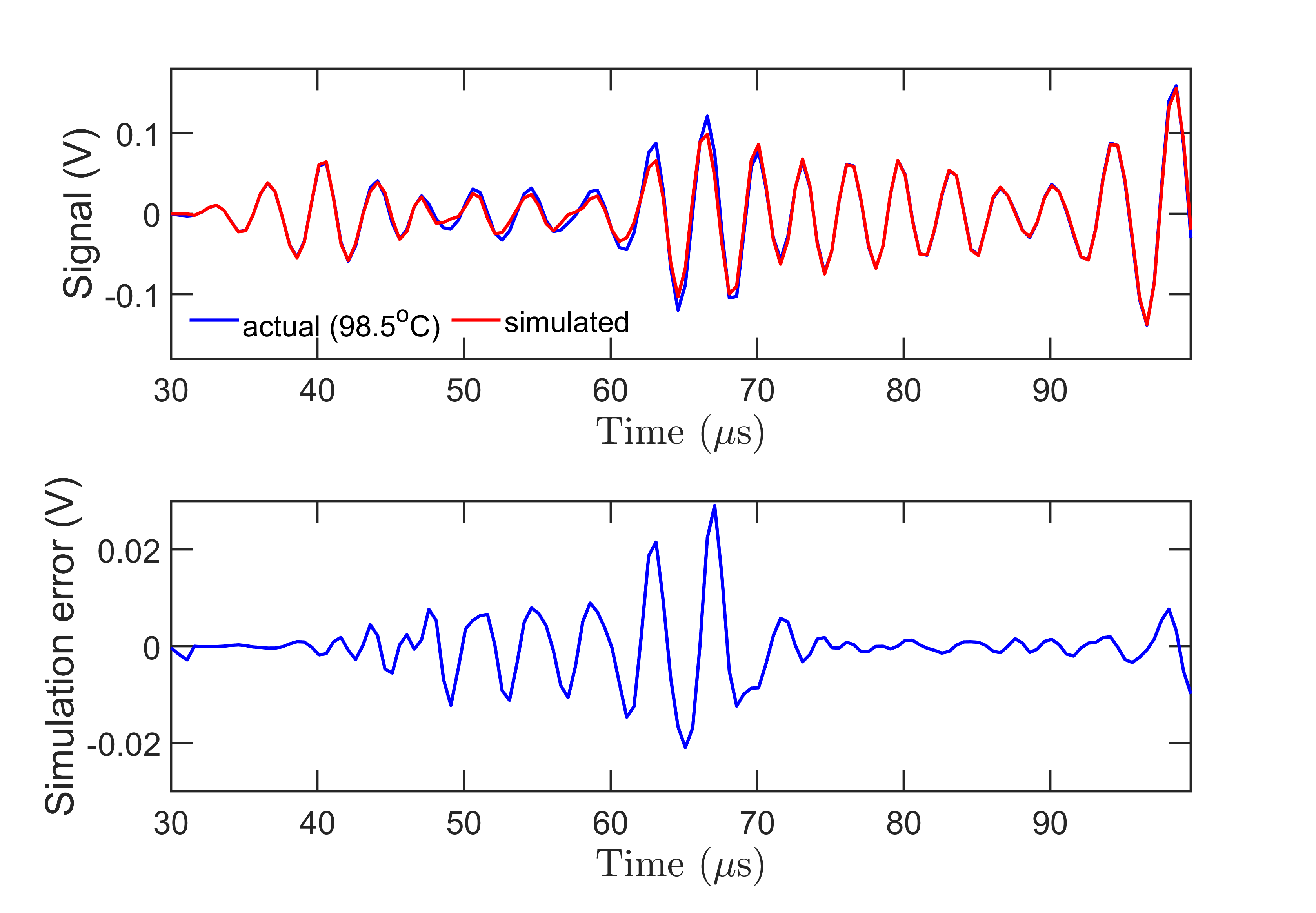}}
    
    \put(-55,140){ \large \textbf{(a)}}
    \put(200,140){ \large \textbf{(b)}}
    \put(-55,-50){\large \textbf{(c)}}   
    \put(200,-50){\large \textbf{(d)}}
    \end{picture}
    \vspace{8cm}
    \caption{ Simulated FEM signal after identifying a single model structure and interpolation (extrapolation) of the model parameter: comparison of simulation performed by RML-TARX$(4,4)_{0.5}$ and actual FEM simulation at (a) $29^o$C (extrapolation); (b) $62.5^o$C (interpolation); (c) $87^o$C (interpolation); and (d) $98.5^o$C (extrapolation).} 
\label{fig: FEM surrogate} \vspace{-12pt}
\end{figure}
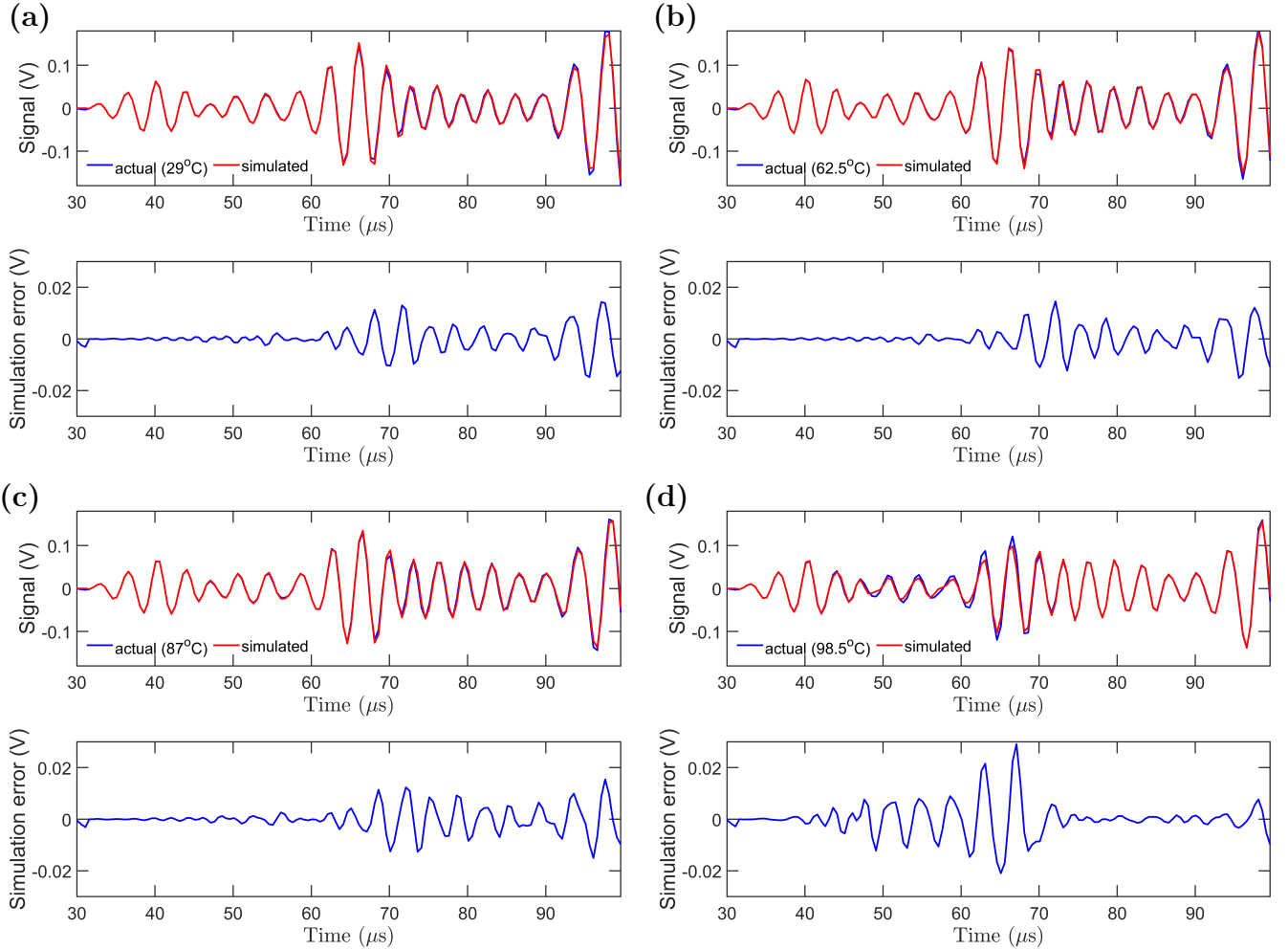 
%

In order to show that the RML-TARX models (after appropriate identification) can be used as surrogate models for FEM, four additional FEM simulations were performed at some random temperatures such as $29, 67.5, 87, 98.5^o$C. Then interpolation (or extrapolation) of the model parameters were performed at these temperatures and guided wave signals were simulated with the RML-TARX$(4,4)_{0.5}$ model. From Table \ref{tbl: simulation}, it can be observed that the ESS/SSS values at these temperatures are quite low except $98.5^o$C (far away from the temperature range) for three data sets. 

Figure \ref{fig: FEM surrogate}(a), (b), (c), and (d) show the comparison between the FEM simulations and RML-TARX-based simulations for guided wave propagation signals at temperatures $29^o$C, $67.5^o$C, $87^o$C, and $98.5^o$C, respectively. In order to perform RML-TARX based simulation at these specific temperatures, 7 data sets were used at temperatures $\{30, 40, 50 \cdots , 90^o \}$C. Then, these given FEM guided wave data sets were used to get the model parameters at these specified temperatures using the RML-TARX$(4,4)_{0.5}$ model. Once the model parameters are determined at these temperatures, a linear interpolation scheme was used to get model parameters at temperatures $\{25, 25.5, 26, 26.5 \cdots , 99.5, 100^o\}$C. Once the model parameters are available at these temperatures, simulation can be performed at any of these temperatures. It can be observed from Figure \ref{fig: FEM surrogate} that the RML-TARX-based simulation closely matches the FEM signals. 

The advantage of using the RML-TARX as surrogate models becomes evident when looked into the computational time. In order to simulate a single guided wave signal with FEM takes about 6 hours. However, an RML-TARX simulation takes about a fraction of a second. This is a significant reduction of computational time and will pave the way for online damage detection in the context of guided wave-based SHM.

\begin{table}[t]
\centering
\caption{ESS/SSS(\%) values obtained for finite element simulations using 3, 6, and 7 data sets}
{\small
\begin{tabular}[t]{lccccccccc}
\toprule
\multicolumn{1}{| c |}{Temperature} & \multicolumn{3}{| c |}{3 data sets} & \multicolumn{3}{| c |}{6 data sets}& \multicolumn{3}{| c |}{7 data sets}\\
\hline
&spline & linear & v5cubic & spline & linear & v5cubic& spline & linear & v5cubic\\
\midrule
25 & --&1.3390  & -- & 1.5434 & 1.1935 & -- & 1.5434 & 1.1935 &--\\
29 & --& 0.7225 & -- &0.8067  &0.7498  & -- & 0.8070 & 0.7498 &--\\
30 & \multicolumn{9}{c}{\textbf{data set used in the training phase}} \\
35 & --&1.5901 &1.3725 & 0.6614 & 0.8036 & 0.6796 & 0.6614 & 0.8036 &0.6796\\
40 & --&2.7707 &2.0829 & \multicolumn{6}{c}{\textbf{data set used in the training phase}} \\ 
45 & --&3.3093 & 1.8949 & 0.6576 & 0.7404 & 0.6611 & 0.6576 & 0.7404 &0.6611\\
50 & --&2.7646 & 1.1327 &  \multicolumn{6}{c}{\textbf{data set used in the training phase}} \\
55 &-- & 1.4776 & 0.5267 & 0.6096  & 0.6718 & 0.5994 & 0.6096 & 0.6718 & 0.5994\\
60 & \multicolumn{9}{c}{\textbf{data set used in the training phase}} \\
62.5 & -- & 1.5071&0.9229  & 0.5944 &0.6850  & 0.6015 & 0.5942 & 0.6850 & 0.6015\\
65 &-- &2.6686 & 1.3727 & 0.5883  & 0.7248  & 0.5785 & 0.5883 & 0.7248  & 0.5785\\
70 &--  &4.3863  & 2.3238 &  \multicolumn{6}{c}{\textbf{data set used in the training phase}} \\ 
75 &-- &4.7298 & 2.8079 & 0.6094 & 0.8699 & 0.6026  & 0.6094 & 0.8699 & 0.6026\\
80 &-- & 3.7379 & 2.4449 & \multicolumn{6}{c}{\textbf{data set used in the training phase}} \\ 
85 &-- &1.9546 & 1.3528  & 0.3038 & 0.8620 & -- & 0.7086 & 0.9873 & 0.7635\\
87 & --&1.2691 & 0.8912 & 0.3800 & 1.9170 & -- & 0.7334 & 0.8677 &0.7512 \\
90 & \multicolumn{3}{c}{\textbf{training data}}  & 42.4956 &  7.6313  &--& \multicolumn{3}{c}{\textbf{training data}} \\
95 & --&2.0214 & -- & -- & 116.0974 & -- & 0.6371  & 0.8573  & --\\
98.5 &-- &7.7010  & -- & -- &  & -- & 3.8549 &1.8318  &--\\
100 &-- &13.1786 & -- & -- &  & -- & 19.3018 & 3.2264 &--\\
\bottomrule
\end{tabular}}
\label{tbl: simulation}
\end{table}%

\section{Concluding Remarks} \label{sec:conclusions}

In this study, at first, a rigorous investigation on how guided wave propagation changes under varying temperatures inside an environmental chamber was performed. A series of laboratory experiments were conducted inside an environmental chamber from $50^o$C to $100^o$C with an increment of $5^o$C, and the corresponding temperature of the individual guided wave signal was also recorded. It was found that although the experiment was conducted within a controlled environment, still there is variation among the individual signal record. It was observed that with the increase in temperature, the signal shifts to the right, and the amplitude decreases with the exception of sudden increase at $65^o$C which may be due to the effect of adhesives that were used in mounting the PZTs. The amount of change of the signal due to the temperature variation is higher in the reflected part of the signal than in the $S_0$ and $A_0$ mode.

In addition, a high-fidelity finite element model was established to model the guided wave propagation under varying temperatures, more specifically, from $25^o$C to $100^o$C with an increment of $5^o$C. It was found that with the increase in temperature, the signal shifts to the right, similar to the experimental case. However, at the beginning, the amplitude of the signal increases up to $80^o$C for $S_0$ mode and then starts decreasing, which is in contrast to the experimental case. For $A_0$ mode, the amplitude of the signal decreases with increasing temperature.

After that, non-stationary time-varying time series models such as the RML-TAR and RML-TARX models were employed to model the non-stationary guided wave propagation signal. RML-TAR models are referred to as the unstructured parameter evolution methods as no restriction is imposed on the evolution of its time-varying parameters. In the context of guided-wave based acousto-ultrasound SHM, the RML-TAR models are used to capture the stochasticity in guided wave propagation under varying temperatures and the applicability, effectiveness and the benefits of the non-stationary RML-TAR models under varying temperatures are clearly demonstrated. Again the capability of the RML-TAR models for recovering the underlying dynamics and the response characteristics of the structure, based on a single non-stationary guided wave signal response are also demonstrated. A single RML-TAR model structure, RML-TAR$(6)_{0.6}$, was identified which can perform one-step ahead prediction of the guided wave signal at different temperatures. In addition, RML-TAR$(6)_{0.6}$-based ``frozen-time" FRF and modal parameter estimates are presented under varying temperatures. They are both in very good agreement with their non-parametric frozen-configuration counterparts. The RML-TAR$(6)_{0.6}$-based frozen-time FRF is of high resolution, smooth (in agreement with the physics of the problem), and clearly depicting the modal information. It is shown that the modal parameters shift to the right with the increase in temperature.

Based on the presented analysis on non-stationary guided wave signals under varying temperatures, the parametric non-stationary analysis methods are shown to achieve high performance characteristics and benefits that are not generally available by their non-parametric counterparts. Yet, the two families of methods should be largely viewed as complementary, as the use of non-parametric methods may guide the subsequent, more elaborate, use of their parametric counterparts in an actual application. In fact, this is often the recommended approach. The general price paid for the extra benefits of parametric methods is higher modeling complexity implying increased conceptual and computational complexity, as well as, user expertise. However, compared to the physics-based FEM model of guided wave propagation, this data-based modeling approach is much faster, computationally inexpensive, and easy to use.

In addition to performing one-step-ahead prediction with the RML-TAR model under varying temperatures, the simulation of guided wave propagation was also performed with the help of the RML-TARX model under varying temperatures. An appropriate RML-TARX model structure was selected at each different temperature following the ESS/SSS criteria to simulate guided wave signals at that temperature. A single model structure was also identified that can simulate guided wave signals at a wide range of temperature. The simulation includes $S_0$ and $A_0$ mode as well as the reflected part of the signal. That is, by using a data-driven model, it is possible to simulate an arbitrary length of the signal, while in the literature, simulation of only $S_0$ and $A_0$ mode was achieved under varying temperature. This is a significant achievement towards the modeling of guided wave propagation under EOC, in the current case, under temperature variation which significantly affects damage detection performance in the context of guided wave-based SHM.

It is also shown that by using only a limited number of data sets at a few selected temperatures, it is possible to simulate the guided wave propagation signal at temperatures where data is not available. This is accomplished by interpolating the model parameters at the required temperature. By performing FEM simulations and comparing with the RML-TARX-based simulation, it is concluded that RML-TARX models can be used as a surrogate for FEM simulations under varying temperatures. In addition, the RML-TARX-based simulations are extremely fast compared to the FEM simulations and computationally inexpensive, which can be leveraged to perform online SHM.

The RML-TARX-based simulation approach presented herein will pave the way for the postulation of an automated and online damage diagnosis scheme taking into account the stochastic nature of guided wave propagation. This will increase the robustness and subsequent reliability of the existing guided wave-based SHM technologies to perform damage detection under varying temperatures and to compensate the effect of temperature. This approach may also be used in the context of digital twin, where a computational model will be used to update and monitor the physical entity. In this respect, a low-fidelity, yet faster simulation may play a significant role. As a result, the RML-TARX-based simulation approach may be well suited in this context. In future work, these models will be used, in the context of guided wave-based SHM, to perform damage diagnosis of a real structures under uncertainty and under varying temperatures.

\section*{Acknowledgment}

This work is supported by the  U.S. Air Force Office of Scientific Research (AFOSR) grant ``Formal Verification of Stochastic State Awareness for Dynamic Data-Driven Intelligent Aerospace Systems'' (FA9550-19-1-0054) with Program Officer Dr. Erik Blasch.


\bibliographystyle{aiaa} 

\bibliography{references} 

\appendix

\section{Material properties and equations}

\begin{table}[t] \begin{minipage}{\columnwidth} 
\centering
\caption{Nominal material property values at $25 \ ^0$C}
\label{tbl:matprop}
\begin{tabular}{lll} \hline \hline
Materials & Property name & Values \\ \hline
 
Piezo-electric: PZT-5A & Density ($\rho$) & $7750$ $kg/m^3$ \\
 & Young's modulus (GPa) &$E_{11} =E_{22} = 60.97$\\ $E_{33}=53.19$  \\
 & Poisson ratio & $\nu_{13} = \nu_{23} = 0.4402, \nu_{12}=0.35$ \\
 & \makecell{Piezo-elctric charge constant (m/V)} & \makecell{$d_{31}= d_{32}= 171e-12, d_{33}=374e-12,$\\ $d_{15}=d_{24}=558e-12$} \\
 & Dielectric constant &$\epsilon_{11}=\epsilon_{22}=15.32e-9,\epsilon_{33}=15e-9$ \\
Aluminum & Density ($\rho$) ($Kg/m^3$)  & $2700$ \\
& Young's modulus ($E$ GPa)& $68.9$ \\
& Poisson ratio & $0.33$ \\
Adhesive & Density ($\rho$) ($Kg/m^3$)  & $1100$ \\
& Young's modulus ($E$ GPa) & $2.19$ \\
& Poisson ratio & $0.30$ \\ \hline \hline
\end{tabular} \end{minipage}
\end{table}

The established functional relationships are outlined in the following. 

Properties of piezoelectric materials:
$$\frac{\partial \rho} {\partial T} = 7751.80-7.26e-02T$$
$$E_{11}=E_{22} = 60.45+2.09e-02T$$
$$E_{33}=52.95+9.8e-03T$$
$$\nu_{13}=\nu_{23}=0.43+3e-04T-3e-06T^2-1e-09T^3$$
$$\nu_{12}=0.35+2e-04T-8e-07T^2+2e-09T^3$$
$$d_{31}=d_{32}=170.78-7.1e-03T+6e-04T^2+2e-16T^3$$
$$d_{33}=369.12+1.49e-01T+1.9e-03T^2-4e-09T^3$$
$$d_{15}=d_{24}=556+4.9e-02T+2e-06T^2-2e-09T^3$$
$$ \varepsilon_{11}=\varepsilon_{22}\\=14.9e-09+1.42e-11T+9.74e-14T^2+4.43e-17T^3$$
$$ \varepsilon_{33}\\=14.60e-09+1.47e-11T+1.18e-13T^2-5.31e-18T^3 $$

Properties of aluminum:
$$E_{Al} = 69.62-2.63e-02T$$ 
$$ \frac{\partial \rho} {\partial x} = 2794.60 - 1.84e-01T$$
$$ \nu_{Al} = 0.32+3e-04T$$
Properties of the adhesive:
$$E_{adh}=3.2-0.065T+1.18e-03T^2-7.72e-06T^3$$
$$G_{adh}=1+0.001T-4e-05T^2$$


\end{document}